\documentclass[final,5p,times,twocolumn]{elsarticle}
 \biboptions{comma,sort&compress}
\usepackage{graphicx}
\usepackage{here}
\usepackage[dvips]{epsfig}
\def\beq{\begin{equation}}
\def\eeq{\end{equation}}
\def\bea{\begin{eqnarray}}
\def\eea{\end{eqnarray}}
\def\bq{\begin{quote}}
\def\eq{\end{quote}}

\def\nnb{\nonumber}
\def\ga{\left(}
\def\dr{\right)}

\def\lrar{\Longrightarrow}
		
\def\nnb{\nonumber}
\def\la{\langle}
\def\ra{\rangle}
\def\nin{\noindent}
\def\ba{\vspace*{-0.2cm}\begin{array}}
\def\ea{\end{array}\vspace*{-0.2cm}}

\def\b{$\bullet~$}
\def\als{\alpha_s}

\def\sGs{\lag\bar{s} G s\rag}
\def\gg2{ \la\alpha_s G^2 \ra}
\def\gg3{g^3f_{abc}\la G^aG^bG^c \ra}
\def\ggg4{\la\als^2G^4\ra}

\def\beq{\begin{equation}}
\def\enq{\end{equation}}
\def\beqa{\begin{eqnarray}}
\def\enqa{\end{eqnarray}}
\def\nnb{\nonumber}

\def\qq{\lag\bar{q}q\rag}

\def\ss{\lag\bar{s}s\rag}
\def\qqs{\lag\bar{s}s\rag}
\def\mix{\lag\bar{q}g\si Gq\rag}
\def\mixs{\lag\bar{s}g\si Gs\rag}
\def\Gd{\lag G^2 \rag}
\def\gG{\lag g^2 G^2 \rag}
\def\GGG{\lag g^3G^3\rag}

\def\si{\sigma}

\def\al{\alpha}
\def\be{\beta}
\def\alma{\alpha_{max}}
\def\almi{\alpha_{min}}
\def\bemi{\beta_{min}}
\def\lb{\label}

\def\Fabs{m_Q^2(\al+\be)-\al\be s}
\def\Has{m_Q^2-\al(1-\al) s}
\newcommand{\rag}{\rangle}
\newcommand{\lag}{\langle}

\newcommand{\ImF}[1]{\mbox{$\:{\cal F}_{#1} $}}
\newcommand{\ImH}[1]{\mbox{$\:{\cal H}_{#1} $}}

\journal{Elsevier}

\begin{document}
\pagestyle{myheadings}
\begin{frontmatter}

\title{$1^{--}$ and $0^{++}$ heavy four-quark and molecule states in QCD}
 \author[label1,label2]{R.M. Albuquerque\corref{cor1}}
  \cortext[cor1]{FAPESP CNPq-Brasil PhD student fellow.}
\ead{rma@if.usp.br}
\address[label1]{Instituto de F\'{\i}sica, Universidade de S\~{a}o Paulo, 
C.P. 66318, 05389-970 S\~{a}o Paulo, SP, Brazil}
\address[label2]{Laboratoire
Particules et Univers de Montpellier, CNRS-IN2P3, 
Case 070, Place Eug\`ene
Bataillon, 34095 - Montpellier, France.}
 \author[label3]{F. Fanomezana\corref{cor2}}
  \cortext[cor2]{PhD student.}
\ead{fanfenos@yahoo.fr}
\address[label3]{Institute of High-Energy Physics of Madagascar (iHEP-MAD), University of Antananarivo, 
Madagascar}
 \author[label2]{S. Narison\fnref{fn1}}
   \fntext[fn1]{Corresponding author.}
    \ead{snarison@yahoo.fr}

 \author[label3]{A. Rabemananjara\corref{cor2}}
\ead{achris\_01@yahoo.fr}

\begin{abstract}
\nin
We estimate the masses  of the $1^{--}$ heavy  four-quark and molecule states by combining  exponential Laplace (LSR) and finite energy (FESR) sum rules known perturbatively to lowest order (LO) in $\alpha_s$  but including non-perturbative terms up to the complete dimension-six condensate contributions. This approach allows to fix  more precisely the value of the QCD continuum threshold (often taken ad hoc) at which the optimal result is extracted. 
 We use double ratio of sum rules (DRSR) for determining the $SU(3)$ breakings terms.  
We also study the effects of the heavy quark mass definitions on these LO results.  
 The $SU(3)$ mass-splittings of about (50 -- 110) MeV and the ones of about (250 -- 300) MeV between the lowest ground states and their 1st radial excitations
 are (almost) heavy-flavour  independent. 
 The mass predictions summarized in Table \ref{tab:res} are compared with the ones in the literature (when available) and with the three $Y_c(4260,~4360,~4660) $  and $Y_b(10890)$ $1^{--}$ experimental candidates. We conclude (to this order approximation) that the lowest observed state  cannot be a {\it pure} $1^{--}$ four-quark nor a {\it pure} molecule but may result from their mixings. We extend the above analyzes to the $0^{++}$ four-quark and molecule states which are  about (0.5-1) GeV heavier than the corresponding $1^{--}$ states, while the splittings between the $0^{++}$ lowest ground state and the 1st radial excitation is about (300-500) MeV. We complete the analysis by estimating the decay constants of the $1^{--}$ and $0^{++}$ four-quark states which are tiny and which exhibit a $1/M_Q$ behaviour. 
 Our predictions  can be further tested using some alternative non-perturbative approaches or/and at LHC$_b$ and  some other  hadron factories.
 \end{abstract}
\end{frontmatter}
\section{Introduction and a short review on the $1^{++}$ channel}
A large amount of  exotic hadrons which differ from the ``standard" $\bar cc$ chamonium and $\bar bb$ bottomium radial excitation states have been recently discovered in $B$-factories through $J/\psi\pi^+\pi^-$ and $\Upsilon\pi^+\pi^-$ processes and have stimulated different
theoretical interpretations.  Most of them have been assigned as four-quarks and/or molecule states \cite{SWANSON}.
In previous papers \cite{X1,X2}, some of us have studied,  using exponential QCD spectral sum rules (QSSR) \cite{SVZ}\,\footnote{For reviews, see e.g. \cite{SNB,RRY}.} and the double ratio of sum rules (DRSR) \cite{DRSR}\,\footnote{For some other successful applications, see 
\cite{SNGh,HBARYON,SNFORM}.}, the nature of the $X(3872)$ $1^{++}$ states found by Belle \cite{BELLEX}
and confirmed by Babar
\cite{BABARX}, CDF \cite{CDF} and D0 \cite{D0}.
If it is a $(cq)(\overline{cq})$ four-quark or $D-D^*$ molecule state, one finds for $m_c=1.23$ GeV \cite{X1} \footnote{The two configurations give almost a degenerate mass-value \cite{X2}.}:
\beq
{X_c}=(3925\pm 127)~{\rm MeV}~,
\lb{eq:x1}
\eeq
corresponding to a $t_c$-value common solution of the exponential Laplace (LSR) and Finite Energy (FESR) sum rules:
\beq
\sqrt{t_c}=(4.15\pm 0.03)~{\rm GeV}~,
\eeq
while in the $b$-meson channel, using $m_b=4.26$ GeV,  one finds\,\cite{X1}:
\beq
X_b=(10144\pm 104)~\rm {MeV}~~{\rm with}~~\sqrt{t_c}=(10.4\pm 0.02)~{\rm GeV},
\eeq
where a similar result has been found in \cite{ZHU} using another choice of interpolating current. 
However, in the case of the $X_c(3872)$, the previous two configurations
are not favoured by its narrow hadronic width ($\leq$ 2.3 MeV), which has lead some of us to propose that it could be, instead, a $\lambda-J/\psi$-type molecule \cite{X2} described by the current:
\beq
J_\mu^\lambda =\ga {g\over\Lambda}\dr^2_{\rm eff}(\bar c\lambda^a\gamma^\mu c)(\bar q\lambda_a\gamma_5q)~,
\eeq
where $\lambda_a$ is the colour matrix , while $g$ and $\Lambda$ are coupling and scale associated to an effective Van Der Vaals force. In this case, the narrow width of the $X_c$ is mainly due to the extra-gluon exchange which gives a suppression of the order $\alpha_s^2$ compared  to the two former configurations, if one evaluates this width using vertex sum rules. 
The corresponding mass is slightly lower than the one in Eq. (\ref{eq:x1}) \cite{X2}:
\beq
r\equiv {X_c^{\lambda}\over X_c^{mol}}=0.96\pm 0.03~~ \lrar X_c^{\lambda}=(3768\pm 127)~{\rm MeV}~.
\eeq
which (within the errors) also agree with the data. 
By assuming that the mass of the radial excitation $X'_Q\approx \sqrt{t_c}$, one can also deduce the mass-splitting:
\beq
 X'_c-X_c\simeq 225~{\rm MeV}\approx X'_b-X_b\simeq 256~{\rm MeV}~,
\eeq
which is much lower than the ones of ordinary charmonium and bottomium states:
\beq
{\psi}(2S)-{\psi}(1S)\simeq 590\approx {\Upsilon}(2S)-{\Upsilon}(1S)\simeq 560~{\rm MeV},
\eeq
and suggests a completely different dynamics for these exotic states. Comparing the previous results with the observed 
 $Z_b(10610)$ and $Z_b(10650)$ states whose quantum numbers have been assigned to be $1^{++}$, one can conclude
 that these observed states are heavier than the 1st radial excitation of the $X_b(10.14)$ expected from QSSR to lowest order in $\alpha_s$ \cite{X1}. 
\section{QCD Analysis of the $1^{--}$ and $0^{++}$ channels}
In the following, we extend the previous analysis to the case of the $1^{--}$ and $0^{++}$ channels and improve some existing  analysis from QCD (spectral) sum rules in the $1^{--}$ channel \cite{RAPHAEL1,RAPHAEL2}. The results will be compared with  the experimental $1^{--}$ candidate states: 
\beq
Y(4260),~~~~~~~ Y(4360), ~~~~~~~ Y(4660)~,~~~~~~~ Y_b(10890)
\label{eq:exp}
\label{eq:exp2}
\eeq
seen by Babar \cite{BABARY} and Belle \cite{BELLEY,BELLEYb} and which decay into $J/\psi\pi^+\pi^-$ and $\Upsilon\pi^+\pi^-$ around the $\Upsilon(5S)$ mass.
These states cannot be identified with standard $\bar cc$ charmonium and $\bar bb$ bottomium radial excitations and have been assigned in the literature to be four-quark or molecule states or some threshold effects.
\subsection*{\b QCD input parameters}
\nin
The QCD parameters which shall appear in the following analysis will be the charm and bottom quark masses $m_{c,b}$, the light quark masses $m_{d,s}$, 
the light quark condensates $\qq$ and $\ss$,  the gluon condensates $ \lag
g^2G^2\rag\equiv \la g^2G^a_{\mu\nu}G_a^{\mu\nu}\ra$ and $\la g^3G^3\ra\equiv \la g^3f_{abc}G^a_{\mu\nu}G^b_{\nu\rho}G^c_{\rho\mu}\ra$, the mixed condensate $\mix\equiv {\la\bar qg\sigma^{\mu\nu} (\lambda_a/2) G^a_{\mu\nu}q\ra}$ and the four-quark 
 condensate $\rho\la\bar qq\ra^2$, where
 $\rho$ indicates the violation of the four-quark vacuum 
saturation. Their values are given in Table {\ref{tab:param}} and we shall work with the running
light quark parameters:
\bea
{\bar m}_s(\tau)&=&{
{\hat m}_s \over \ga-\log{ \sqrt{\tau}\Lambda}\dr^{-2/{
\beta_1}}
}\nnb\\
{\la\bar qq\ra}(\tau)&=&-\hat \mu_q^3 \ga-\log{ \sqrt{\tau}\Lambda}\dr^{-2/{
\beta_1}}\nnb\\
{\la\bar qg\sigma Gq\ra}(\tau)&=&-M_0^2{\hat \mu_q^3} \ga-\log{ \sqrt{\tau}
\Lambda}\dr^{-1/{3\beta_1}}~,
\eea
where $\beta_1=-(1/2)(11-2n/3)$ is the first coefficient of the $\beta$ function 
for $n$ flavours; $\hat m_s$ and $\hat\mu_q$ are renormalization group invariant light quark mass and condensate \cite{FNR,TARRACH}.
{\scriptsize
\begin{table}[hbt]
\setlength{\tabcolsep}{0.2pc}
 \caption{
QCD input parameters. For the heavy quark masses, we use 
the range spanned
 by the running $\overline{MS}$ mass $\overline{m}_Q(M_Q)$  and the on-shell mass 
from QCD (spectral) sum rules compiled in pages
 602 and 603 of the book in \cite{SNB} and recently obtained in Ref. \cite{SNH10}. The values of $\Lambda$ and $\hat\mu_q$ have 
been obtained from $\alpha_s(M_\tau)=0.325(8)$ \cite{SNTAU} and from the running 
masses:  $(\overline{m}_u+\overline{m}_d)(2)=7.9(3)$ MeV \cite{SNmass}. The 
original errors have been multiplied by 2 for a conservative estimate of the 
errors.     }
    {\small
\begin{tabular}{lll}
&\\
\hline
Parameters&Values& Ref.    \\
\hline
$\Lambda(n_f=4)$& $(324\pm 15)$ MeV &\cite{SNTAU,BNP,PDG}\\
$\Lambda(n_f=5)$& $(194\pm 10)$ MeV &\cite{SNTAU,BNP,PDG}\\
$\hat m_s$&$(0.114\pm0.021)$ GeV &\cite{SNB,SNmass,PDG}\\
$m_c$&$(1.26\sim1.47)$ GeV &\cite{SNB,SNmass,SNHmass,PDG,SNH10,IOFFE}\\
$m_b$&$(4.17\sim4.70)$ GeV &\cite{SNB,SNmass,SNHmass,PDG,SNH10}\\
$\hat \mu_q$&$(263\pm 7)$ MeV&\cite{SNB,SNmass}\\
$\kappa\equiv \la \bar ss\ra/\la\bar uu\ra$& $(0.74\pm 0.06)$&\cite{HBARYON}\\
$M_0^2$&$(0.8 \pm 0.2)$ GeV$^2$&\cite{JAMI2,HEID,SNhl}\\
$\la\alpha_s G^2\ra$& $(7\pm 2)\times 10^{-2}$ GeV$^4$&
\cite{SNTAU,LNT,SNI,fesr,YNDU,SNHeavy,BELL,SNH10,SNG}\\
$\la g^3  G^3\ra$& $(8.3\pm 1.0)$ GeV$^2\times\la\alpha_s G^2\ra$&
\cite{SNH10}\\
$\rho\equiv \la \bar qq \bar qq\ra/\la \bar qq\ra^2$&$(2\pm 1)$&\cite{SNTAU,LNT,JAMI2}\\
\hline
\end{tabular}
}
\label{tab:param}
\end{table}
}
\subsection*{\b Interpolating currents}
\nin
We assume that the  $Y$  state is described either by the 
lowest dimension  (without derivative terms)  four-quark and molecule $\bar D_sD^*_s$ vector currents $J_\mu$ given in Tables \ref{tab:4q} and \ref{tab:mol}. Unlike the case of baryons where both positive and parity states can couple to the same operator \cite{MANNEL}, the situation is simpler here as the vector and axial-vector currents have a well-defined quantum numbers to which are associated  the $1^{--}$ (resp. $1^{++}$) states for the transverse part and the $0^{++}$ 
 (resp. $0^{--}$) states for the longitudinal part. In the case of four-quark currents, we can have two-types of lowest derivative vector operators which can mix through the mixing parameter $b$\,\footnote{The $1^{++}$ four-quark state described by the axial-vector current has been analyzed in \cite{X1,X2}.}. Another possible mixing can occur through the renormalization of operators \cite{TARRACH,KREMER} though this type of mixing will only induce an overall effect due to the anomalous dimension which will be relevant at higher order in $\alpha_s$ but will disappear in the ratio of sum rules used in this paper. 
For the molecule current, we choose the product of local bilinear current which has the quantum number of the corresponding meson state. In this sense, we have only an unique interpolating current.
Observed states can be a mixing of different states associated to each choice of operators and their selection can only be done through the analysis of their decays \cite{X2} but this is beyond the scope of this paper. 
\subsection*{\b The two-point function in QCD}
\nin
The two-point functions of the $Y_Q~(Q\equiv c,b)$ 
 (assumed to be a $1^{--}$ vector meson)  is defined as:
\bea
\Pi^{\mu\nu}(q)&\equiv&i\int d^4x ~e^{iq.x}\lag 0
|T[j^\mu(x){j^\nu}^\dagger(0)]
|0\rag\nnb\\
&=&-\Pi^{(1)}(q^2)(g^{\mu\nu}-{q^\mu q^\nu\over q^2})+\Pi^{(0)}(q^2){q^\mu
q^\nu\over q^2}~,
\lb{2po}
\eea
where $J^\mu$ are the interpolating vector currents given Tables \ref{tab:4q} and \ref{tab:mol}.
We assume that the  $Y$  state is described either by the 
lowest dimension  (without derivative terms)  four-quark and molecule $\bar D_sD^*_s$ currents given in Tables \ref{tab:4q} and \ref{tab:mol}.
The two invariants, $\Pi^{(1)}$ and $\Pi^{(0)}$, appearing in 
Eq.~(\ref{2po}) are independent and have respectively the quantum numbers 
of the spin 1 and 0 mesons. We can extract $\Pi_Q^{(1)}$ and $\Pi^{(0)}(q^2)$ or the corresponding  spectral functions from the complete expression of $\Pi^{\mu\nu}_Q(q)$ by applying respectively to  
it the projectors:
\beq
{\cal P}^{(1)}_{\mu\nu}= - \frac{1}{3} \left( g^{\mu\nu} - \frac{q^\mu q^\nu}{q^2} \right)~~~~~{\rm and}~~~~~{\cal P}^{(0)}_{\mu\nu}= {q_\mu q_\nu\over q^2}~.
\eeq
Due to its analyticity, the correlation function, $\Pi^{(1,0)}(q^2)$ in Eq.~(\ref{2po}), 
obeys the dispersion relation:
\beq
\Pi^{(1,0)}(q^2)={1\over\pi}\int_{4m_c^2}^\infty ds {{\rm Im}\:\Pi^{(1,0)}(s)\over s-q^2-i\epsilon}+\cdots \;,
\lb{ope}
\enq
where $\mbox{Im}\:\Pi^{(1,0)}(s)$ are the spectral functions.  
 The QCD expressions of these spectral functions are given in Tables \ref{tab:4q} and \ref{tab:mol}. $1/q^2$ terms discussed in \cite{CNZ,ZAK},which are dual
to higher order terms of the QCD series will not be included here as we work to leading order.

\begin{table*}[hbt]
{\small
\setlength{\tabcolsep}{2.pc}
\newlength{\digitwidth} \settowidth{\digitwidth}{\rm 0}
\catcode`?=\active \def?{\kern\digitwidth}
 \vspace*{-0.5cm}
\caption{QCD expression of the Four-Quark Spectral Functions to lowest order in $\alpha_s$ and up to dimension-six condensates: $Q\equiv c,b$ is the heavy quark field. }
\begin{tabular*}{\textwidth}{@{}l@{\extracolsep{\fill}}l}
\hline
&\vspace{-0.2cm}\\
{\bf
Current }& \hspace{0.5cm}$j^\mu_{4q} = \frac{\epsilon_{abc} \epsilon_{dec}}{\sqrt{2}}  \Bigg\{ \Big{[} \left( s^T_a {\cal C} \gamma_5 Q_b \right) 
	\left( \bar{s}_d \gamma^\mu \gamma_5 {\cal C} \bar{Q}^T_e \right) + 
	\left( s^T_a {\cal C} \gamma_5 \gamma^\mu Q_b \right) 
	\left( \bar{s}_d \gamma_5 {\cal C} \bar{Q}^T_e \right) \Big{]} + b  \Big{[} \left( s^T_a {\cal C} Q_b \right) 
	\left( \bar{s}_d \gamma^\mu {\cal C} \bar{Q}^T_e \right) + 
	\left( s^T_a {\cal C} \gamma^\mu Q_b \right) 
	\left( \bar{s}_d {\cal C} \bar{Q}^T_e \right) \Big{]} \Bigg\}$\\
	& \vspace{-0.2cm}\\
\boldmath$1^{--}$& {\bf Spectral function}~~~~\boldmath ${1\over\pi}\mbox{Im} \:\Pi^{(1)}(s)$\\
&\vspace{-0.4cm}\\
Pert&$-\frac{1}{3\cdot2^{10} \:\pi^6} 
	\int\limits^{\alpha_{max}}_{\alpha_{min}} \!\frac{d\alpha}{\alpha^3} 
	\int\limits^{1-\alpha}_{\beta_{min}} \!\frac{d\beta}{\beta^3} \:(1 \!-\! \al \!-\! \be) \left[ 
	2m_Q^2(1-b^2)(1\!-\!\al\!-\!\be)^2 \ImF{3}  - 3(1+b^2) (1 \!+\!  \al \!+\! \be) \ImF{4} + 12 b^2 m_Q m_s (1 \!-\! \al \!-\! \be)(\al+\be) \ImF{3} \right] ~,$ \\ 
${\qqs}$&$\frac{\qqs}{2^{5} \:\pi^4} 
	\int\limits^{\alpha_{max}}_{\alpha_{min}} \!\frac{d\alpha}{\al^2} \bigg{\{} 
	\frac{m_s(1+b^2) \:\al}{(1-\al)} \ImH{2} -
	\int\limits^{1-\al}_{\be_{min}} \!\frac{d\beta}{\beta^2} ~
	\bigg[ b^2 m_Q(1-\al-\be)(\al+\be) \ImF{2} + m_s \al\:\be ~\bigg( m_Q^2 \left( 5-\al-\be + b^2(3+\al+\be) \right)\ImF{1} + 2\ImF{2} \bigg) \bigg] ~\bigg{\}}~,$\\
${\la G^2\ra}$&$-\frac{\gG}{3^2 \cdot 2^{11} \:\pi^6} 
	\!\!\!\!\int\limits^{\alpha_{max}}_{\alpha_{min}} \!\!\!\!\frac{d\alpha}{\al}
	\!\!\int\limits^{1-\alpha}_{\beta_{min}} \!\!\frac{d\beta}{\beta^3} \Bigg{\{}
	2m_Q^4(1\!-\!b^2) \al(1\!-\!\al\!-\!\be)^3 - 3 m_Q^2 (1\!-\!\al\!-\!\be)\bigg[ 8\al(1 \!+\! \al \!+\! \be) - (1 \!-\! b^2) 
	\bigg(2 \!-\! \be + (6 \alpha \!-\! \beta ) (\alpha \!+\! \beta )  \bigg)\bigg] \ImF{1} + 6 (1 \!+\! b^2)\be \left( 1 \!-\! 2\al \!-\! 2\be \right)\ImF{2} \Bigg{\}}~,$\\
$\sGs$&$\frac{\sGs}{3 \cdot 2^{7} \:\pi^4} \Bigg{\{}
	3m_Q \!\!\int\limits^{\alpha_{max}}_{\alpha_{min}} \!\!\frac{d\al}{\al} \!\!\int\limits^{1-\al}_{\be_{min}} \!\!\frac{d\be}{\be^2}
	\bigg[ 2\be(1 \!-\! 2\al \!-\! 2\be) + (1 - b^2) \bigg( 2\al(\al+\be) - \be(1 \!-\! 3\al \!-\! 3\be) \bigg) \bigg] \ImF{1} - 
	m_s \int\limits^{\alpha_{max}}_{\alpha_{min}} \!\!{d\al} \Bigg{[} 
	\frac{2}{\al} \bigg( 8m_Q^2\al(1+b^2) + \left( 1-\al +b^2(1-7\al) \right) \ImH{1} \bigg)$\\
	& \hspace{1cm} $  + \int\limits^{1-\al}_{\be_{min}} \!\!\frac{d\be}{\be} 
	\bigg( m_Q^2 \left( 6 \!+\! 3\al \!-\! 5\be + b^2(6\!-\!3\al\!+\!5\be) \right) + 3(7+b^2)\ImF{1} \bigg)  \Bigg] ~\Bigg\} $\\
${\qqs^2}$&$-\frac{\rho\qqs^2}{3 \cdot 2^3 \:\pi^2} 
	\int\limits^{\alpha_{max}}_{\alpha_{min}} \!d\alpha
	\left[ 4m_Q^2 - (1-b^2)( 2m_Q^2 - \ImH{1}) + m_s m_Q b^2\right] $\\
${\la G^3\ra}$& $\frac{\GGG}{3^3\cdot 2^{11} \:\pi^6}  \Bigg{\{}
	3 \int\limits^{\alpha_{max}}_{\alpha_{min}} \!\frac{d\alpha}{\al} \int\limits^{1-\al}_{\be_{min}} \!\frac{d\be}{\be^3} 
	(1\!-\! \al \!-\! \be) \Bigg[ m_Q^2 \Bigg( 12\al^2(1 \!+\! \al \!+\! \be) - (1-b^2) 
	\bigg( 3\al(1 \!+\! 3\al^2 \!+\! 3\be^2) + 2\be \left(14 \!+\! 4\al \!+\! 11\al^2 - (1+\al)(9 \!+\! 4\al \!+\! 9\be) \right) \bigg) \Bigg) $ \\
	& \hspace{1cm} $ + 3\al(1+b^2)(1 \!+\! \al \!+\! \be)\ImF{1} \Bigg] 
	+ m_Q^4(1-b^2) \int\limits^{1}_{0} \!\!\frac{d\al}{\al} \int\limits^{1}_{0} \!\!\frac{d\be}{\be^4} (1 \!-\! \al \!-\! \be)^2(3\al+4\be) 
	~\delta\left(s - \frac{(\al+\be)}{\al \be} m_Q^2 \right)  \Bigg\} ~.$\\
	&\\

\boldmath$0^{++}$& {\bf Spectral function}~~~~\boldmath ${1\over\pi}\mbox{Im} \:\Pi^{(0)}(s)$\\
&\vspace{-0.4cm}\\
Pert&$-\frac{1}{3\cdot2^{10} \:\pi^6} 
	\int\limits^{\alpha_{max}}_{\alpha_{min}} \!\frac{d\alpha}{\alpha^3} 
	\int\limits^{1-\alpha}_{\beta_{min}} \!\frac{d\beta}{\beta^3} \:(1-\al-\be) \Bigg[ 
	12m_Q^4(1- b^2)(\al+\be)(1 \!-\! \al \!-\! \be)^2 \ImF{2}  - 2m_Q^2 (1 \!-\! \al \!-\! \be) \bigg( 7 \!-\! 19\al \!-\! 19\be -
	b^2 (7+5\al+5\be) \bigg) \ImF{3} $\\
	& \hspace{1cm} $ - 3(1+b^2)(7 \!-\! 9\al \!-\! 9\be) \ImF{4} + 12 b^2 m_Q m_s (1 \!-\! \al \!-\! \be)^2 (\alpha+\beta) 
	\bigg(6 m_Q^2 (\al+\be) \ImF{2} - 7 \ImF{3} \bigg) \Bigg] ~,$ \\ 
${\qqs}$&$-\frac{\qqs}{2^{5} \:\pi^4} 
	\int\limits^{\alpha_{max}}_{\alpha_{min}} \!\frac{d\alpha}{\al^2} \bigg{\{} 
	\frac{m_s(1+b^2) \:\al}{(1-\al)} \ImH{2} +
	\int\limits^{1-\al}_{\be_{min}} \!\frac{d\beta}{\beta^2} ~
	\bigg[ b^2 m_Q(1-\al-\be)(\al+\be) \bigg( 4m_Q^2 (\al+\be) \ImF{1} - 5\ImF{2} \bigg) $\\
	& \hspace{1cm} $ + m_s \al\:\be ~\bigg( 2m_Q^4(1-b^2)(1-\al-\be)(\al+\be) - 
	m_Q^2 \left( 7-11\al-11\be + b^2(1+3\al+3\be) \right)\ImF{1} - 10\ImF{2} \bigg) \bigg] ~\bigg{\}}~,$\\
${\la G^2\ra}$&$-\frac{\gG}{3^2 \cdot 2^{11} \:\pi^6} \Bigg\{
	\int\limits^{\alpha_{max}}_{\alpha_{min}} \!\!\frac{d\alpha}{\al}
	\int\limits^{1-\alpha}_{\beta_{min}} \!\!\frac{d\beta}{\beta^3} \Bigg{[}
	2m_Q^4(1\!-\!\al\!-\!\be)^2 \bigg( 24\al(\al+\be) + (1\!-\!b^2) \bigg( \al(5\!-\!17\al\!-\!17\be) + 3 \be (2\!+\!\al\!+\!\be) \bigg) \bigg) $\\
	& \hspace{1cm} $ - 3 m_Q^2 (1 \!-\! \al \!-\! \be) \bigg( 8\al(1 \!-\! 3\al \!-\! 3\be) - 32\be(\al +\be) + (1-b^2) \bigg( 
	6 - 2\al(8 \!-\! 9\al \!-\! 9\be) - \be(3 \!-\! 13\al \!-\! 13\be) \bigg) \bigg) \ImF{1} $\\
	& \hspace{1cm} $ - 6(1+b^2) \be (9 \!-\! 10\al \!-\! 10\be) \ImF{2} \Bigg] 
	- 4m_Q^6 (1-b^2) \int\limits^{1}_{0} \!\!\frac{d\alpha}{\al}	\!\!\int\limits^{1}_{0} \!\!\frac{d\beta}{\beta^4} 
	(\al+\be)(1 \!-\! \al \!-\! \be)^3 ~\delta\left(s - \frac{(\al+\be)}{\al \be} m_Q^2 \right) \Bigg\}
	~,$\\
$\sGs$&$\frac{\sGs}{3 \cdot 2^{7} \:\pi^4} \Bigg\{
	3m_Q \int\limits^{\alpha_{max}}_{\alpha_{min}} \!\!\frac{d\al}{\al} \int\limits^{1-\al}_{\be_{min}} \!\!\frac{d\be}{\be^2} 
	\bigg[ \bigg(2m_Q^2 (\al+\be) - 3 \ImF{1} \bigg)	\bigg( 2\be(1 \!-\! 2\al \!-\! 2\be) - (1-b^2) \left(2\al(1 \!-\! \al \!-\! \be) + 
	\be(1 \!-\! 3\al \!-\! 3\be) \right) \bigg) - 2\al(1-b^2) \ImF{1} \bigg]$\\
	& \hspace{1cm} $ - m_s \int\limits^{\alpha_{max}}_{\alpha_{min}} \!\!{d\al} \Bigg{[} 
	\frac{2}{\al} \bigg( 2m_Q^2\al(1-5b^2) - \left( 2 - (1-b^2)(1-9\al) \right) \ImH{1} \bigg) +
	\int\limits^{1-\al}_{\be_{min}} \!\!\frac{d\be}{\be} \Bigg( m_Q^2 \left(4(3 \!-\! 4\al \!-\! 4\be) - (1-b^2)(3\!+\!\al\!+\!3\be) \right) $\\
	&\hspace{1cm} $ + 3(7+b^2)\ImF{1} \Bigg)  \Bigg] 
	- 2 m_s m_Q^4(1-b^2) \int\limits^{1}_{0} \!\!\frac{d\al}{\al} \int\limits^{1}_{0} \frac{d\be}{\be^2} ~
	(3 \!-\! 3\al \!-\! 5\be) (\al+\be) ~\delta\left(s - \frac{(\al+\be)}{\al \be} m_Q^2 \right)  \Bigg\}~,$\\
${\qqs^2}$&$\frac{\rho\qqs^2}{3 \cdot 2^3 \:\pi^2} \Bigg\{
	\int\limits^{\alpha_{max}}_{\alpha_{min}} \!d\alpha
	\bigg[ 4m_Q^2 - (1-b^2)(4m_Q^2 -3 \ImH{1}) + m_s m_Q b^2 \bigg] 
	+ 2m_s m_Q^3 b^2 \int\limits^{1}_{0} \!\frac{d\alpha}{\al(1-\al)} ~\delta\left(s - \frac{m_Q^2}{\al(1-\al)} \right)  \Bigg\}~,$\\
${\la G^3\ra}$& $-\frac{\GGG}{3^3\cdot 2^{11} \:\pi^6}  \Bigg{\{}
	3 \int\limits^{\alpha_{max}}_{\alpha_{min}} \!\frac{d\alpha}{\al^2} \int\limits^{1-\al}_{\be_{min}} \!\frac{d\be}{\be^3} 
	(1\!-\! \al \!-\! \be) \Bigg[ m_Q^2 \Bigg( 12\al^2 \left( \al(2-\be) + \be(1-\be) \right) - (1-b^2) \left( 3\al^2 (1 \!+\! \al)^2
	+\be^2(10-\al(5\al+8) - 2\be(6-\be)) \right) \Bigg) $ \\
	& \hspace{1cm} $ - 3\al^2(1+b^2)(1 \!-\! 3\al \!-\! 3\be)\ImF{1} \Bigg] 
	+ m_Q^4 \int\limits^{1}_{0} \!\!\frac{d\al}{\al^2} \int\limits^{1}_{0} \!\!\frac{d\be}{\be^5} (1 \!-\! \al \!-\! \be)^2
	~\delta\left(s - \frac{(\al+\be)}{\al \be} m_Q^2 \right) \bigg[ 72 \:\al^2 \be(\al+\be) $\\
	& \hspace{1cm} $+  (1-b^2) \:\be\left( 3(7-19\al) \al^2 +2(6-5\al)\be^2 + (34-91\al) \al\be \right)
	-2m_Q^2 \tau (1-b^2)(1-\al-\be) (\al+\be)(3\al+4\be) \bigg] \Bigg\} ~.$\\
	with : &$\ImF{k} = \left[ \Fabs \right]^k,~~\ImH{k} = \left[ \Has \right]^k ~,~~ 
 \bemi={\al m_Q^2/( s\al-m_Q^2)}~,$\\
 &$\almi={1\over 2}({1-v}),~~~~ \alma={1\over 2}({1+v})~, ~~v$ the $Q$-quark velocity:
$v\equiv  \sqrt{1-4m_Q^2/s}~~{\rm and}~~z \equiv \frac{m_Q^2 (\al+\be)}{\al\be}$\\
&\\
\hline
\end{tabular*}
\label{tab:4q}}
\end{table*}
\nin
\begin{table*}[hbt]
\setlength{\tabcolsep}{1.pc}
\caption{QCD expression of the Molecule Spectral Functions to lowest order in $\alpha_s$ and up to dimension-six condensates: $Q\equiv c,b$ is the heavy quark field, while $g'$ and $\Lambda'$ are coupling and scale associated to an effective Van Der Vaals force.  }
\begin{tabular*}{\textwidth}{@{}l@{\extracolsep{\fill}}l}
&\\
\hline
&\\
{\bf
Current }&$j^\mu_{mol}={1\over \sqrt{2}}\ga {g'\over\Lambda'^2}\dr^2_{\rm eff}\Big{[} \ga\bar s\gamma^\mu Q\dr \ga \bar Q s\dr+\ga\bar Q\gamma^\mu s\dr\ga\bar s Q\dr\Big{]}$\\
	&\\
\boldmath$1^{--}$& {\bf Spectral function}~~~~\boldmath ${1\over\pi}\mbox{Im} \:\Pi^{(1)}(s)$\\
&\\
Pert&
\small $-\frac{1}{2^{12} \:\pi^6} 
	\int\limits^{\alpha_{max}}_{\alpha_{min}} \!\frac{d\alpha}{\alpha^3} 
	\int\limits^{1-\alpha}_{\beta_{min}} \!\frac{d\beta}{\beta^3}
	(1-\al-\be) {\cal F}_3
	\left[ 2 m_Q^2 \left(1 - \al - \be \right)^2 -3(1+\al+\be) {\cal F}_1 \right]$~, \\
${\qqs}$&\small $\frac{3m_s \qqs}{2^{7} \:\pi^4} \Bigg{\{}
	\int\limits^{\alpha_{max}}_{\alpha_{min}} \!\frac{d\alpha}{\al(1-\al)} \ImH{2}- 
	\int\limits^{\al_{max}}_{\al_{min}} \!\frac{d\al}{\al} \int\limits^{1-\al}_{\be_{min}} \!\frac{d\beta}{\beta}
	{\cal F}_1 \left[ m_Q^2(5-\al-\be) + 2 {\cal F}_1 \right] \Bigg{\}}$~,\\
${\Gd}$&\small $-\frac{\gG}{3 \cdot 2^{12} \:\pi^6} 
	\int\limits^{\alpha_{max}}_{\alpha_{min}} \!\frac{d\alpha}{\al}
	\int\limits^{1-\alpha}_{\beta_{min}} \!\frac{d\beta}{\beta^3} \Bigg{\{}
	m_Q^4 \al(1-\al-\be)^3  +	3 m_Q^2 (1-\al-\be) [ 1- \al(4+\al+\be) + \be(1-2\al-2\be) ] {\cal F}_1
	+6\be \left( 1-2\al-2\be \right) {\cal F}_2 \Bigg{\}}$~,\\
$\sGs$&\small$\frac{\mixs}{2^{8} \:\pi^4}
	\Bigg\{
	3 m_Q \!\!\int\limits^{\alpha_{max}}_{\alpha_{min}} \!\!\!\frac{d\alpha}{\al^2}
	\!\!\int\limits^{1-\al}_{\be_{min}} \!\!\!\frac{d\be}{\be} \bigg( \al^2 \!-\! \al(1 \!+\! \be) - 2\be^2 \bigg)  \!\ImF{1} 
	-m_s \Bigg[ \int\limits^{\alpha_{max}}_{\alpha_{min}} \!\!\!\frac{d\alpha}{\al} 
	\bigg( 8 m_Q^2 \al + (2 \!-\! \al) \ImH{1} \bigg) -\!\!\!\! \int\limits^{\alpha_{max}}_{\alpha_{min}} \!\!\!\!d\alpha
	\!\!\!\int\limits^{1-\al}_{\be_{min}} \!\!\!\frac{d\be}{\be} 
	\bigg( m_Q^2 (9 \!-\! 3\al \!-\! 4\be) + 7 \!\ImF{1} \!\bigg) \Bigg]  \Bigg\}$~,\\	
${\qqs^2}$&\small$-\frac{\rho\qqs^2}{2^{6} \:\pi^2} 
	\int\limits^{\alpha_{max}}_{\alpha_{min}} \!d\alpha
	\left[ 3m_Q^2 - \al(1-\al)s \right]$~,\\
${\la G^3\ra}$&\small$-\frac{\GGG}{5\cdot3 \cdot 2^{16} \:\pi^6} \Bigg{\{}
	5\int\limits^{\alpha_{max}}_{\alpha_{min}} \!d\alpha
	\int\limits^{1-\al}_{\be_{min}} \!\frac{d\be}{\be^3} (1-\al-\be) \Bigg[ m_Q^2 \bigg(5 \alpha^2 - \alpha (37 - 19 \beta) +\!\! 14 (1 - \beta)^2 \bigg) - 3 (7 + 9 \alpha + 9 \beta) \ImF{1} \Bigg] $\\
	&\small$+ m_Q^4 \int\limits^{1}_{0} \!d\alpha
	\int\limits^{1}_{0} \!\frac{d\be}{\be^5} ~e^{-\frac{(\al+\be)}{\al\be} m_Q^2 \tau} (1-\al-\be) \Bigg[ 2 m_Q^2 \tau (1 - \alpha - \beta)^2 -\beta \bigg( 50 \alpha^2 \!-\! \alpha (61 \!-\! 85 \beta) \!+\! 
	35 (1 \!-\! \beta)^2 \!\bigg) \Bigg] \Bigg\}$~.\\	
&\\
\boldmath$0^{++}$& {\bf Spectral function}~~~~\boldmath ${1\over\pi}\mbox{Im} \:\Pi^{(0)}(s)$\\
&\\
Pert&
\small$-\frac{1}{2^{12} \:\pi^6} 
	\int\limits^{\alpha_{max}}_{\alpha_{min}} \!\!\!\frac{d\alpha}{\alpha^3} 
	\int\limits^{1-\alpha}_{\beta_{min}} \!\!\!\frac{d\beta}{\beta^3} (1-\al-\be)  \Bigg{[} 12 m_Q^4 (\al+\be) (1-\al-\be)^2 \ImF{2} 
	-\!\! 2m_Q^2 (1-\al-\be)(7-19\al-19\be) \ImF{3} 
	-\!\! 3(7-9\al-9\be) \ImF{4} \Bigg{]}~,$\\	
${\qqs}$&\small $-\frac{3m_s \qqs}{2^{7} \:\pi^4} \Bigg{\{}
	\int\limits^{\alpha_{max}}_{\alpha_{min}} \!d\alpha\frac{\ImH{2}}{\al(1-\al)} + \int\limits^{\al_{max}}_{\al_{min}} \!\!\!\frac{d\al}{\al} 
	\int\limits^{1-\al}_{\be_{min}} \!\!\!\frac{d\beta}{\beta}
	\Bigg[ 2m_Q^4(\al+\be) (1\!-\! \al \!-\! \be)  - m_Q^2 (7-11\al-11\be) \ImF{1} - 10\ImF{2} \Bigg] \Bigg{\}}~,$\\
${\Gd}$&\small $-\frac{\gG}{3 \cdot 2^{12} \:\pi^6} \Bigg{\{}
	\int\limits^{\alpha_{max}}_{\alpha_{min}} \!\frac{d\alpha}{\al}
	\int\limits^{1-\alpha}_{\beta_{min}} \!\frac{d\beta}{\beta^3}  \Bigg[ m_Q^4 (1\!-\! \alpha \!-\! \beta)^2 \bigg( 7 \alpha^2 + \alpha (5 + 19 \beta) + 6 \beta (1 + 2 \beta) \bigg)  $\\
	&\small$ +3 m_Q^2 (1 - \alpha - \beta)  \bigg( 3 \alpha^2 \!+\! \alpha (4 \!+\! 25 \beta) \!-\! 
	\beta (3 \!-\! 22 \beta) \!-\! 3 \bigg)\ImF{1}  - \!\! 6 \beta (9 \!-\! 10 \alpha \!-\! 10 \beta) \ImF{2} \Bigg] - 2 m_Q^6 \int\limits^{1}_{0} 
	\!\!d\alpha \!\!\int\limits^{1}_{0} \!\!d\beta \bigg[\frac{(\al+\be)(1-\al-\be)^3}{\al \beta^4}\bigg] 
	~e^{-\frac{(\al+\be)}{\al\be} m_Q^2 \tau} \Bigg{\}}~,$\\
$\sGs$&\small$-\frac{\mixs}{2^{8} \:\pi^4} \Bigg{\{}
	3m_Q \int\limits^{\alpha_{max}}_{\alpha_{min}} \!\frac{d\alpha}{\al^2}
	\int\limits^{1-\al}_{\be_{min}} \!\frac{d\be}{\be} \Bigg{[} 2m_Q^2(1-\al-\be)(\al+\be)(\al-2\be) +
	 \bigg(3\al^2-3\al(1+\be) + 2\be(2-3\be) \bigg) \!\!\ImF{1} \Bigg]$\\
	&\small $+m_s \int\limits^{\alpha_{max}}_{\alpha_{min}} \!\frac{d\alpha}{\al} 
	\left[ 2m_Q^2 \al  - (2+9\al) \ImH{1} \right]  + m_s \!\!\int\limits^{\alpha_{max}}_{\alpha_{min}} \!\!\!d\alpha
	\int\limits^{1-\al}_{\be_{min}} \!\!\!\frac{d\be}{\be} 
	\bigg( m_Q^2 (9 \!-\! 17\al \!-\! 18\be) +\! 21 \ImF{1} \bigg) + 2m_s m_Q^4 \!\int\limits^{1}_{0} 
	\!\!d\alpha \!\!\int\limits^{1}_{0} \!\!d\beta  \frac{(\al+\be)(3-3\al-4\be)}{\al \beta^2}~e^{-\frac{(\al+\be)}{\al\be} m_Q^2 \tau}  \Bigg{\}}~,$\\
${\qqs^2}$&\small$\frac{3\rho\qqs^2}{2^5 \:\pi^2} 
	\int\limits^{\alpha_{max}}_{\alpha_{min}} \!d\alpha
	\left[ m_Q^2 - \al(1-\al)s \right]~,$\\
${\la G^3\ra}$&\small$ \frac{\GGG}{5\cdot3\cdot 2^{16} \:\pi^6}  \Bigg\{
	5 \!\!\int\limits^{\alpha_{max}}_{\alpha_{min}} \!\!\!d\alpha
	\!\!\int\limits^{1-\al}_{\be_{min}} \!\!\!\frac{d\be}{\be^3} (1 \!-\! \al \!-\! \be) 
	\Bigg[ m_Q^2 \bigg( 59 \alpha^2 - \alpha (91 - 127 \beta) +14 -
	 2\be(41 \!-\! 34 \beta) \bigg) \!+\! (33 \!-\! 81 \alpha \!-\! 81 \beta) \ImF{1} \Bigg] $\\
	 &\small $	+
	  m_Q^4 \int\limits^{1}_{0} \!\!\frac{d\al}{\al} 
	\!\!\int\limits^{1}_{0} \!\!\frac{\:d\be}{\be^6} (1 \!-\! \al \!-\! \be) ~ 
	e^{-\frac{(\al+\be)}{\al\be} m_Q^2\tau}
	 \Bigg{[} 4 m_Q^4 \tau^2 (\al \!+\! \be) (1 \!-\! \al \!-\! \be)^2 + 2 m_Q^2 \tau \be  (1- \al- \be) \bigg( 53 \al^2 - \al (38 - 88 \be) -\! 35 \beta (1 \!-\! \beta) \!\bigg) $\\
	&\small$+ \beta^2 \bigg( \!100 \alpha^3 \!+\! 140 \beta (1 \!-\! \beta)^2 -\!\! 13 \alpha^2 (23 \!-\! 25 \beta) \!+\! 
	5 \alpha (1 \!-\! \beta) (35 \!-\! 73 \beta) \bigg) \Bigg] \Bigg\}~. $\\

 &\\
\hline
\end{tabular*}
\label{tab:mol}
\end{table*}
\nin
\section{$1^{--}$  four-quark state mass $Y_{Qq}$ from QSSR}
In the following, we shall estimate the mass of the $1^{--}$ four-quark state $(\overline{Qq}) (Qq)$ ($Q\equiv c,~b$ and $q\equiv u,~d$ quarks), hereafter denoted by $Y_{Qd}$. In so doing, we shall use the ratios of the Laplace (exponential) sum rule:
\beq
{\cal R}^{LSR}_{Qd}(\tau)\equiv -{d\over d\tau}{\rm Log}{\int_{t_<}^{t_c} dt e^{-t\tau }{1\over\pi}{\rm Im}\Pi^{(1)}(t)}~,
\label{eq:lsr}
\eeq
and of FESR:
\beq
 {\cal R}_{Qd}^{FESR}\equiv {\int_{t_<}^{t_c}dt ~t^n{1\over\pi}{\rm Im}\Pi^{(1)}(t)\over \int_{t<}^{t_c} dt ~t^{n-1} {1\over\pi}{\rm Im}\Pi^{(1)}(t)}~~~~:~~ n=1~,
 \label{eq:fesr}
\eeq
where $t_<$ is the hadronic (quark) threshold. 
Within the usual duality ansatz ``one resonance" + $\theta(t-t_c)\times QCD~continuum$ parametrization of the spectral function, the previous ratios of sum rules give:
\beq
{\cal R}^{LSR}_{Qd}(\tau)\simeq M_{Y_{Qd}}^2 \simeq {\cal R}_{Qd}^{FESR}~.
\eeq
For a discussion more closed to the existing literature which we shall test the reliability in the following, we start to work with  the current corresponding to  $b=0$. We shall discuss the more general choice of current when $b$ is a free parameter at the end of this section.
\subsection*{\b The $Y_{cd}$ mass from LSR and FESR for the case $b$=0}
\nin
\begin{figure}[hbt] 
\begin{center}
\centerline {\hspace*{-7cm} a) }\vspace{-0.6cm}
{\includegraphics[height=30mm]{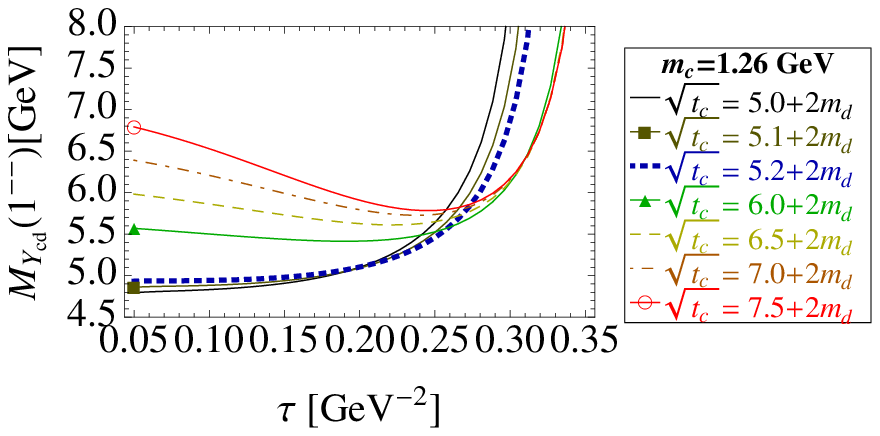}}
\centerline {\hspace*{-7cm} b) }\vspace{-0.6cm}
{\includegraphics[height=30mm]{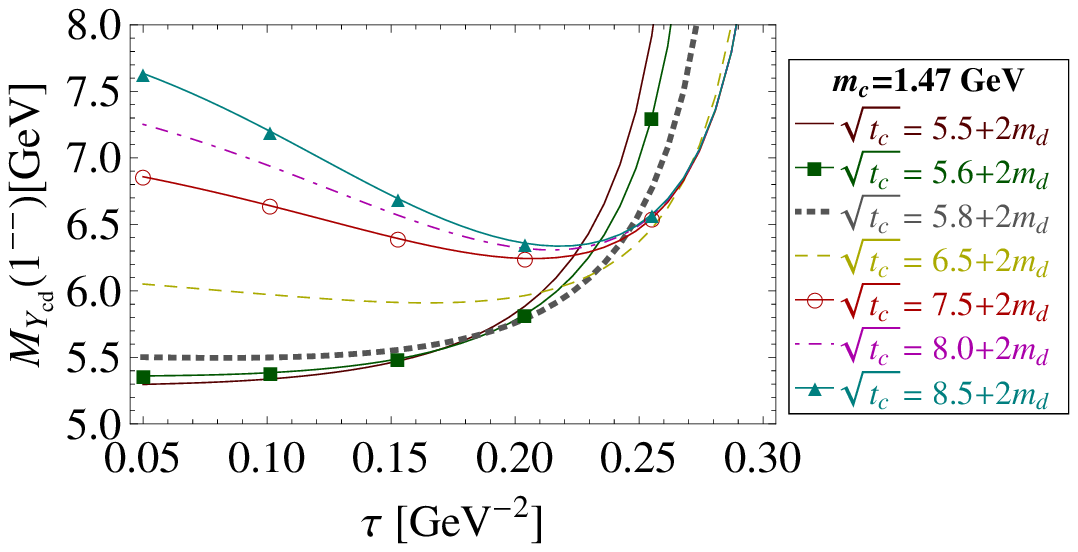}}
\centerline {\hspace*{-7cm} c) }\vspace{-0.6cm}
{\includegraphics[height=30mm]{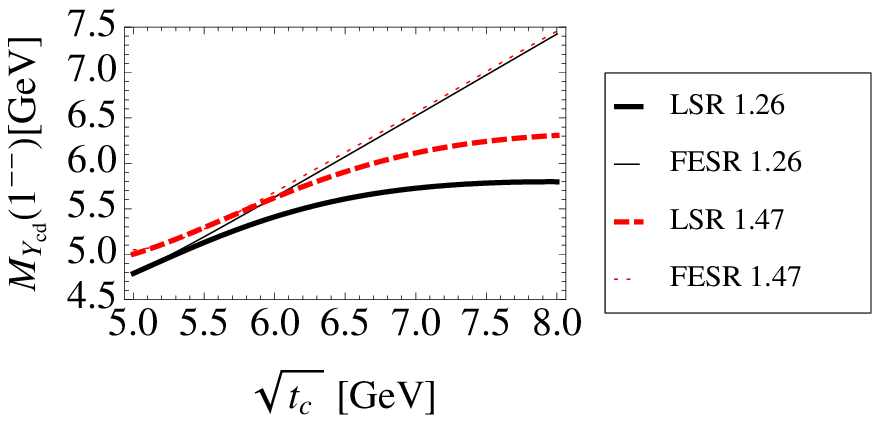}}
\caption{\scriptsize 
{\bf a)} $\tau$-behaviour of $M_{Y_{cd}}(1^{--})$ from ${\cal R}_{cd}^{LSR}$  for the current mixing parameter $b=0$, for different values of $t_c$
and for $m_c=1.26$ GeV;  {\bf b)} The same as {\bf a)} but for $m_c=1.47$ GeV;  {\bf c)} $t_c$-behaviour of the LSR results obtained at the $\tau$-stability points and comparison with the ones from  $ {\cal R}^{FESR}_{cd}$ for  $m_c=1.26$ and 1.47 GeV.}
\label{fig:yc} 
\end{center}
\end{figure} 
Using the QCD inputs in Table \ref{tab:param}, we show the $\tau$-behaviour of $M_{Y_{cd}}$ from ${\cal R}^{LSR}_{cd}$ in Fig. \ref{fig:yc}a for $m_c=1.26$ GeV and for different values of $t_c$. One can notice from Fig. \ref{fig:yc}a that the $\tau$-stability is obtained from $\sqrt{t_c}\geq 5.1$ GeV, while the $t_c$-stability is reached for $\sqrt{t_c}=7$ GeV. The most conservative prediction from the LSR
is obtained in this range of $t_c$-values for $m_c=1.26$ GeV and gives in units of GeV:
\bea
4.79 \leq M_{Y_{cd}}\leq 5.73~~{\rm for}~~5.02 \leq \sqrt{t_c}\leq 7~{\rm and}~ m_c=1.26,\nnb\\
5.29 \leq M_{Y_{cd}}\leq 6.11~~{\rm for}~~5.5 \leq \sqrt{t_c}\leq 7~{\rm and}~m_c=1.47.
\eea
We compare in Fig. \ref{fig:yc}b), the $t_c$-behaviour of the LSR results obtained at the $\tau$-stability points with the ones from  $ {\cal R}^{FESR}_{cd}$ for  the charm quark mass $m_c$=1.23 GeV (running) and 1.47 GeV (on-shell). One can deduce the common solution in units of GeV:
\bea
M_{Y_{cd}}&=& 4.814~~~~{\rm for}~~~~ \sqrt{t_c}=5.04(5)~~  {\rm and}~~m_c=1.26,\nnb\\
&=& 5.409~~~~{\rm for}~~~~ \sqrt{t_c}=5.6~~  {\rm and}~~m_c=1.47~.
\label{eq:ycd}
\eea
In order to fix the values of $M_{Y_{cd}}$ obtained at this lowest order PT calculations, we can also refer to the predictions of the $J/\psi$ mass using the LSR at the same lowest order PT calculations and including the condensate contributions up to dimension-six. We 
observe that the on-shell $c$-quark mass value tends to overestimate $M_{J/\psi}$ \cite{X2,SNH10}. The same feature happens for the
evaluation of the $X(1^{++})$ four-quark state mass \cite{X1}. Though this observation may not be
rigorous as the strength of the radiative corrections is channel dependent, we are tempted to take as a final result in this paper the prediction  obtained by using the running mass $\overline{m}_c(m_c)=1262(17)$ MeV within which it is known, from different examples in the literature, that the PT series converge faster \cite{SNH10}\,\footnote{We plan to check this conjecture in a future publication when PT radiative corrections are included.}. Including different sources of errors, we deduce in MeV\,\footnote{We consider this result as an improvement (smaller error) of the one e.g. in \cite{RAPHAEL1,RAPHAEL2} where only exponential sum rules have been used. However, the present error and the existing ones in the literature may have been underestimated due to the non-inclusion of the unknown PT radiative corrections and some eventual systematics of the approach.}:
\bea
M_{Y_{cd}}&=&4814(50)_{t_c}(14)_{m_c}(2)_\Lambda (17)_{\bar uu}(2)_{G^2}(4)_{M^2_0}(13)_{G^3}(6)_\rho\nnb\\
&=&4814(57)~.
\label{eq:ycd2}
\eea
Using the fact that the 1st FESR moment  gives a correlation between the mass of the lowest ground state and the onset of continuum threshold $t_c$, where its value coincide approximately
with the value of the 1st radial excitation mass (see e.g. ref. \cite{fesr} and some other examples in \cite{SNB}), we shall approximately identify its value with the one of the radial excitation. In order to take into account the systematics of the approach and  some  eventual small local duality violation advocated by \cite{SHIF} which can only be detectable in a high-precision analysis like the extraction of $\alpha_s$ from $\tau$-decay \cite{SNTAU,PERIS}, we have allowed $t_c$ to move around this intersection point. 
Assuming that the mass of the radial excitation is approximately $\sqrt{t_c}$, one can deduce the mass-splitting:
\beq
M'_{Y_{cd}}-M_{Y_{cd}}\approx 226~{\rm MeV}~,
\eeq
which is similar to the one obtained for the $X(1^{++})$ four-quark state \cite{X1}. This  splitting is much lower than the one intuitively used in the current literature:
\beq
M_{\psi}(2S)-M_{\psi}(1S)\simeq 590~{\rm MeV}~,
\eeq
for fixing the arbitrary value of $t_c$ entering in different Borel (exponential) sum rules of the four-quark and molecule states. This difference may signal some new dynamics for the exotic states compared with the usual $\bar cc$ charmonium states and need to be tested from some other approaches such as potential models, heavy quark symmetry, AdS/QCD and lattice calculations.
\begin{figure}[hbt] 
\begin{center}
\centerline {\hspace*{-7cm} a) }\vspace{-.6cm}
{\includegraphics[height=30mm]{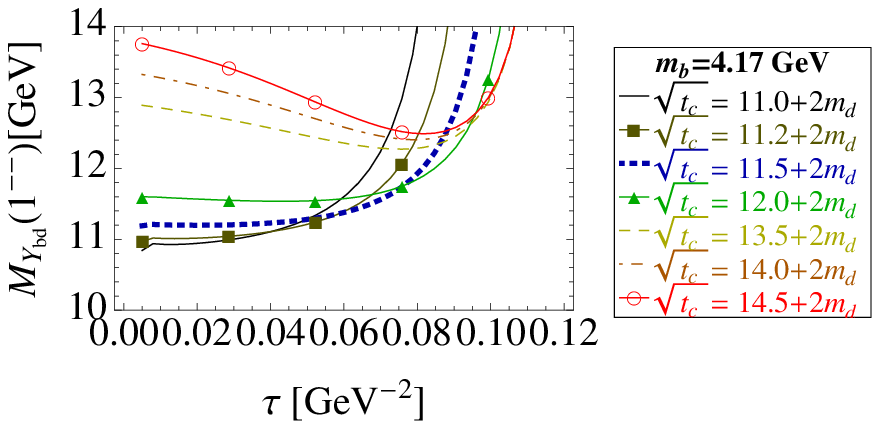}}\\
\centerline {\hspace*{-7cm} b) }\vspace{-0.6cm}
{\includegraphics[height=30mm]{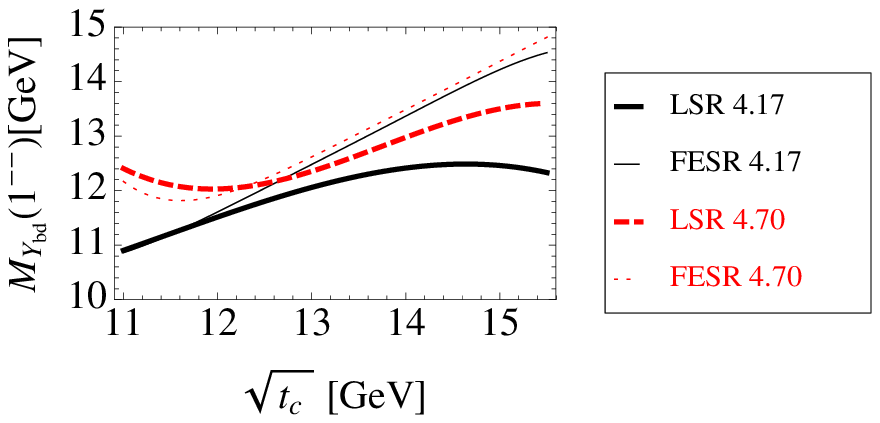}}
\caption{\scriptsize 
{\bf a)} $\tau$-behaviour of $M_{Y_{bd}}(1^{--})$ from ${\cal R}^{LSR}_{bd}$ for the current mixing parameter $b=0$, for different values of $t_c$
and for $m_b=4.17$ GeV.  {\bf b)} $t_c$-behaviour of the LSR results obtained at the $\tau$-stability points and comparison with the ones from  $ {\cal R}^{FESR}_{bd}$ for  $m_b=4.17$ and 4.70 GeV.}
\label{fig:yb} 
\end{center}
\end{figure} 
\nin
\subsection*{\b The $Y_{bd}$ mass from LSR and FESR for the case $b=0$}
\nin
Using similar analysis for the $b$-quark, we show the $\tau$-behaviour of ${\cal R}^{LSR}_{bd}(\tau)$ in Fig. \ref{fig:yb}a for $m_b=4.17$ GeV and for different values of $t_c$. In  Fig. \ref{fig:yb}b, the same analysis is shown for $m_b=4.70$ GeV. The most conservative result from the LSR is (in units of GeV) is:
\bea
11.0 \leq M_{Y_{bd}}\leq 12.4~{\rm for}~11.2 \leq \sqrt{t_c}\leq 14.5~{\rm and}~ m_b=4.17,\nnb\\
12.1 \leq M_{Y_{bd}}\leq 13.4~{\rm for}~12. 2\leq \sqrt{t_c}\leq 15.5~{\rm and}~ m_b=4.70,\nnb
\eea
where the lower (resp. higher) values of $t_c$ correspond to the beginning of $\tau$ (resp. $t_c$)-stability. 
We compare in Fig. \ref{fig:yb}b), the $t_c$-behaviour of the LSR results obtained at the $\tau$-stability points with the ones from  $ {\cal R}^{FESR}_{bd}$ for  the $b$ quark mass $m_b$=4.17 GeV (running) and 4.70 GeV (on-shell). One can deduce the common solution in units of~GeV:
\bea
M_{Y_{bd}}&=& 11.26~~~~{\rm for}~~~~ \sqrt{t_c}=11.57(7)~  {\rm and}~m_b=4.17\nnb\\
&=& 12.09~~~~{\rm for}~~~~ \sqrt{t_c}=12.2~  {\rm and}~m_b=4.70.
\label{eq:ybd}
\eea
One can notice, like in the case of the charm quark that the value of the on-shell quark mass tends to give a higher value of 
$M_{Y_{bd}}$ within this lowest order PT calculations. Considering, like in the case of charm, as a final estimate the one from the running $b$-quark mass $\overline{m}_b(m_b)=4177(11)$ MeV \cite{SNH10}, we deduce in MeV:
\bea
M_{Y_{bd}}&=&11247(45)_{t_c}(8)_{m_b}(2)_\Lambda (15)_{\bar uu}(1)_{G^2}(1)_{M^2_0}(1)_{G^3}(5)_\rho\nnb\\
&=&11256(49)~.
\label{eq:ybd2}
\eea
From the previous result, one can deduce the approximate value of the mass-splitting between the 1st radial excitation and the lowest mass ground state:
\beq
M'_{Y_{bd}}-M_{Y_{bd}}\approx M'_{Y_{cd}}-M_{Y_{cd}}\approx 250~{\rm MeV}~,
\eeq
which are (almost) heavy-flavour independent and  also smaller than the one of the bottomium splitting:
\beq
M_{\Upsilon}(2S)-M_{\Upsilon}(1S)\simeq 560~{\rm MeV}~.
\eeq
\subsection*{ \b Effect of the current mixing $b$ on the mass}
\begin{figure}[hbt] 
\begin{center}
\centerline {\hspace*{-7cm} a) }\vspace{-0.7cm}
{\includegraphics[height=30mm]{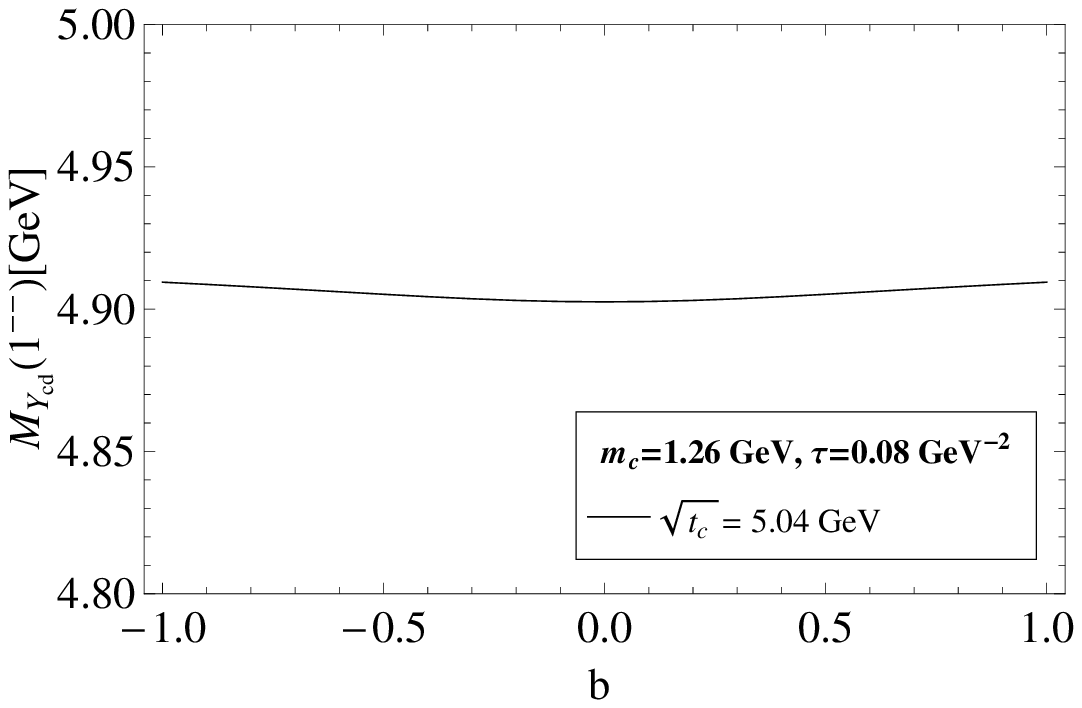}}\\
\centerline {\hspace*{-7cm} b) }\vspace{-0.7cm}
{\includegraphics[height=30mm]{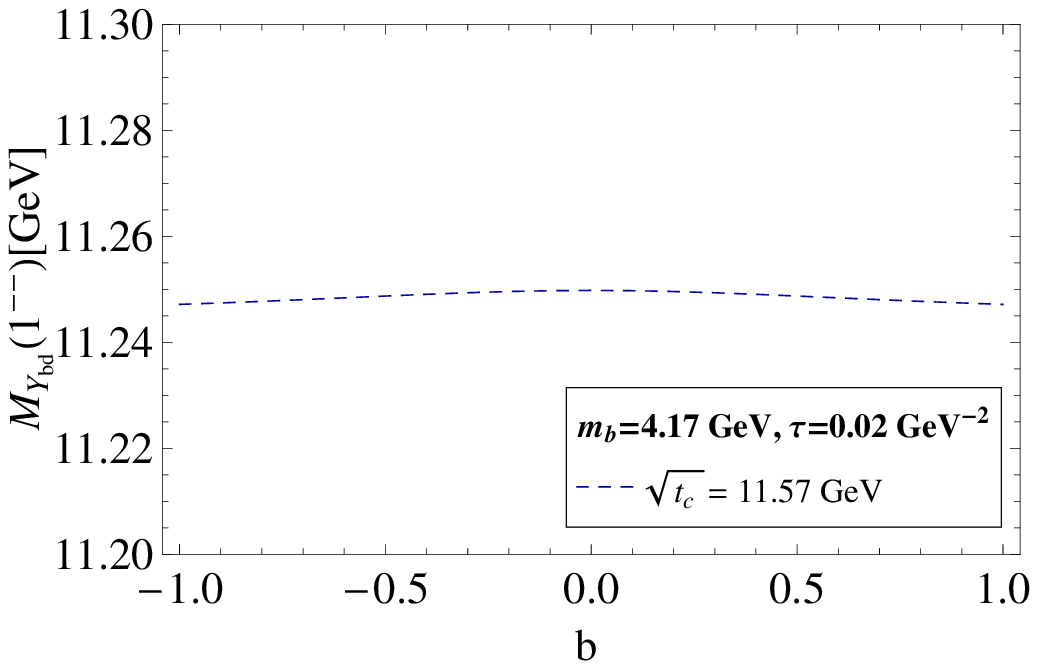}}
\caption{\scriptsize 
{\bf a)} $b$-behaviour of $M_{Y_{cd}}$  for given values of $\tau$ and $t_c$ 
and for $m_c=1.26$ GeV; {\bf b)} the same as a) but for $M_{Y_{bd}}$
and for $m_b=4.17$ GeV.}
\label{fig:b1} 
\end{center}
\end{figure}

In the following, we shall let the current mixing parameter $b$ defined in Table \ref{tab:4q}
free and study its effect on the results obtained in Eqs. (\ref{eq:ycd2}) and (\ref{eq:ybd2}). In so doing, we fix the values of $\tau$ around the $\tau$-stability point and $t_c$ around the intersection point of the LSR and FESR. The results of the analysis are shown in Fig. \ref{fig:b1}.
We notice that the results are optimal at the value $b=0$ which a posteriori justifies the
results obtained previously for $b=0$. 
\subsection*{ \b Effect of the current mixing  $b$ on the decay constant $f_{Y_{Qd}}$}
\begin{figure}[hbt] 
\begin{center}
\centerline {\hspace*{-7cm} a) }\vspace{-0.5cm}
{\includegraphics[height=30mm]{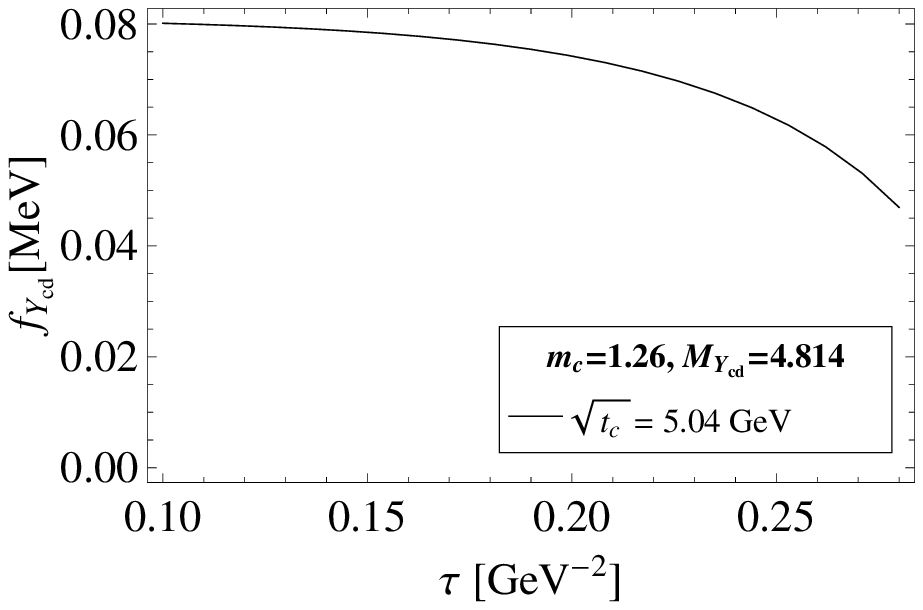}}\\
\centerline {\hspace*{-7cm} b) }\vspace{-0.5cm}
{\includegraphics[height=30mm]{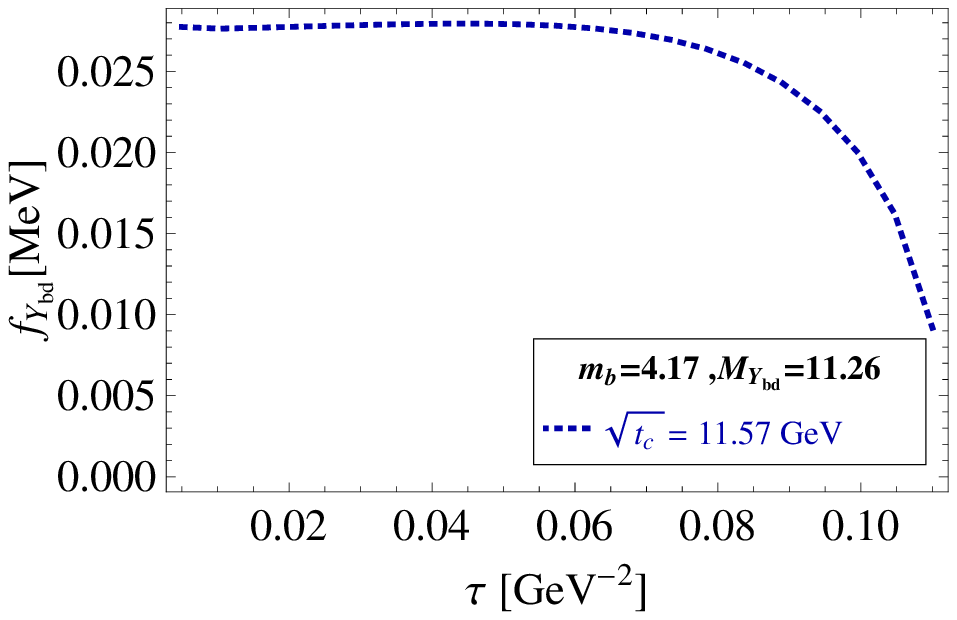}}
\centerline {\hspace*{-7cm} c) }\vspace{-0.5cm}
{\includegraphics[height=30mm]{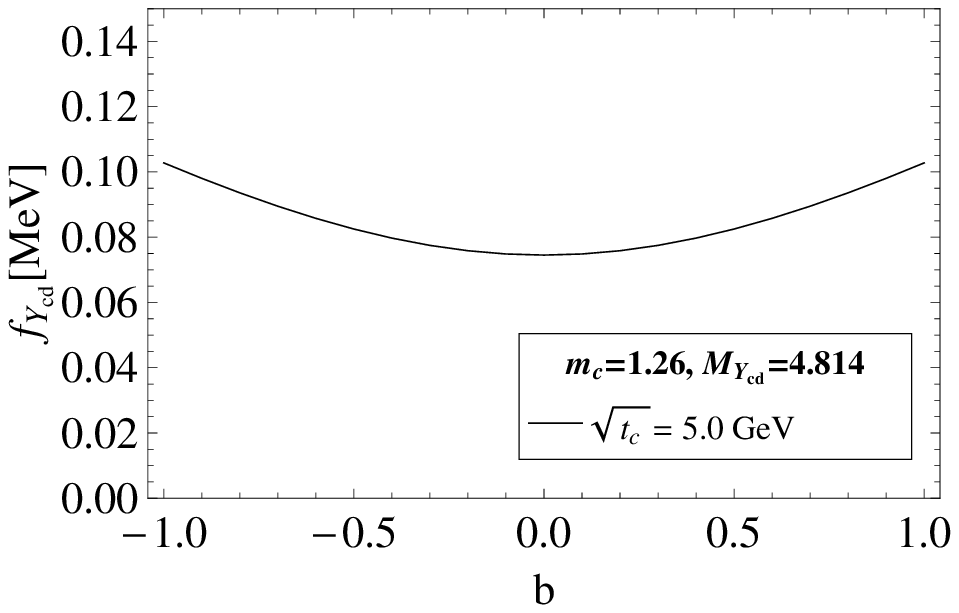}}\\
\centerline {\hspace*{-7cm} d) }\vspace{-0.5cm}
{\includegraphics[height=30mm]{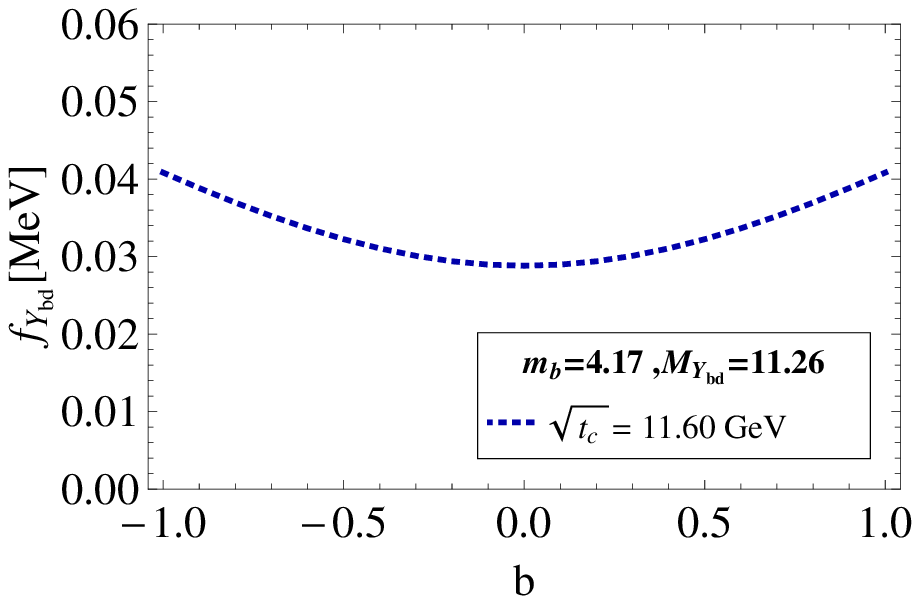}}
\caption{\scriptsize 
{\bf a)} $\tau$-behaviour of $f_{Y_{cd}}$  for given values of $b=0$ and $t_c$ 
and for $m_c=1.26$ GeV; {\bf b)} the same as a) but for $f_{Y_{bd}}$
and for $m_b=4.17$ GeV; 
{\bf c)} $b$-behaviour of $f_{Y_{cd}}$  for given values of $\tau$ at the stability and $t_c$
 {\bf d)} the same as c) but for $f_{Y_{bd}}$
}
\label{fig:fb1} 
\end{center}
\end{figure}

For completing the analysis of the effect of $b$, we also study the decay constant $f_{Y_Qd}$
defined as:
\beq
\la 0\vert j^\mu_{4q}\vert Y_{Qd}\ra=f_{Y_Qd}M^4_{Y_{Qd}}\epsilon^\mu~.
\eeq
We show the analysis in Fig. \ref{fig:fb1} giving  $M_{Y_{Qd}}$ and the corresponding $t_c$ obtained above. One can deduce the optimal values at $b=0$:
\beq
f_{Y_{cd}} \simeq 0.08~{\rm MeV}~~~~{\rm and} ~~~~~f_{Y_{bd}} \simeq 0.03~{\rm MeV}~,
\label{eq:f1}
\eeq
which are much smaller than $f_\pi=132$ MeV, $f_\rho\simeq 215$ MeV and $f_D\simeq f_B$=203 MeV \cite{SNFB}. On can also note that
the decay constant decreases like $1/M_Q$ which can be tested in HQET or/and lattice QCD.
\subsection*{ \b $SU(3)$ breaking for $M_{Y_{Qs}}$ from DRSR}
\begin{figure}[hbt] 
\begin{center}
\centerline {\hspace*{-7cm} a) }\vspace{-0.7cm}
{\includegraphics[height=30mm]{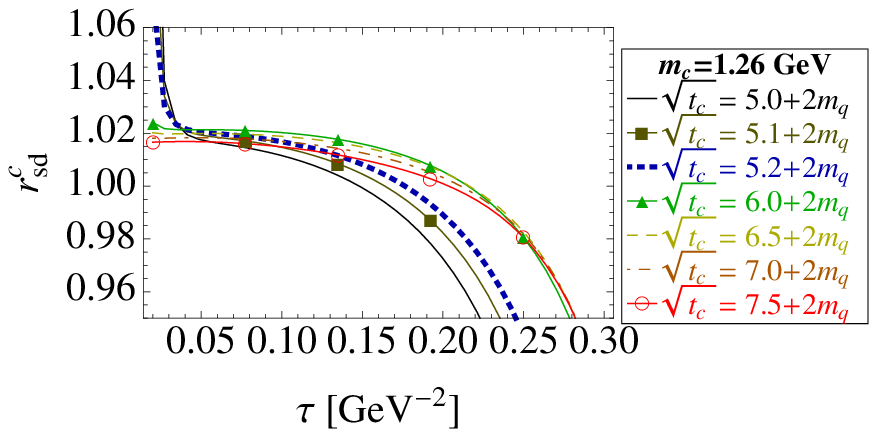}}\\
\centerline {\hspace*{-7cm} b) }\vspace{-0.7cm}
{\includegraphics[height=30mm]{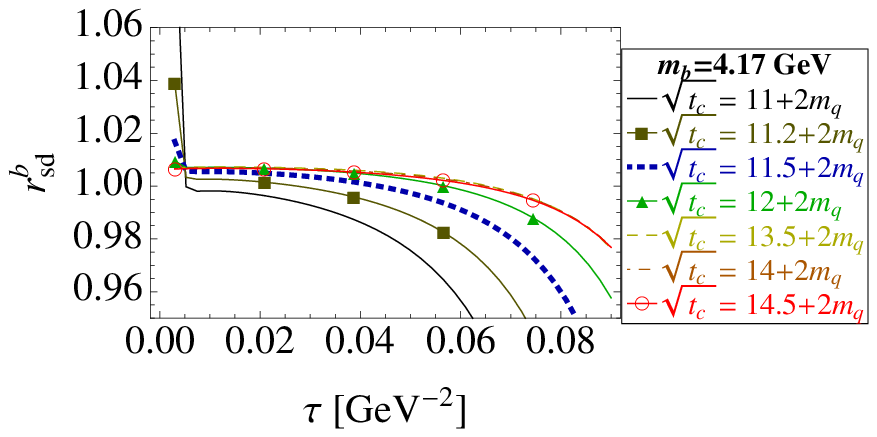}}
\centerline {\hspace*{-7.cm} c) }\vspace{-0.7cm}
{\includegraphics[height=30mm]{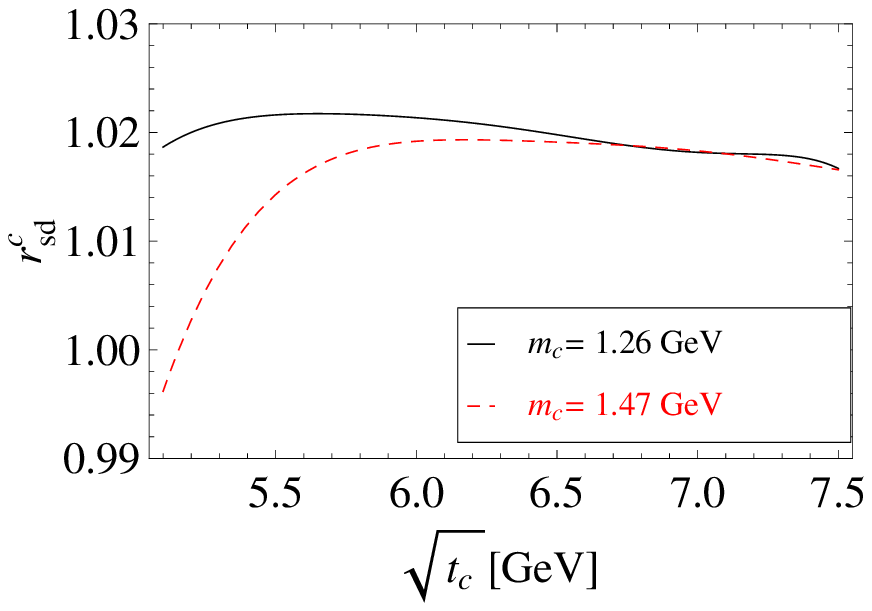}}\\
\centerline {\hspace*{-7.cm} d) }\vspace{-0.7cm}
{\includegraphics[height=30mm]{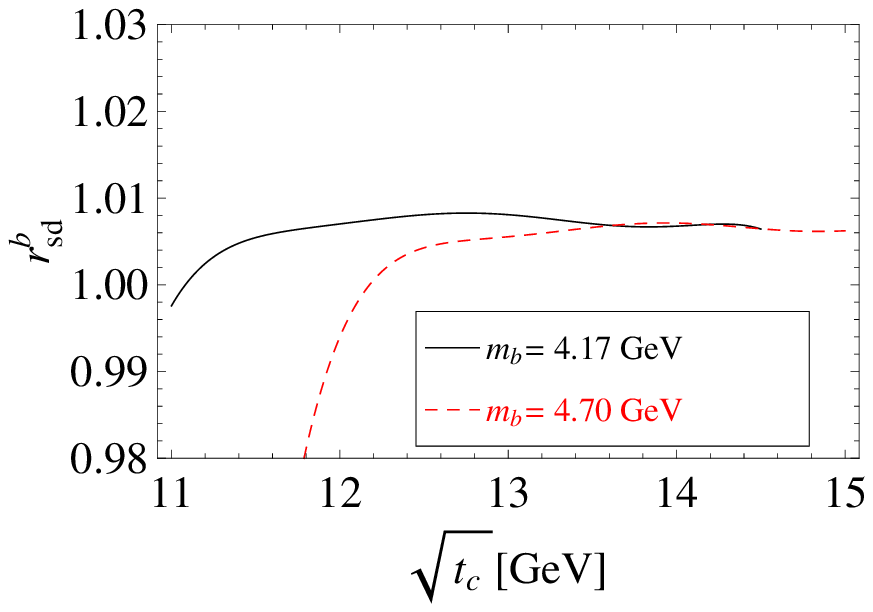}}
\caption{\scriptsize 
{\bf a)} $\tau$-behaviour of $r^c_{sd}$ for the current mixing parameter $b=0$, for different values of $t_c$
and for $m_c=1.26$ GeV; {\bf b)} $\tau$-behaviour of $r^b_{sd}$  for different values of $t_c$
and for $m_b=4.17$ GeV; 
{\bf c)} $t_c$-behaviour of the inflexion points (or minimas) of $r^c_{sd}$  from Fig \ref{fig:rsd}a;  {\bf d)} The same  for the $b$ quark using $r^b_{sd}$  from Fig \ref{fig:rsd}b.}
\label{fig:rsd} 
\label{fig:rsdtc}
\end{center}
\end{figure}

\nin
We study the ratio $M_{Y_{Qs}}/M_{Y_{Qd}}$  using double ratio of LSR (DRSR):
\beq
r^Q_{sd}\equiv {\sqrt{{\cal R}_{Qs}^{LSR}}\over\sqrt{{\cal R}_{Qd}^{LSR}}}~~~~{\rm where}~~~~Q\equiv c,~b~.
\eeq
We show the $\tau$-behaviour of $r^c_{sd}$ and $r^b_{sd}$ respectively in Figs. \ref{fig:rsd}a and \ref{fig:rsd}b for $m_c=1.26$ GeV and  $m_b=4.17$ GeV for different
values of $t_c$. We show, in Fig. \ref{fig:rsdtc}c and Fig. \ref{fig:rsdtc}d, the $t_c$-behaviour of the stabilities or inflexion points for two different values (running and on-shell) of the quark masses. One can see in these figures that the DRSR is very stable versus the $t_c$ variations in the case of the running heavy quark masses. We deduce the corresponding DRSR:
\bea
r^c_{sd}&=&1.018(1)_{m_c}(5)_{m_s} (2)_{\kappa}(2)_{\bar uu}(1)_\rho~,\nnb\\
r^b_{sd}&=&1.007(0.5)_{m_b}(2)_{m_s} (0.5)_{\kappa}(1)_{\bar uu}(0.3)_\rho~,
\label{eq:ratioQs}
\eea
respectively for $\sqrt{t_c}=5.1$ and 11.6 GeV. 
Using the results for  $Y_{Qd}$ in Eqs. (\ref{eq:ycd2}) and  (\ref{eq:ybd2}) and the
values of the $SU(3)$ breaking ratio in Eq. (\ref{eq:ratioQs}), we can deduce the mass of the $Y_{Qs}$ state
in MeV:
\beq
M_{Y_{cs}}=4900(67)~,~~~~~~~~~~~
M_{Y_{bs}}=11334(55)~,
\eeq
leading to the $SU(3)$ mass-splitting:
\beq
\Delta M^{Y_c}_{sd}\approx 87~{\rm MeV}\approx \Delta M^{Y_b}_{sd}\approx 78~{\rm MeV}~,
\lb{eq:split4q}
\eeq
which is also (almost) heavy-flavour independent. 

\section{ $1^{--}$ molecule masses from QSSR }

\subsection*{\b The $\bar D^*_{d(s)}D_{d(s)}$ and $\bar B^*_{d(s)}B_{d(s)}$ molecules\,\footnote{Hereafter, for simplifying notations, $D$ and $B$ denote the scalar $D^*_0$ and $B^*_0$ mesons.}}
\nin
\begin{figure}[hbt] 
\begin{center}
\centerline {\hspace*{-7.cm} a) }\vspace{-0.7cm}
{\includegraphics[height=30mm]{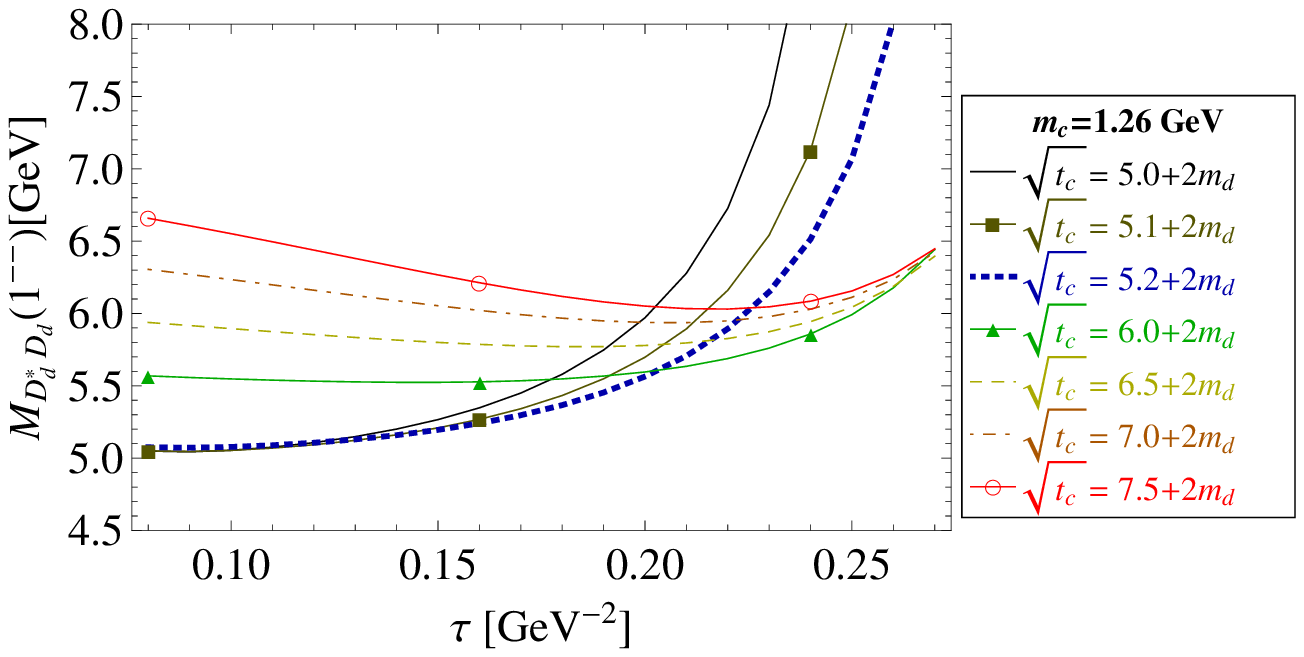}}\\
\centerline {\hspace*{-7.cm} b) }\vspace{-0.7cm}
{\includegraphics[height=30mm]{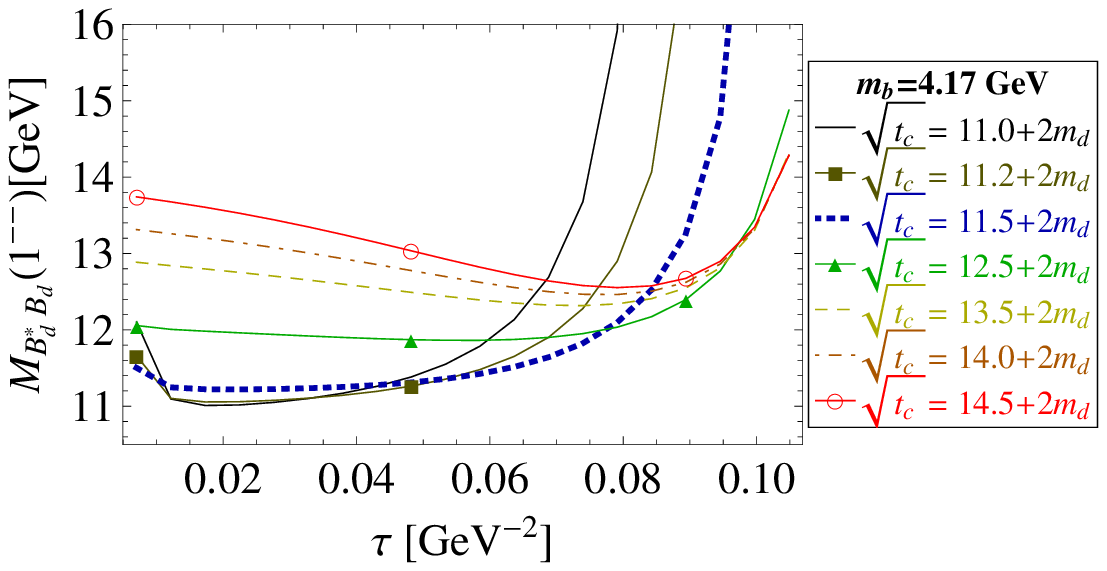}}
\centerline {\hspace*{-7.cm} c) }\vspace{-0.7cm}
{\includegraphics[height=30mm]{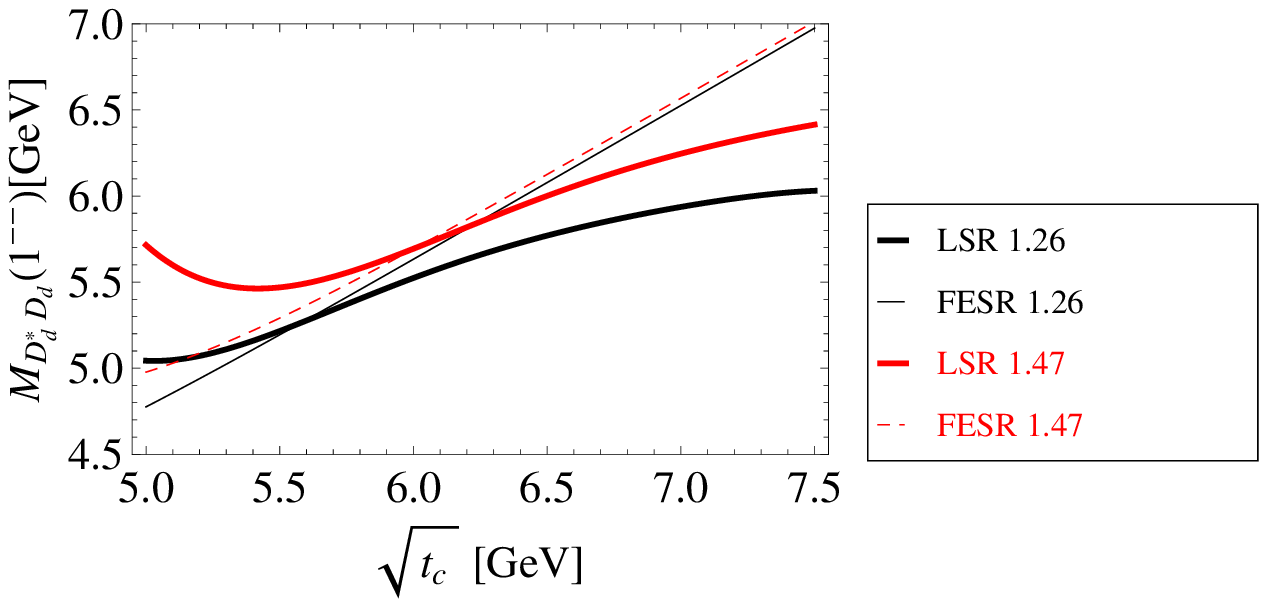}}\\
\centerline {\hspace*{-7.cm} d) }\vspace{-0.7cm}
{\includegraphics[height=30mm]{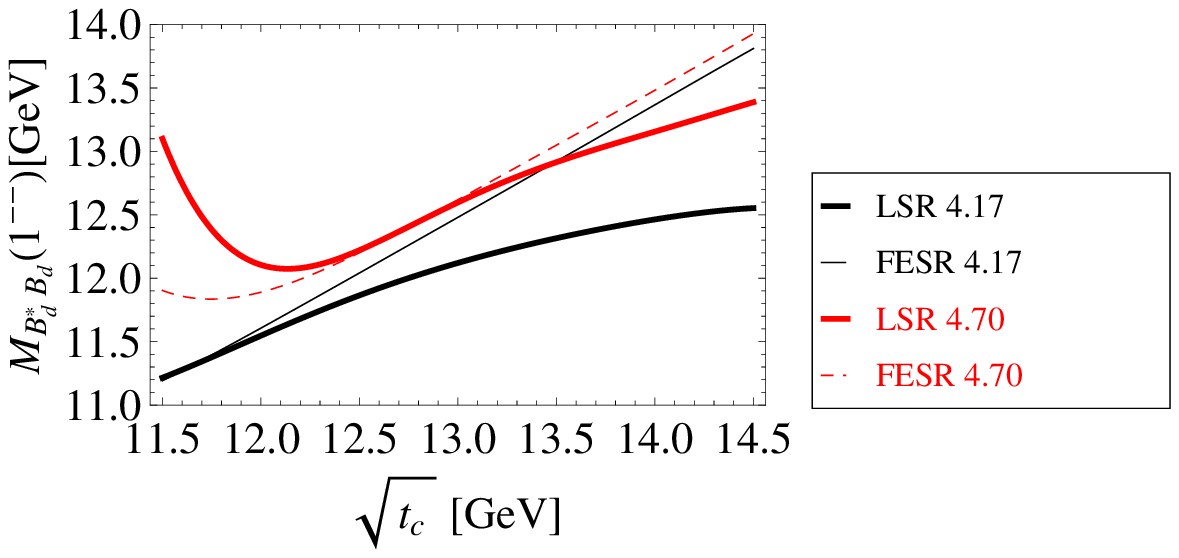}}
\caption{\scriptsize 
{\bf a)} $\tau$-behaviour of $M_{D^*_dD_d}$  for different values of $t_c$
and for $m_c=1.26$ GeV; {\bf b)} $\tau$-behaviour of $M_{B^*_dB_d}$  for different values of $t_c$ and for $m_b=4.17$ GeV;
{\bf c)} $t_c$-behaviour of the extremas  in $\tau$ of $M_{D^*_dD_d}$ and for $m_c=1.26-1.47$ GeV; {\bf d)} the same as c) but for $M_{B^*_dB_d}$ and for $m_b=4.17-4.70$ GeV.}
\label{fig:rmol} 
\end{center}
\end{figure} 
\nin
\begin{figure}[hbt] 
\begin{center}
\centerline {\hspace*{-7cm} a) }\vspace{-0.7cm}
{\includegraphics[height=30mm]{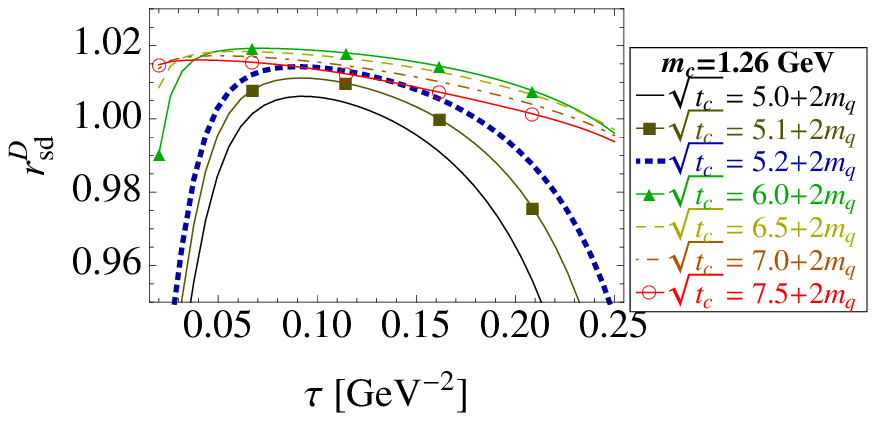}}\\
\centerline {\hspace*{-7cm} b) }\vspace{-0.5cm}
{\includegraphics[height=30mm]{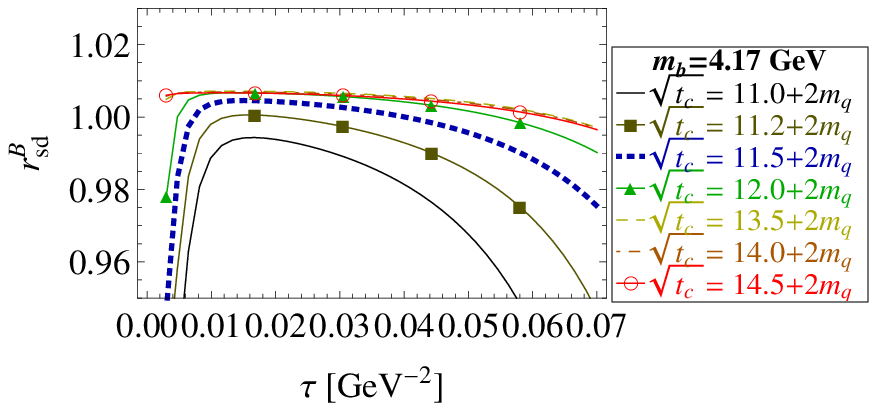}}
\centerline {\hspace*{-7cm} c) }\vspace{-.5cm}
{\includegraphics[height=30mm]{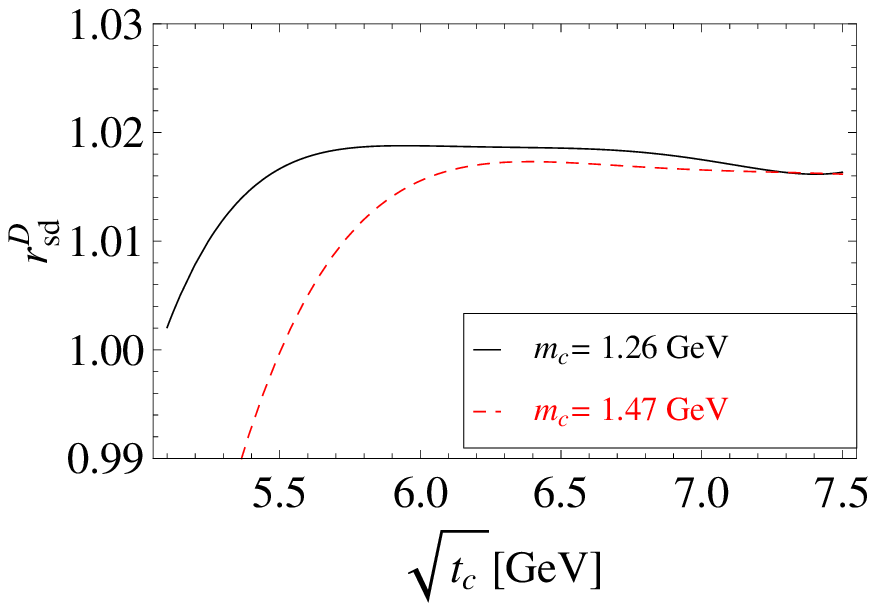}}\\
\centerline {\hspace*{-7cm} d) }\vspace{-0.5cm}
{\includegraphics[height=30mm]{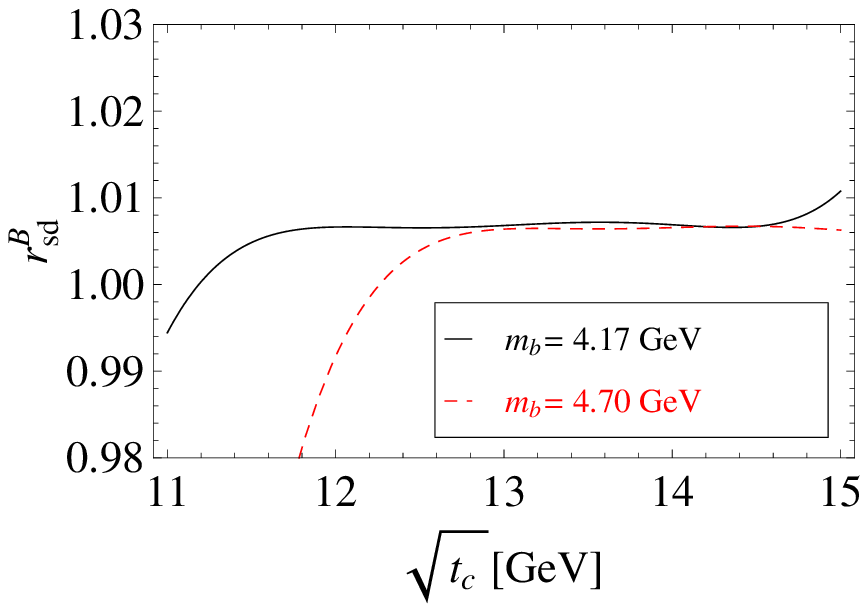}}
\caption{\scriptsize 
{\bf a)} $\tau$-behaviour of $r^D_{sd}$  for different values of $t_c$
and for $m_c=1.26$ GeV; {\bf b)} $\tau$-behaviour of $r^B_{sd}$  for different values of $t_c$
and for $m_b=4.17$ GeV; {\bf c)} $t_c$-behaviour of the inflexion points (or minimas) of $r^D_{sd}$  from Fig \ref{fig:rsd2}a; {\bf d)} the same for the $b$ quark using $r^B_{sd}$  from Fig \ref{fig:rsd2}b.}
\label{fig:rsd2} 
\end{center}
\end{figure} 
\nin
Like in the previous case, we use LSR and FESR for studying the masses of the $\bar D^*_dD_d$ and 
 $\bar B^*_{d}B_{d}$ and DRSR for studying the $SU(3)$ breaking ratios:
\beq
r^D_{sd}\equiv  {M_{D^*_sD_s}\over M_{D^*_dD_d}}~,~~~~~~~~~~~~~~~~~~~r^B_{sd}\equiv  {M_{B^*_sB_s}\over M_{B^*_dB_d}}~.
\eeq
We show their $\tau$-behaviour for different values of $t_c$ and for $m_c=1.26$ GeV and $m_b=4.17$ GeV respectively in Figs. \ref{fig:rmol}a,b and  \ref{fig:rsd2}a,b. The $t_c$-behaviour of the $\tau$-minimas is shown in  Fig. \ref{fig:rmol}c,d for the masses and in Fig.  \ref{fig:rsd2}c,d for the $SU(3)$ breaking ratios. Using the sets ($m_c=1.26$ GeV, $\sqrt{t_c}=5.58~{\rm GeV}$) and ($m_b=4.17~{\rm GeV}$, $\sqrt{t_c}=11.64(3)~{\rm GeV}$) common solutions of  LSR and FESR, one can deduce in MeV:
\bea
M_{{D^*_dD_d}}&=&5268(14)_{m_c}(3)_\Lambda (19)_{\bar uu}(0)_{G^2}(0)_{M^2_0}(2)_{G^3}(5)_\rho,\nnb\\
&=&5268(24)~,\nnb\\
 M_{{B^*_dB_d}}&=&11302 (20)_{t_c}(9)_{m_b}(2)_\Lambda (19)_{\bar uu}(0)_{G^2}(0)_{M^2_0}(1)_{G^3}(5)_\rho \nnb\\
 &=&11302 (30)~,\nnb\\
r^D_{sd}&=&1.018(1)_{m_c}(4){m_s}(0.8)_{\kappa}(0.5)_{\bar uu}(0.2)_{\rho}(0.1)_{G^3}~,\nnb\\
r^B_{sd}&=&1.006(1)_{m_b}(2){m_s}(1)_{\kappa}(0.5)_{\bar uu}(0.2)_{\rho}(0.1)_{G^3}~.
\label{eq:D*D}
\eea
Using the previous results in Eq. (\ref{eq:D*D}), one obtains in MeV :
\beq
M_{{D^*_sD_s}}=5363(33)~,~~~~~ M_{{B^*_sB_s}}=11370(40)~,
\label{eq:D*sDs}
\eeq
corresponding to a $SU(3)$ mass-splitting:
\beq
\Delta M^{DD^*}_{sd}\simeq 95 ~{\rm MeV} \approx \Delta M^{BB^*}_{sd}\simeq 68~{\rm MeV}~.
\lb{eq:splitmol1}
\eeq
These results for $M_{DD^*}$ are in the upper part of the range given in \cite{RAPHAEL1} due both to the smaller values of $m_c=1.23$ GeV and $\sqrt{t_c}$=5.5 GeV  used in that paper. Though the $DD^*$ molecule mass is above the $DD^*$ threshold which is similar to the e.g. the case of the $\pi\pi$ continuum and $\rho$-meson resonance in $e^+e^-$ to the I=1 hadrons channel, one expects that at the $\tau$-stability point or inside the sum rule window, where the QCD continuum contribution is minimum while the OPE is still convergent, the lowest ground state dominates the sum rule.
\begin{figure}[hbt] 
\begin{center}
\centerline {\hspace*{-7cm} a) }\vspace{-0.7cm}
{\includegraphics[height=30mm]{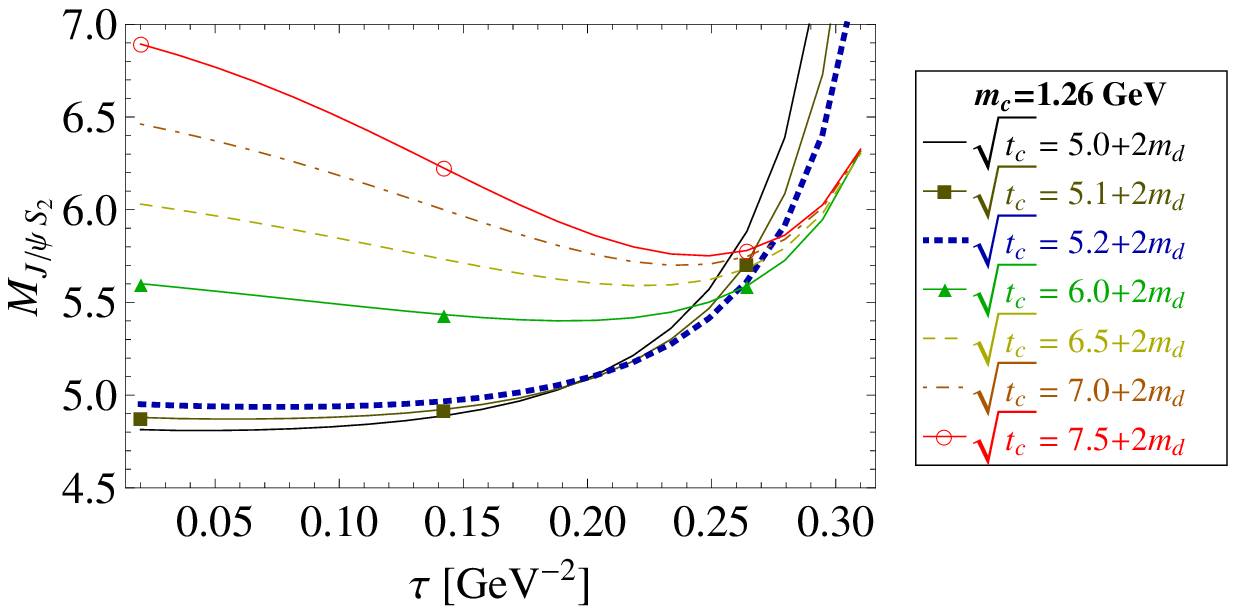}}\\
\centerline {\hspace*{-7cm} b) }\vspace{-0.5cm}
{\includegraphics[height=30mm]{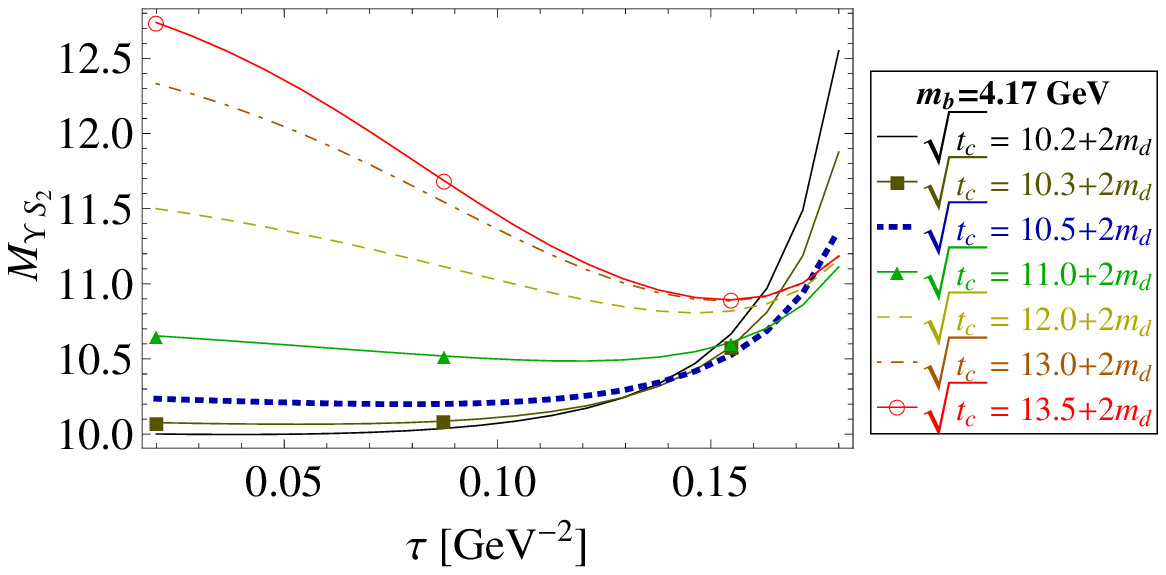}}
\centerline {\hspace*{-7cm} c) }\vspace{-0.7cm}
{\includegraphics[height=30mm]{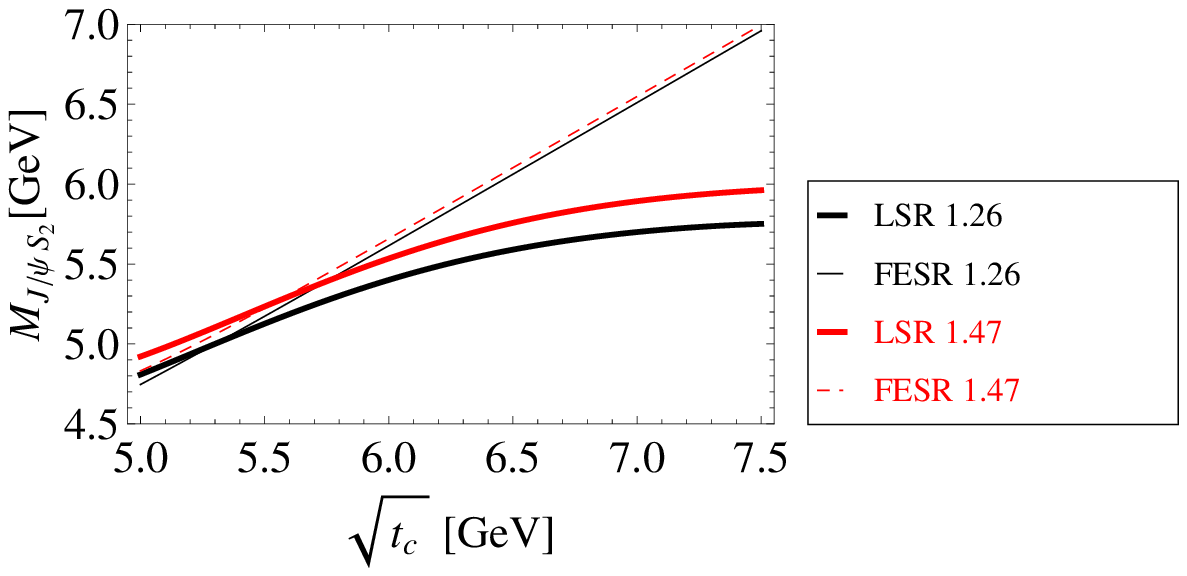}}\\
\centerline {\hspace*{-7cm} d) }\vspace{-0.7cm}
{\includegraphics[height=30mm]{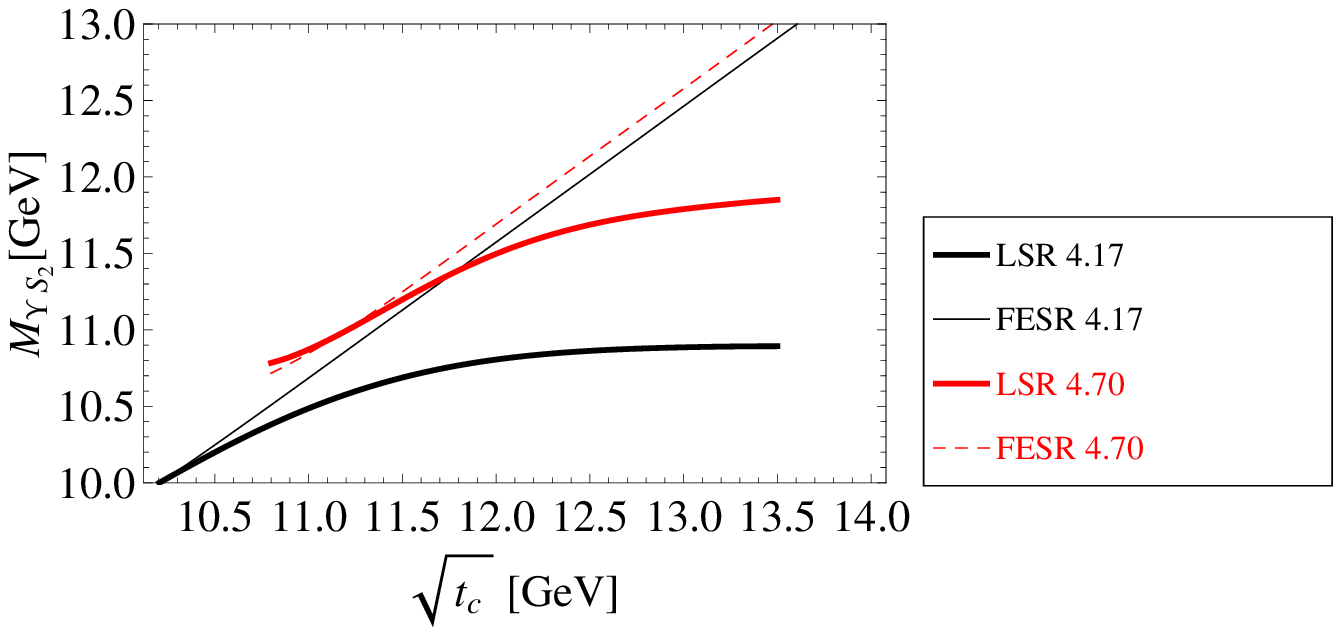}}
\caption{\scriptsize 
{\bf a)} $\tau$-behaviour of $M_{J/\psi S_2}$  for different values of $t_c$
and for $m_c=1.26$ GeV; {\bf b)} $\tau$-behaviour of $M_{\Upsilon S_2}$  for different values of $t_c$ and for $m_b=4.17$ GeV;
{\bf c)} $t_c$-behaviour of the extremas  in $\tau$ of $M_{J/\psi S_2}$  for $m_c=1.26-1.47$ GeV; {\bf d)} the same as c) but for $M_{\Upsilon S_2}$ for $m_b=4.17-4.70$ GeV.}
\label{fig:psimol} 
\end{center}
\end{figure} 
\nin
\begin{figure}[hbt] 
\begin{center}
\centerline {\hspace*{-7cm} a) }\vspace{-0.7cm}
{\includegraphics[height=30mm]{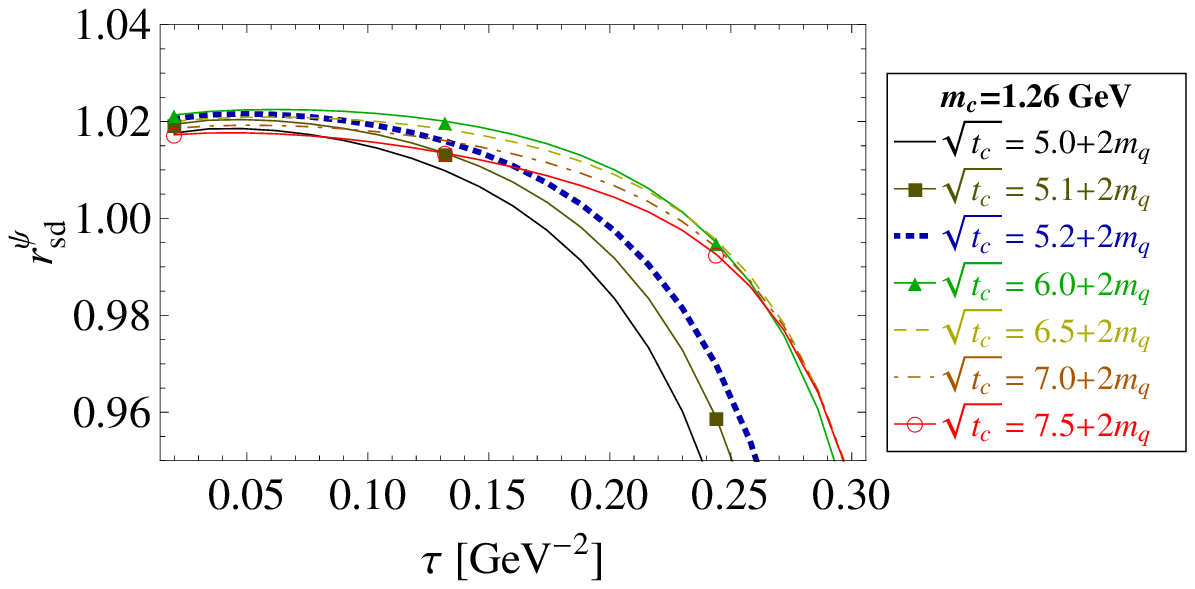}}\\
\centerline {\hspace*{-7cm} b) }\vspace{-0.7cm}
{\includegraphics[height=30mm]{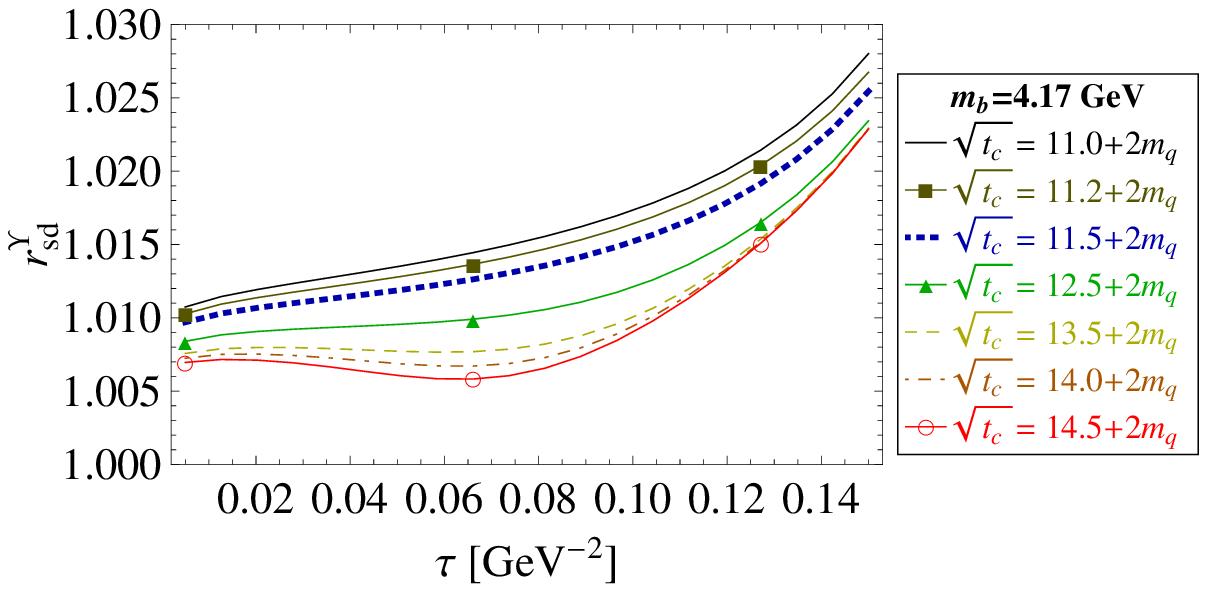}}
\centerline {\hspace*{-7cm} c) }\vspace{-0.7cm}
{\includegraphics[height=30mm]{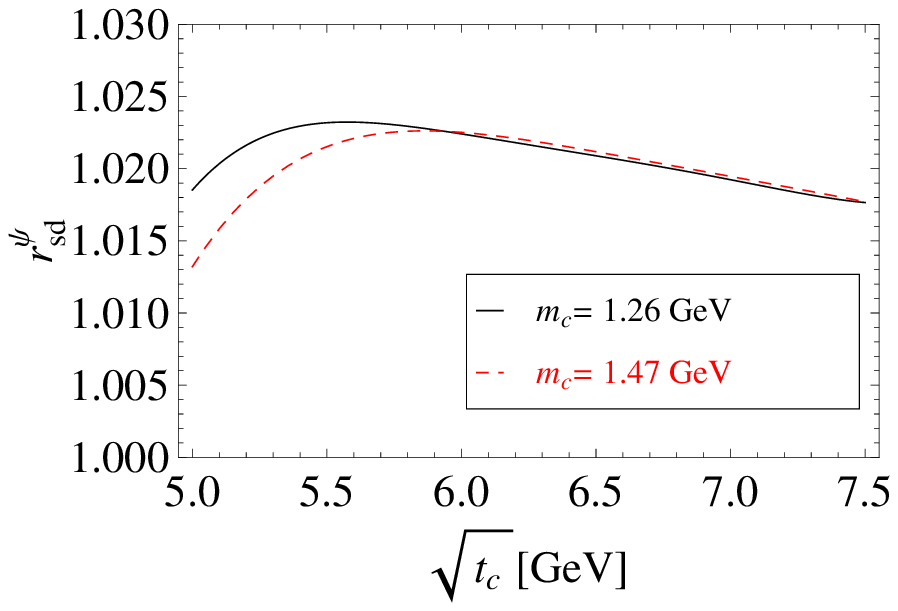}}\\
\centerline {\hspace*{-7cm} d) }\vspace{-0.7cm}
{\includegraphics[height=30mm]{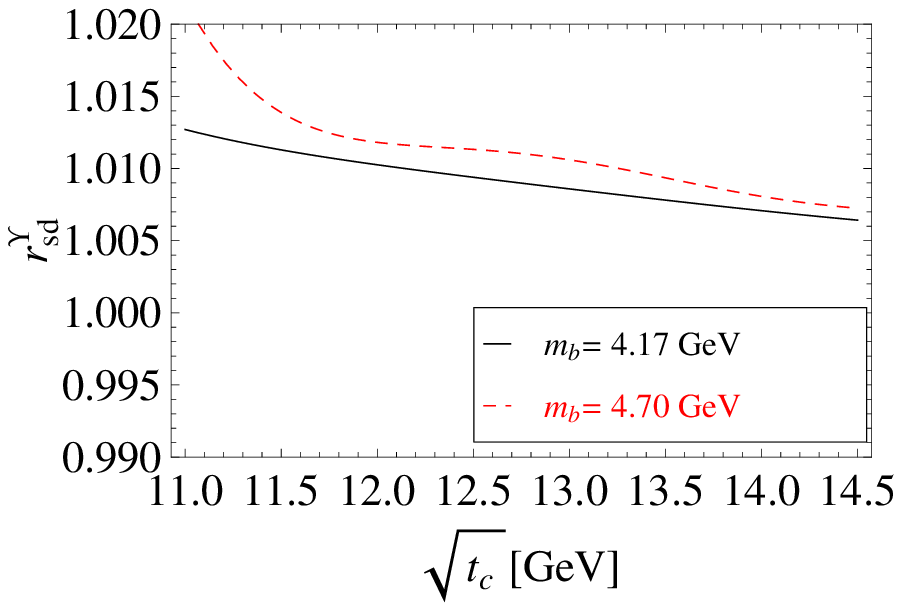}}
\caption{\scriptsize 
{\bf a)} $\tau$-behaviour of $r^\psi_{sd}$  for different values of $t_c$
and for $m_c=1.26$ GeV; {\bf b)} $\tau$-behaviour of $r^\Upsilon_{sd}$  for different values of $t_c$ and for $m_b=4.17$ GeV;
{\bf c)} $t_c$-behaviour of the extremas  in $\tau$ of $r^\psi_{sd}$  for $m_c=1.26-1.47$ GeV; {\bf d)} the same as c) but for $r^\Upsilon_{sd}$ for $m_b=4.17-4.70$ GeV.}
\label{fig:su3psi} 
\label{fig:su3psi2} 
\end{center}
\end{figure} 
\nin
\subsection*{\b The  $J/\psi S_2$ and $\Upsilon S_2$ molecules}
\nin
Combining LSR and FESR, we consider the mass of the  $J/\psi S_2$ and $\Upsilon S_2$  molecules in a colour singlet combination, where $S_2\equiv \bar uu+\bar dd$ is a scalar meson\,\footnote{The low-mass $\pi^+\pi^-$ invariant mass due to the $\sigma$ meson is expected to result mainly from its gluon rather than from its quark component \cite{VENEZIA,MINK} such that an eventual  quark-gluon hybrid meson nature of the $Y_c$ is also possible.}. In so doing, we work with the LO QCD expression obtained in \cite{RAPHAEL2}. We show the results versus the LSR variable $\tau$ in Fig. \ref{fig:psimol}a,b.
The $t_c$-behaviour of different $\tau$-extremas is given in Figs. \ref{fig:psimol}c,d from which we can deduce for the running quark masses for $\sqrt{t_c}= 5.30(2)$ and 10.23(3) GeV in units of MeV:
\bea
M_{J/\psi S_2}&=&5002(20)_{t_c}(8)_{m_c}(2)_\Lambda (19)_{\bar uu}(9)_{G^2}(0)_{M^2_0}(0)_{G^3}(6)_\rho\nnb\\
&=&5002(31)~,\nnb\\
M_{\Upsilon S_2}&=&10015(20)_{t_c}(9)_{m_b}(2)_\Lambda (16)_{\bar uu}(17)_{G^2}(0)_{M^2_0}(0)_{G^3}(5)_\rho\nnb\\
&=&10015(33)~.
\lb{eq:psis2}
\eea
The splitting (in units of MeV) with the first radial excitation approximately given by $\sqrt{t_c}$ is:
\beq
M'_{J/\psi S_2}-M_{J/\psi S_2}\approx 298~,~~~~~~ M'_{\Upsilon S_2}-M_{\Upsilon S_2}\approx 213~.
\eeq
In the same way, we show in Figs. \ref{fig:su3psi} the $\tau$ and $t_c$ behaviours of the $SU(3)$ breaking ratios, from which, we can deduce:
\bea
r^\psi_{sd}&\equiv& {M_{J/\psi S_3}\over M_{J/\psi S_2}}=1.022(0.2)_{m_c}(5)_{m_s}(2)_{\kappa}~, \nnb\\
r^\Upsilon_{sd}&\equiv& {M_{\Upsilon S_3}\over M_{\Upsilon S_2}}=1.011(1)_{m_b}(2)_{m_s}(0.2)_{\kappa}~,
\label{eq:psisu3}
\eea
where $S_3\equiv \bar ss$ is a scalar meson. Then, we obtain in MeV:
\beq
M_{J/\psi S_3}=5112(41)~,~~~~~~~~~M_{\Upsilon S_3}=10125(40)~,
\lb{eq:psis3}
\eeq
corresponding to the $SU(3)$ mass-splittings:
\beq
\Delta M^{J/\psi}_{sd}\simeq  \Delta M^{\Upsilon}_{sd}\approx 110~{\rm MeV}~.
\lb{eq:psis32}
\eeq
The mass-splittings in Eq. (\ref{eq:psis32}) are comparable
with the ones obtained previously.\\
Doing the same exercise for the octet current, we deduce the results in Table \ref{tab:res} where the molecule associated to the octet current is 100 (resp. 250) MeV above the one of the singlet current for $J/\psi$ (resp $\Upsilon$) contrary to the $1^{++}$ case  discussed in \cite{X2}. The ratio of $SU(3)$ breakings are respectively 1.022(5) and 1.010(2) in the $c$ and $b$ channels which are comparable with the ones in Eq. \ref{eq:psisu3}.
 When comparing our results with the ones in Ref. \cite{RAPHAEL2}, 
we notice that the low central value of $M_{J/\psi S_2}$ obtained there (which we reproduce) corresponds to a smaller value of $m_c=1.23$ GeV and mainly to a low value of $\sqrt{t_c}=5.1$ GeV which does not co\"\i ncide with the common solution $\sqrt{t_c}=5.3$ GeV from LSR and FESR. On the opposite, the large value of  $M_{\Upsilon S_2}=10.74$ (resp. 11.09) GeV obtained there corresponds to a  too high value $\sqrt{t_c}=11.3$ (resp. 11.7) GeV compared with the LSR and FESR solution $\sqrt{t_c}=10.23$ (resp. 10.48) GeV for the singlet (resp. octet) current. 

\begin{figure}[hbt] 
\begin{center}
\centerline {\hspace*{-7.cm} a) }\vspace{-0.5cm}
{\includegraphics[height=30mm]{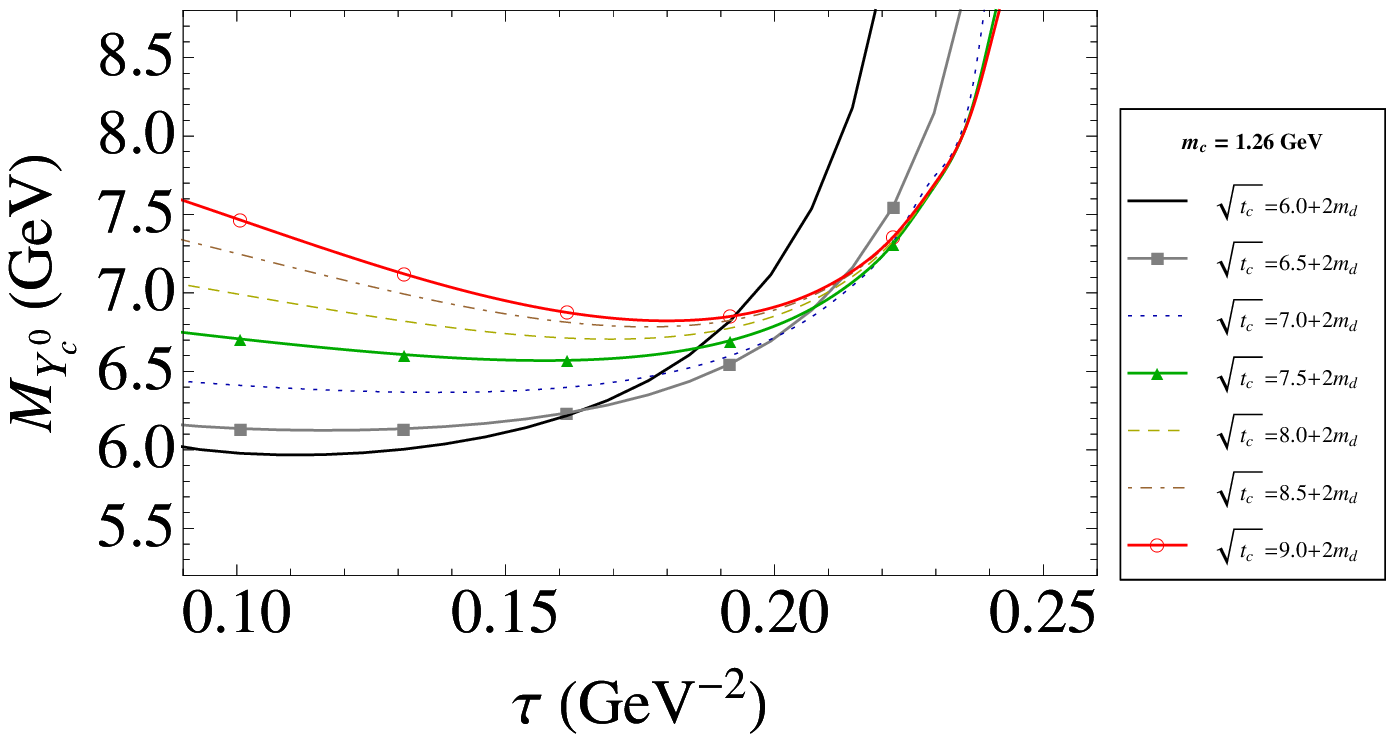}}\\
\centerline {\hspace*{-7.cm} b) }\vspace{-0.5cm}
{\includegraphics[height=30mm]{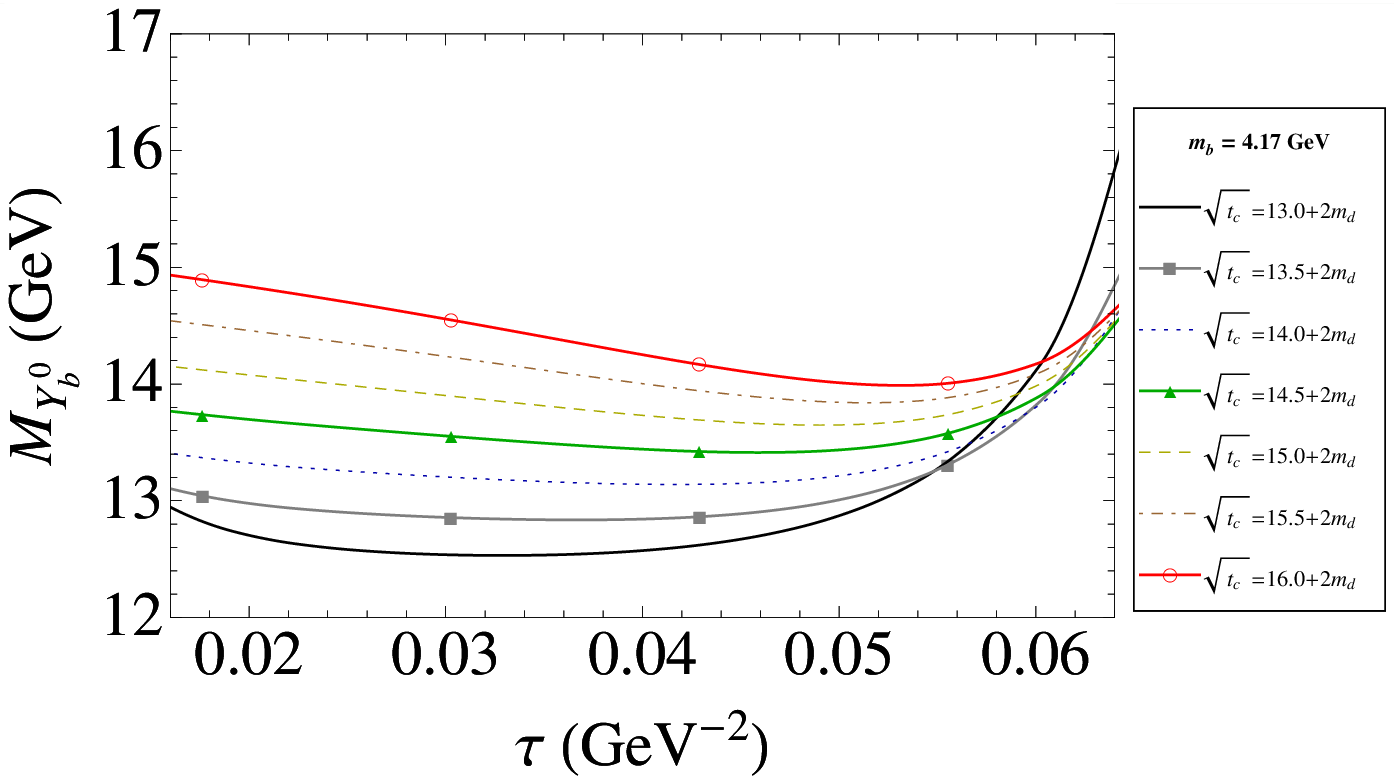}}
\centerline {\hspace*{-7cm} c) }\vspace{-0.5cm}
{\includegraphics[height=30mm]{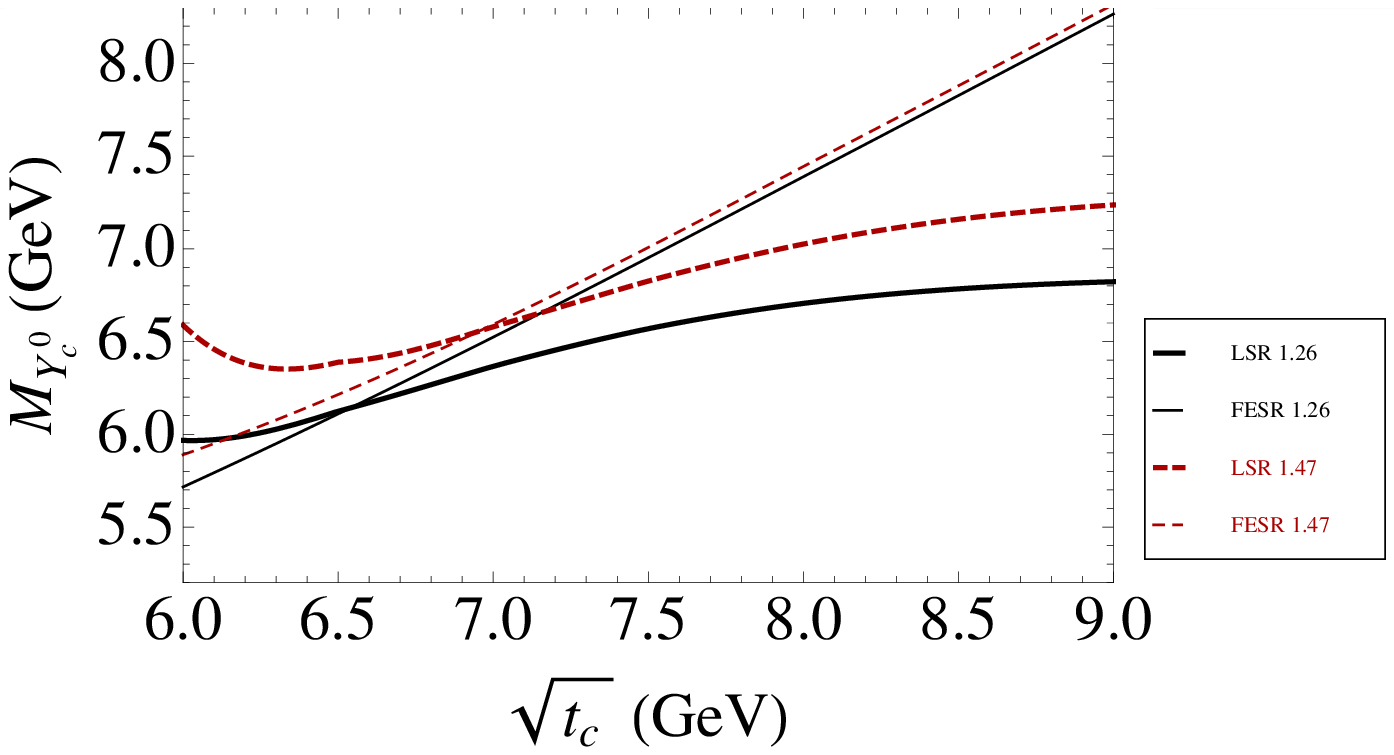}}\\
\centerline {\hspace*{-7cm} d) }\vspace{-0.5cm}
{\includegraphics[height=30mm]{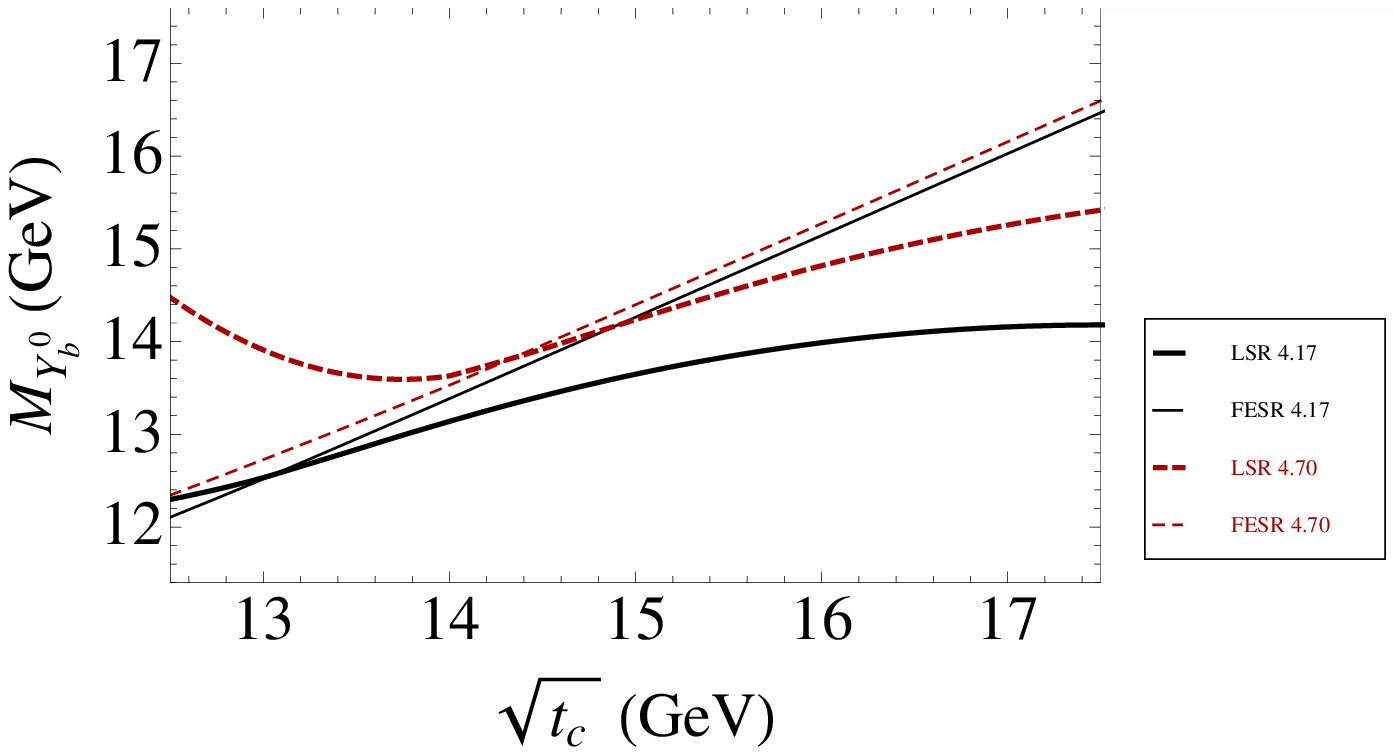}}
\caption{\scriptsize 
{\bf a)} $\tau$-behaviour of ${Y^0_{cd}}$ for the current mixing parameter $b=0$, for different values of $t_c$
and for $m_c=1.26$ GeV
; {\bf b)} $\tau$-behaviour of ${Y^0_{bd}}$  for different values of $t_c$ and for $m_b=4.17$ GeV; {\bf c)} $t_c$-behaviour of the extremas  in $\tau$ of ${Y^0_{cd}}$  for $m_c=1.26-1.47$ GeV; {\bf d)} the same as c) but for ${Y^0_{bd}}$ for $m_b=4.17-4.70$ GeV.}
\label{fig:Y0} 
\label{fig:tcY0} 
\end{center}
\end{figure} 
\begin{figure}[hbt] 
\begin{center}
\centerline {\hspace*{-7cm} a) }\vspace{-0.3cm}
{\includegraphics[height=30mm]{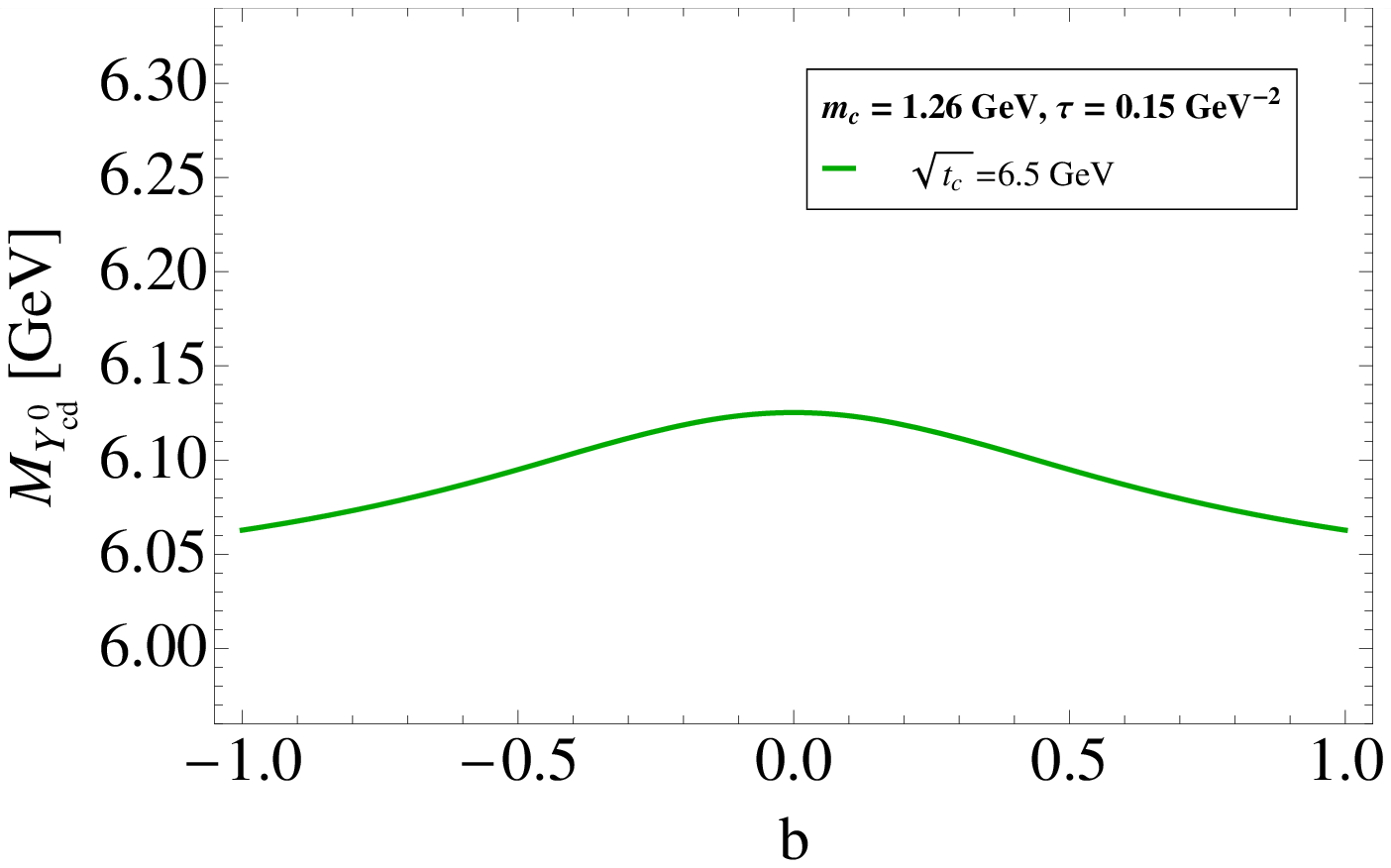}}\\
\centerline {\hspace*{-7cm} b) }\vspace{-0.5cm}
{\includegraphics[height=30mm]{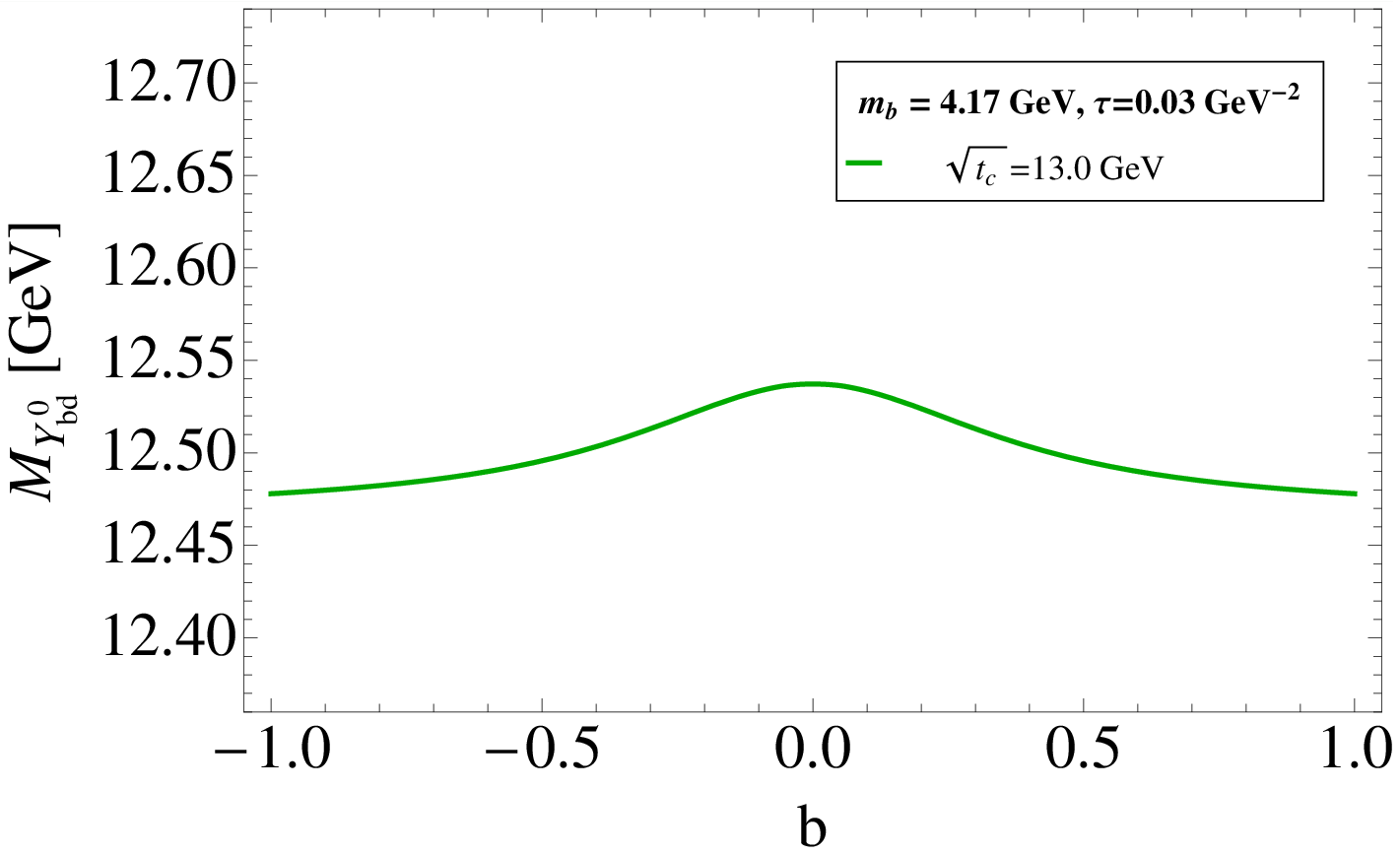}}
\caption{\scriptsize 
{\bf a)} $b$-behaviour of $M_{Y^0_{cd}}$  for given values of $\tau$ and $t_c$ 
and for $m_c=1.26$ GeV; {\bf b)} the same as a) but for $M_{Y^0_{bd}}$
and for $m_b=4.17$ GeV.}
\label{fig:b0} 
\end{center}
\end{figure} 
\nin
\begin{figure}[hbt] 
\begin{center}
\centerline {\hspace*{-7cm} a) }\vspace{-0.5cm}
{\includegraphics[height=30mm]{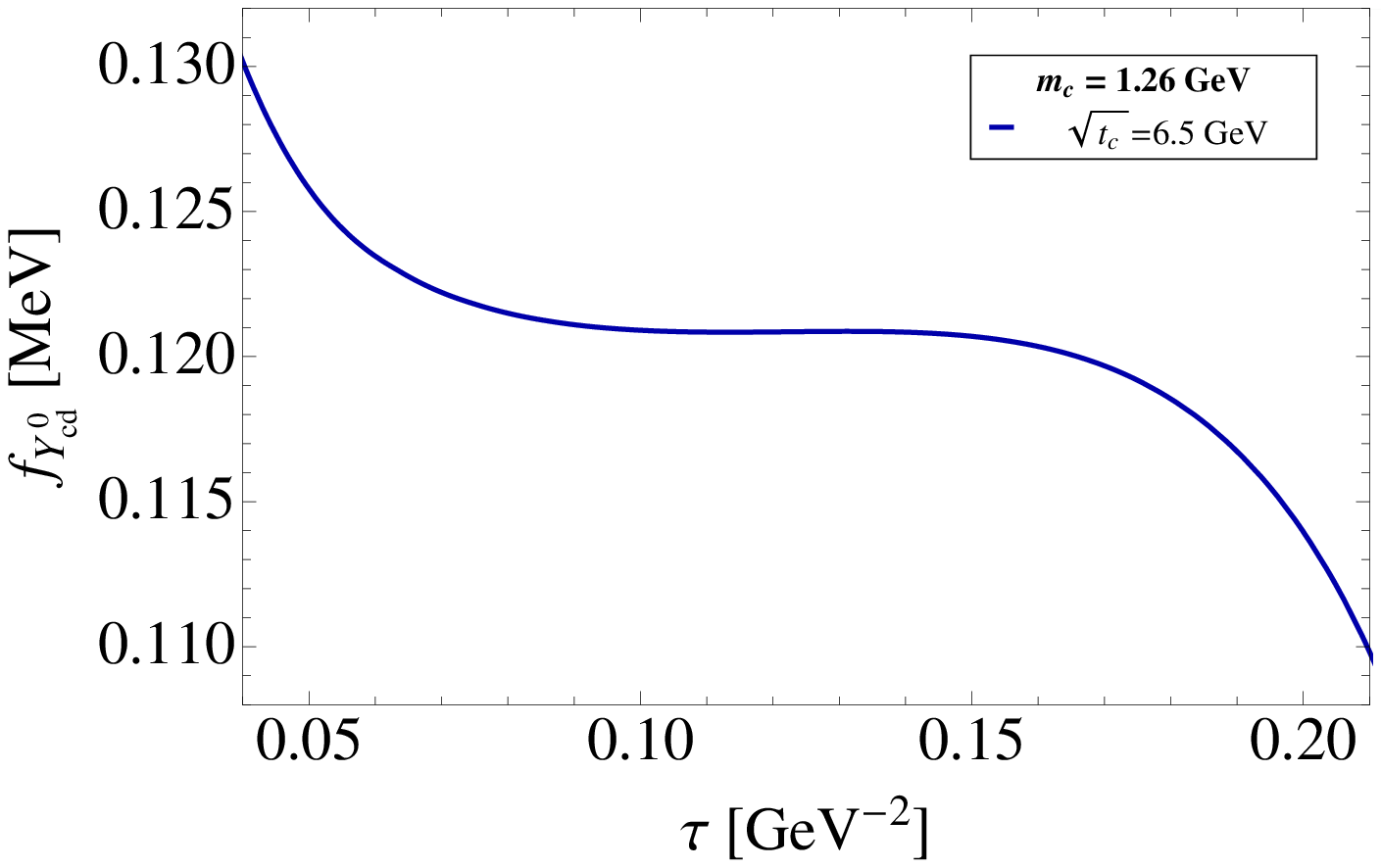}}\\
\centerline {\hspace*{-7cm} b) }\vspace{-0.5cm}
{\includegraphics[height=30mm]{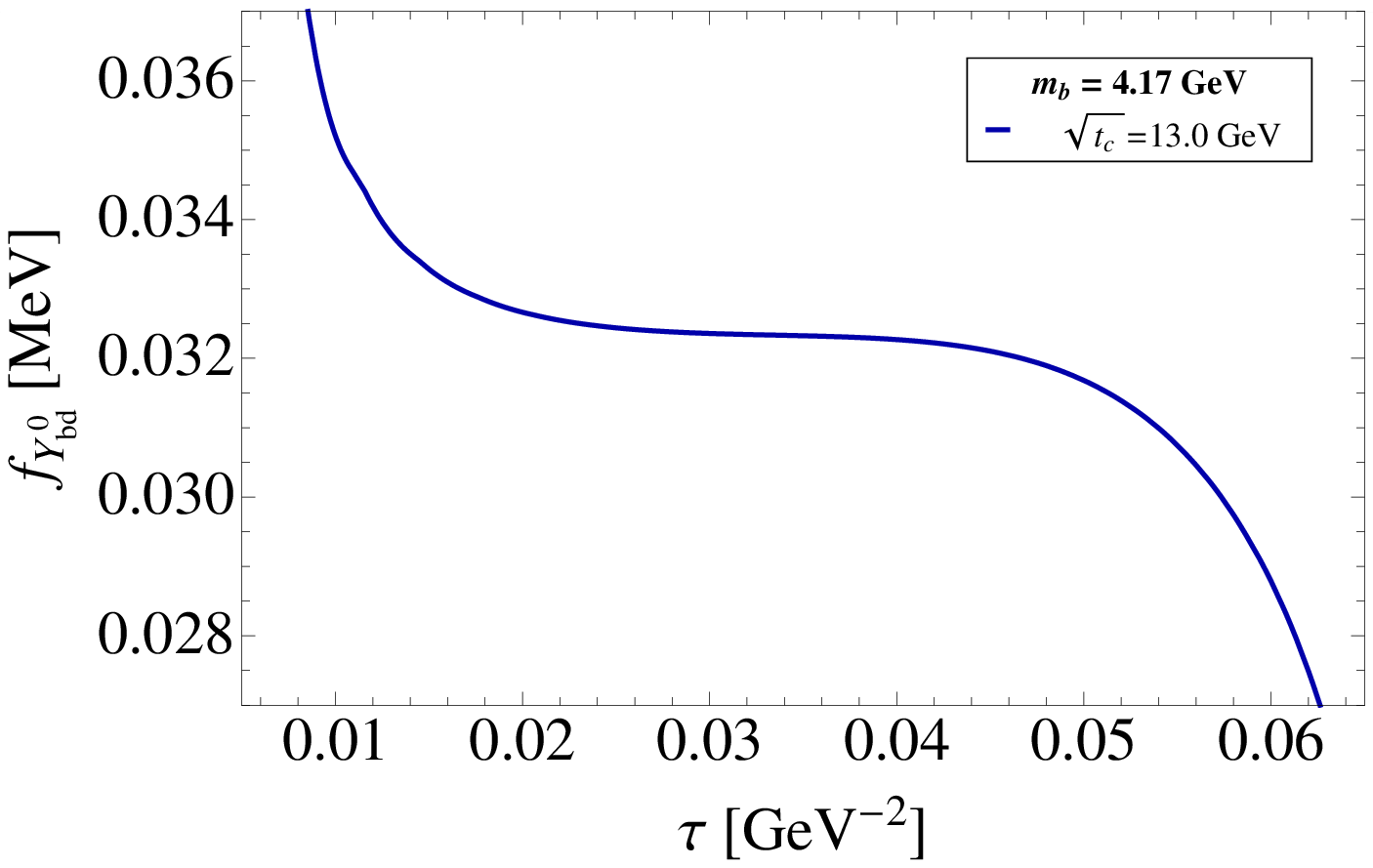}}
\centerline {\hspace*{-7cm} c) }\vspace{-0.4cm}
{\includegraphics[height=30mm]{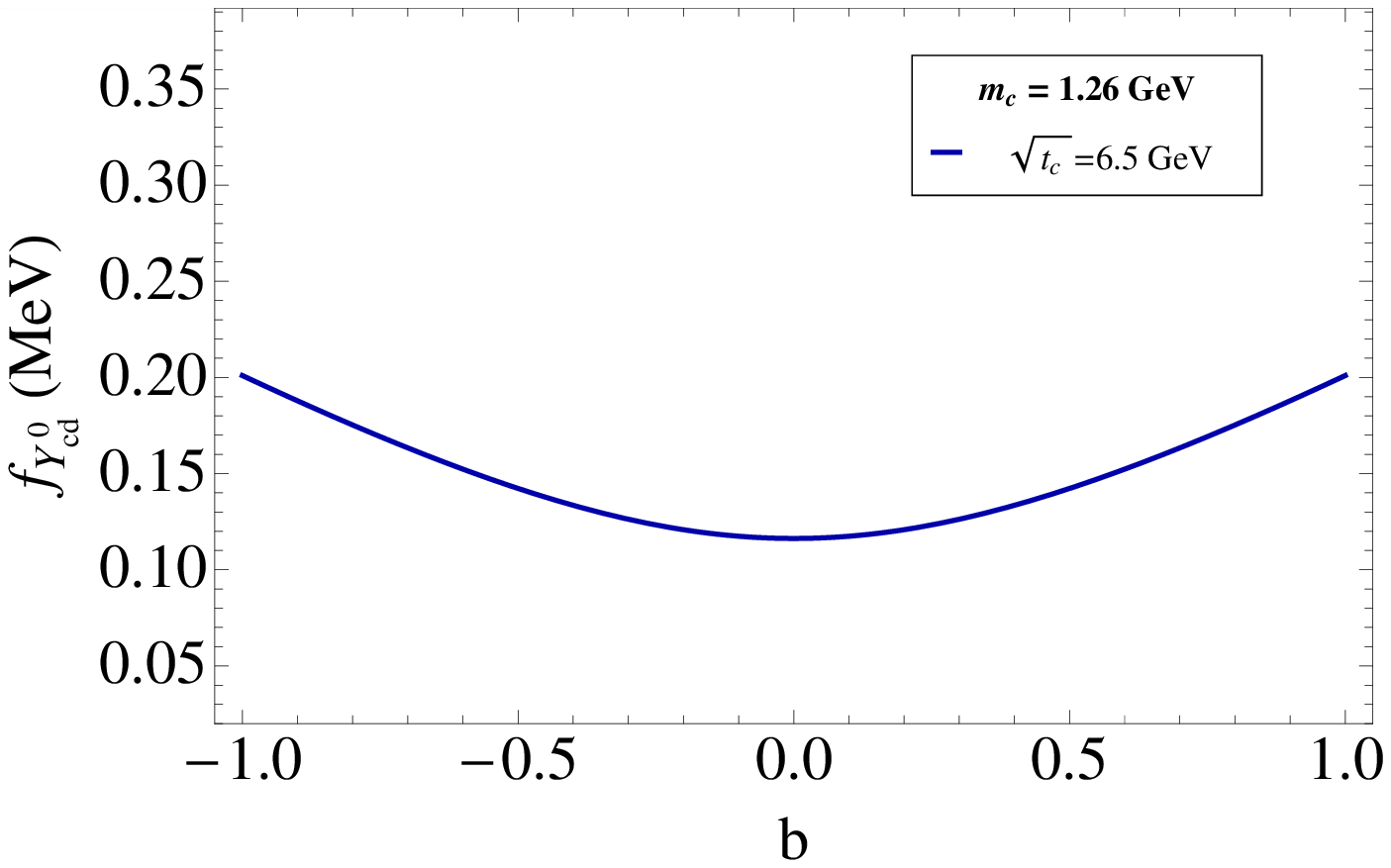}}\\
\centerline {\hspace*{-7cm} d) }\vspace{-0.5cm}
{\includegraphics[height=30mm]{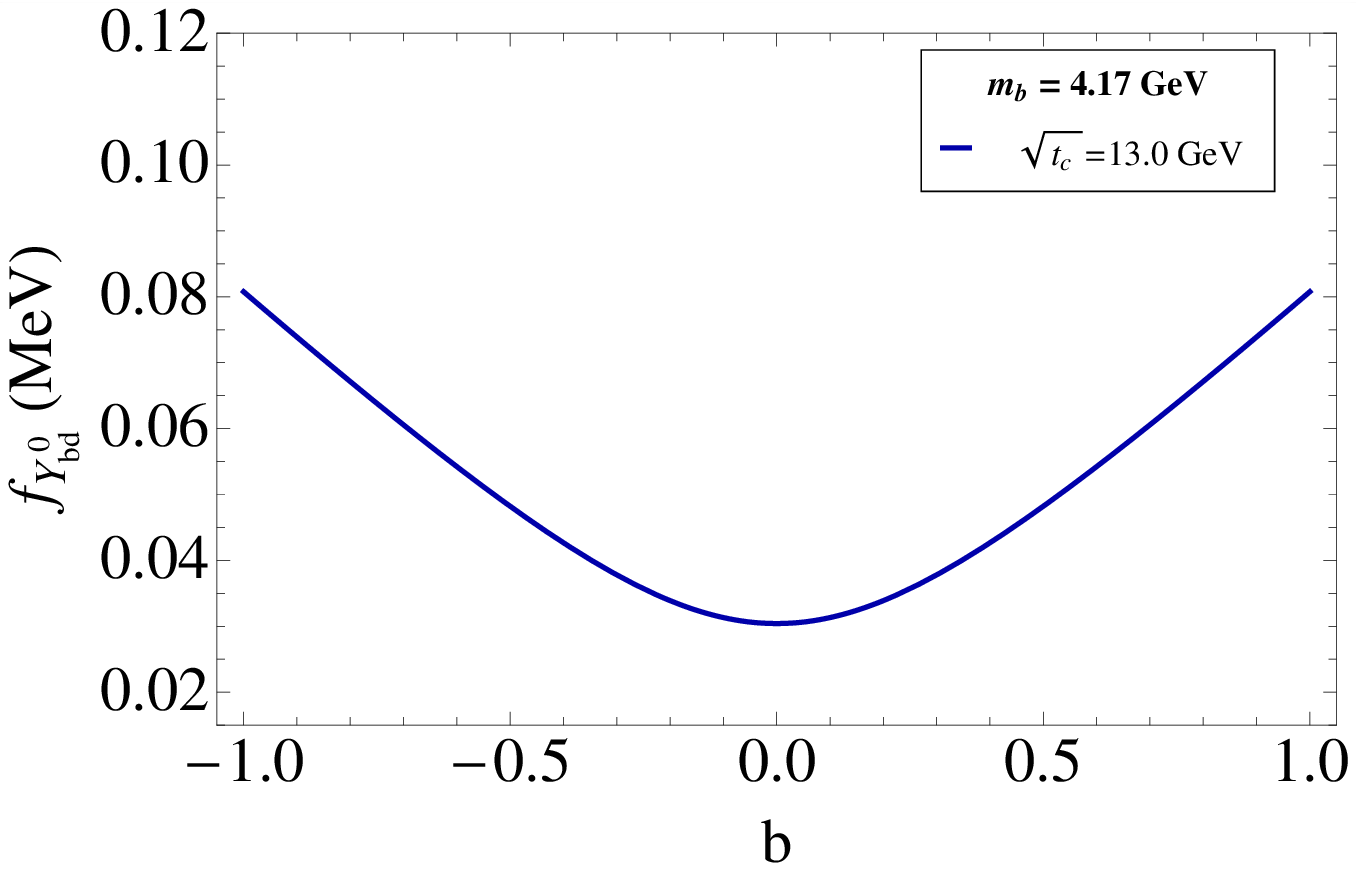}}
\caption{\scriptsize 
{\bf a)} $\tau$-behaviour of $f_{Y^0_{cd}}$  for given values of $b=0$ and $t_c$ 
and for $m_c=1.26$ GeV; {\bf b)} the same as a) but for $f_{Y^0_{bd}}$
and for $m_b=4.17$ GeV; 
{\bf c)} $b$-behaviour of $f_{Y^0_{cd}}$  at the $\tau$-stability and for a given value of $t_c$
 {\bf d)} the same as c) but for $f_{Y^0_{bd}}$
}
\label{fig:fb0} 
\end{center}
\end{figure} 
\vspace*{-0.5cm}
\nin
\begin{figure}[hbt] 
\begin{center}
\centerline {\hspace*{-7cm} a) }\vspace{-0.5cm}
{\includegraphics[height=30mm]{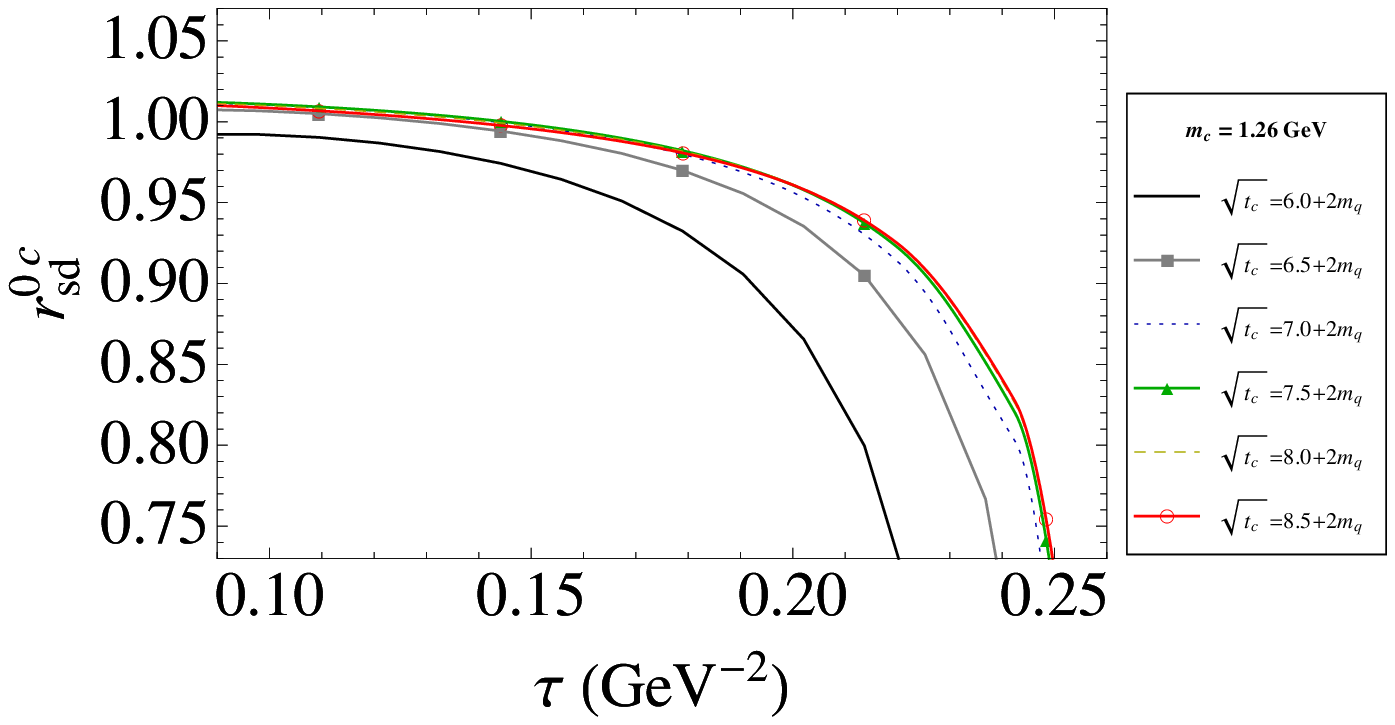}}\\
\centerline {\hspace*{-7cm} b) }\vspace{-0.5cm}
{\includegraphics[height=30mm]{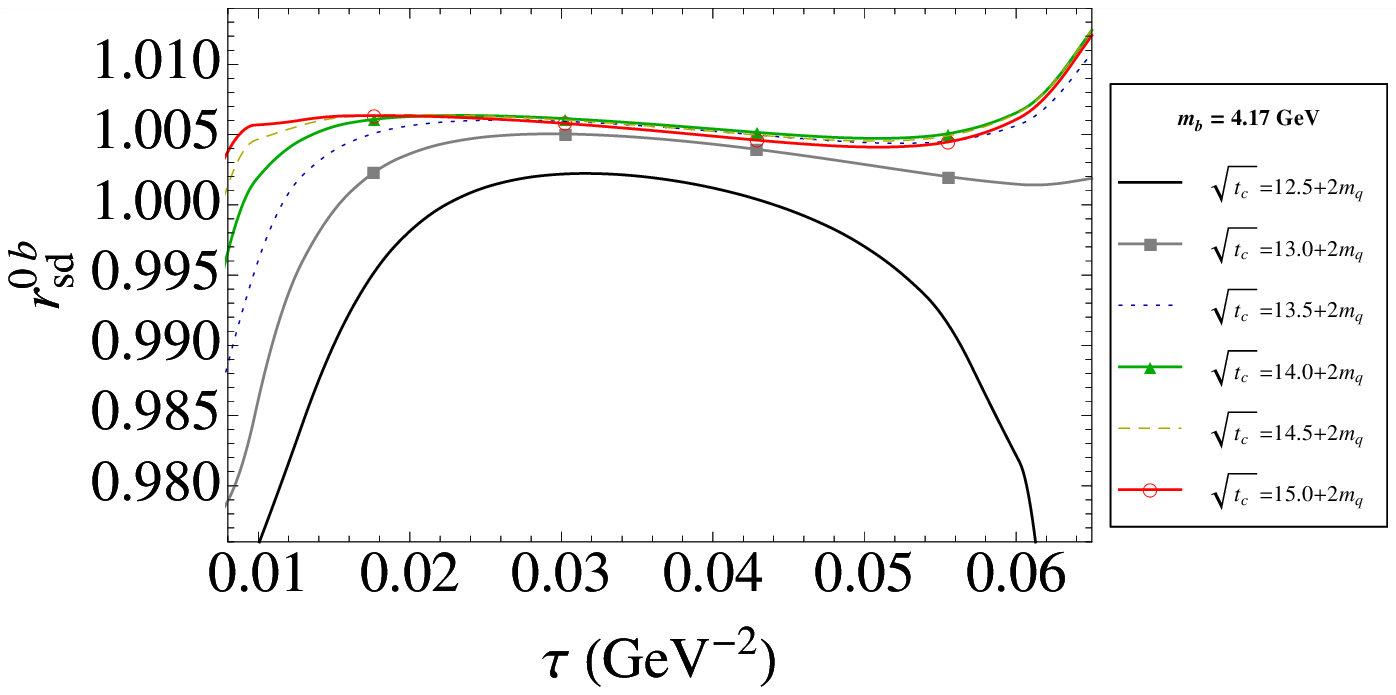}}\\
\centerline {\hspace*{-7cm} c) }\vspace{-0.5cm}
{\includegraphics[height=30mm]{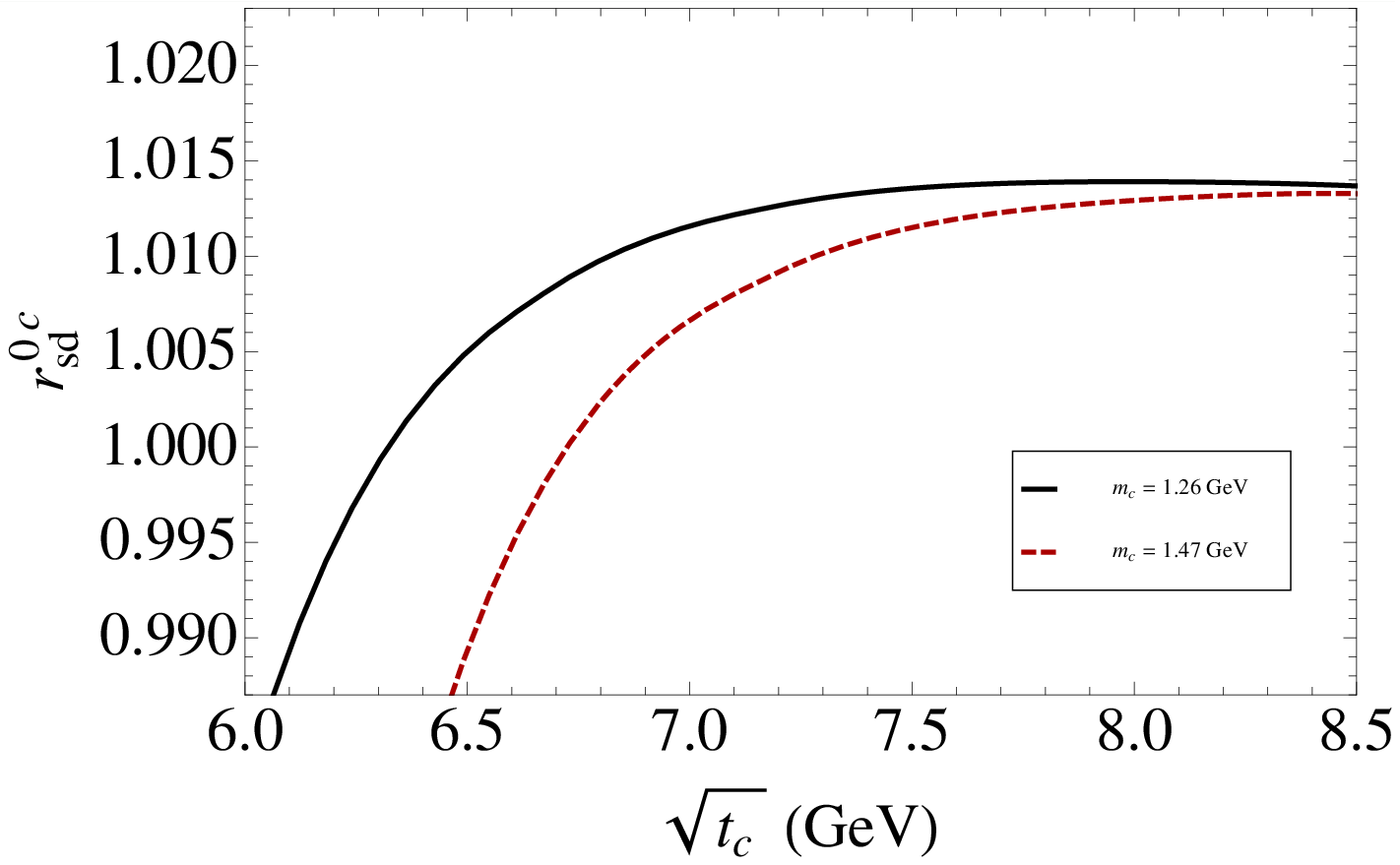}}\\
\centerline {\hspace*{-7cm} d) }\vspace{-0.5cm}
{\includegraphics[height=30mm]{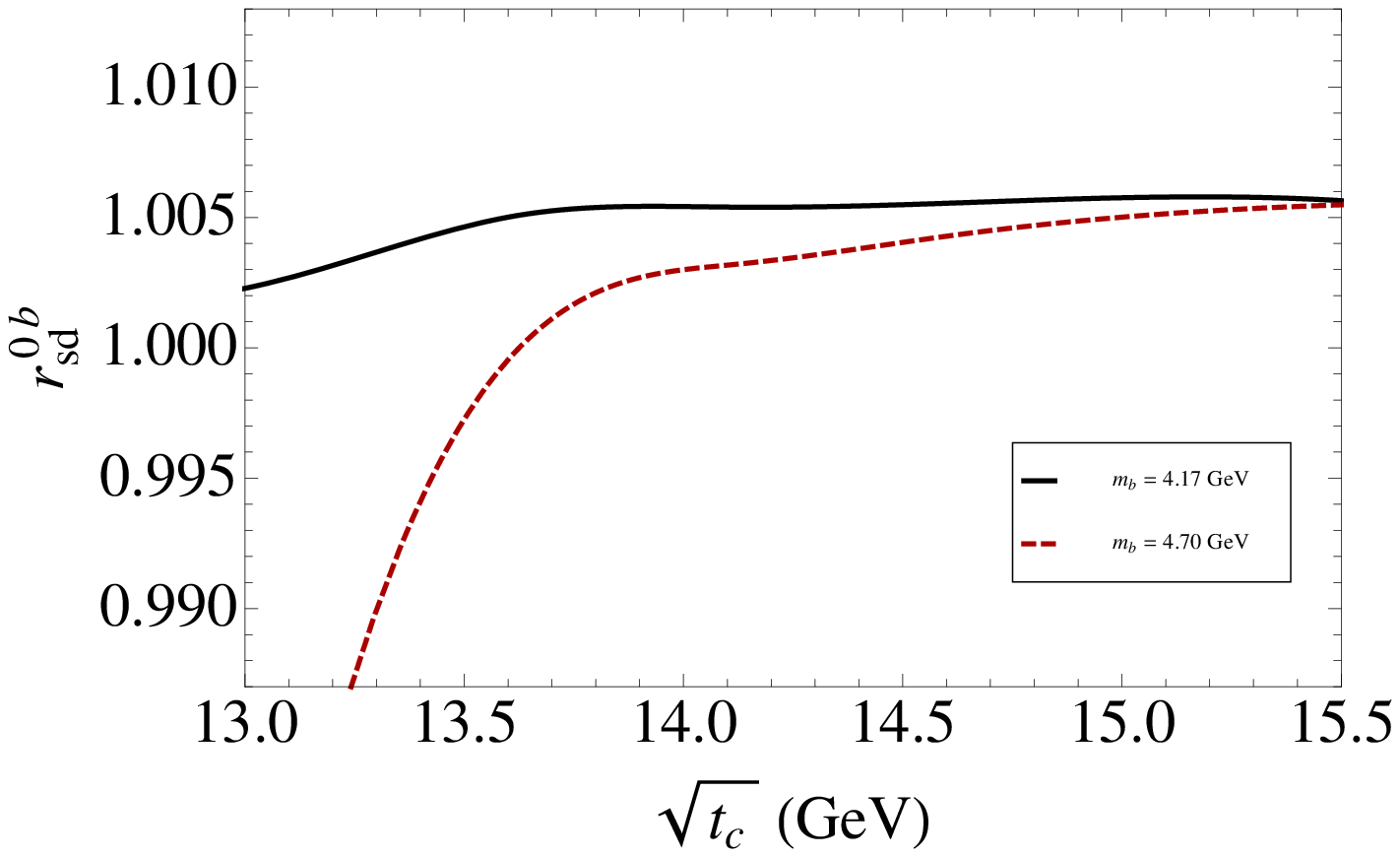}}

\caption{\scriptsize 
{\bf a)} $\tau$-behaviour of $r^{0c}_{sd}$ for the current mixing parameter $b=0$, for different values of $t_c$
and for $m_c=1.26$ GeV; {\bf b)} $\tau$-behaviour of $r^{0b}_{sd}$  for different values of $t_c$ and for $m_b=4.17$ GeV;
{\bf c)} $t_c$-behaviour of the extremas  in $\tau$ of $r^{0c}_{sd}$  for $m_c=1.26-1.47$ GeV; {\bf d)} the same as c) but for $r^{0b}_{sd}$ for $m_b=4.17-4.70$ GeV.}
\label{fig:su3tcY0} 
\label{fig:su3Y0} 
\end{center}
\end{figure} 
\nin
\section{$0^{++}$ four-quark and molecule masses from QSSR}
\nin
In the following, we extend the previous analysis to the case of the $0^{++}$ mesons. 
\begin{figure}[hbt] 
\begin{center}
\centerline {\hspace*{-7cm} a) }\vspace{-0.7cm}
{\includegraphics[height=30mm]{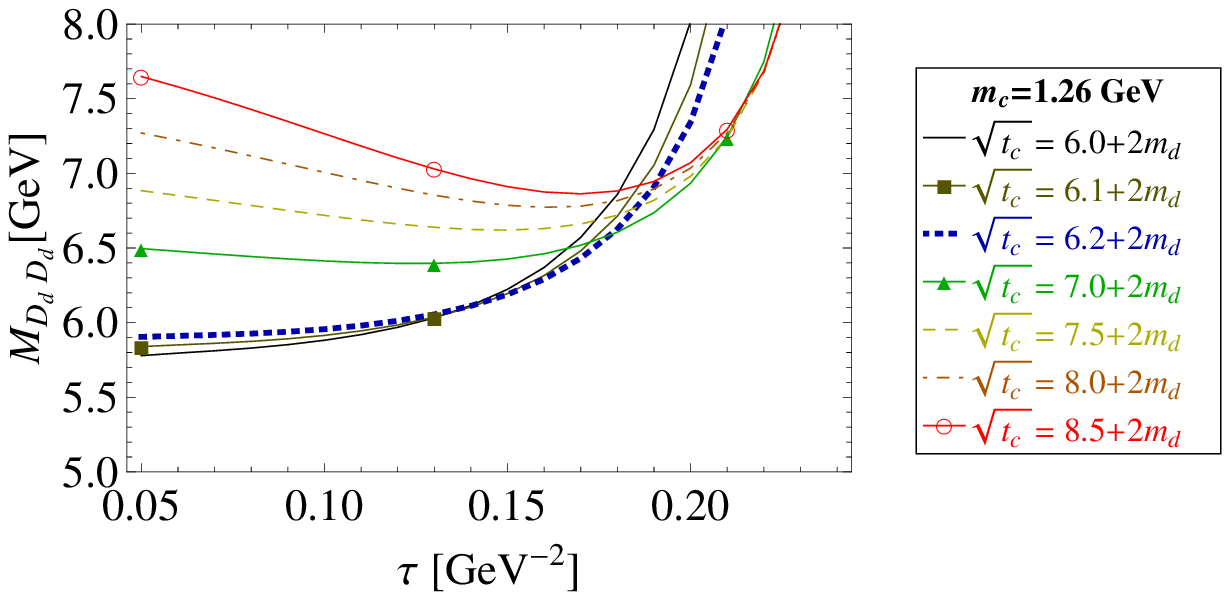}}\\
\centerline {\hspace*{-7cm} b) }\vspace{-0.7cm}
{\includegraphics[height=30mm]{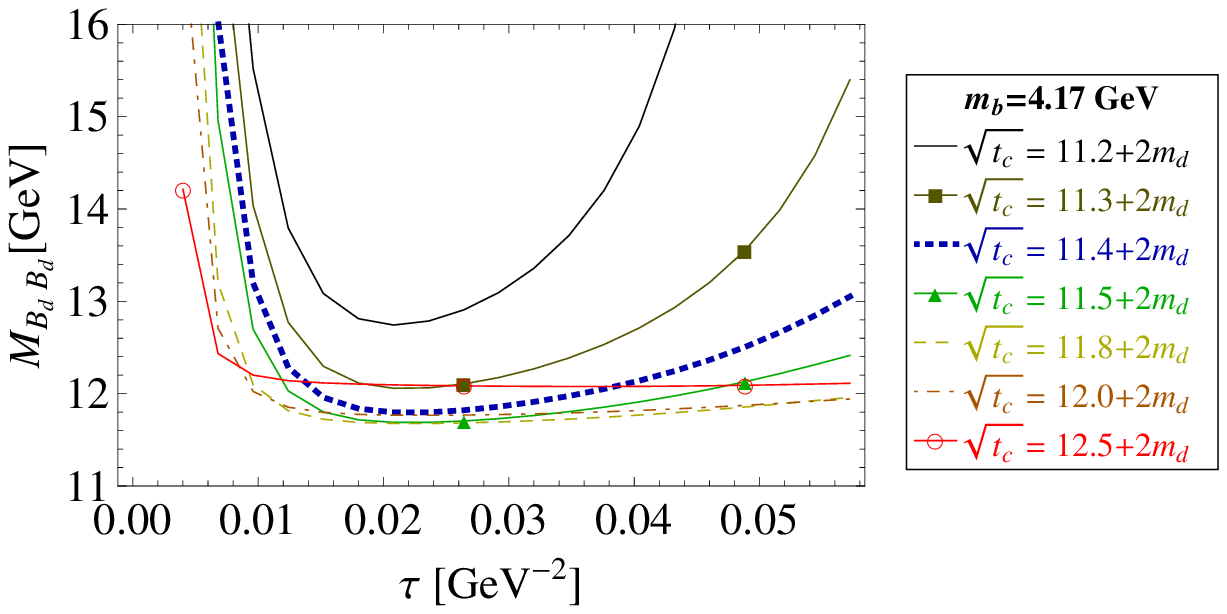}}
\centerline {\hspace*{-7.cm} c) }\vspace{-0.7cm}
{\includegraphics[height=30mm]{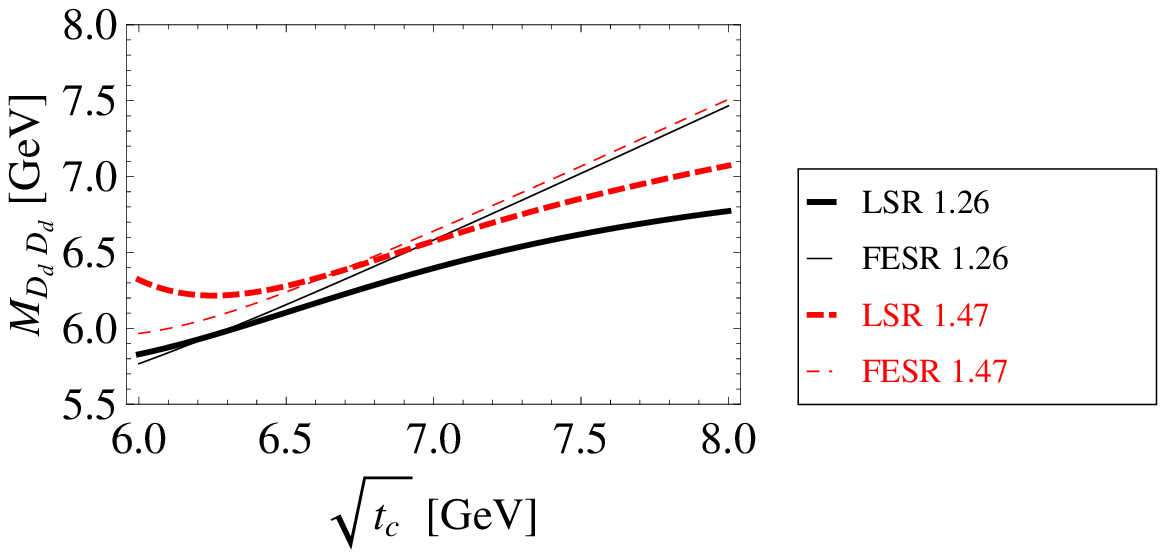}}\\
\centerline {\hspace*{-7.cm} d) }\vspace{-0.7cm}
{\includegraphics[height=30mm]{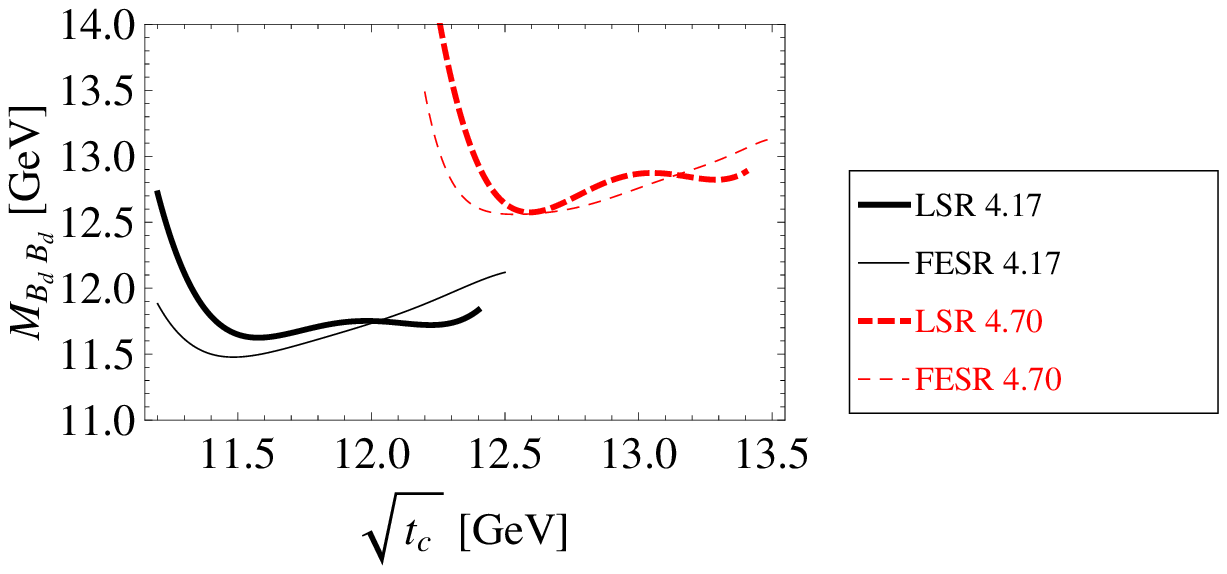}}\\
\caption{\scriptsize 
{\bf a)} $\tau$-behaviour of $M_{D_dD_d}$  for different values of $t_c$
and for $m_c=1.26$ GeV; {\bf b)} $\tau$-behaviour of $M_{B_dB_d}$  for different values of $t_c$ and for $m_b=4.17$ GeV;
{\bf c)} $t_c$-behaviour of the extremas  in $\tau$ of $M_{D_dD_d}$ and for $m_c=1.26-1.47$ GeV; {\bf d)} the same as c) but for $M_{B_dB_d}$ and for $m_b=4.17-4.70$ GeV.}
\label{fig:Dd} 
\label{fig:Dd2} 
\end{center}
\end{figure} 
\nin
\begin{figure}[hbt] 
\begin{center}
\centerline {\hspace*{-7cm} a) }\vspace{-0.7cm}
{\includegraphics[height=30mm]{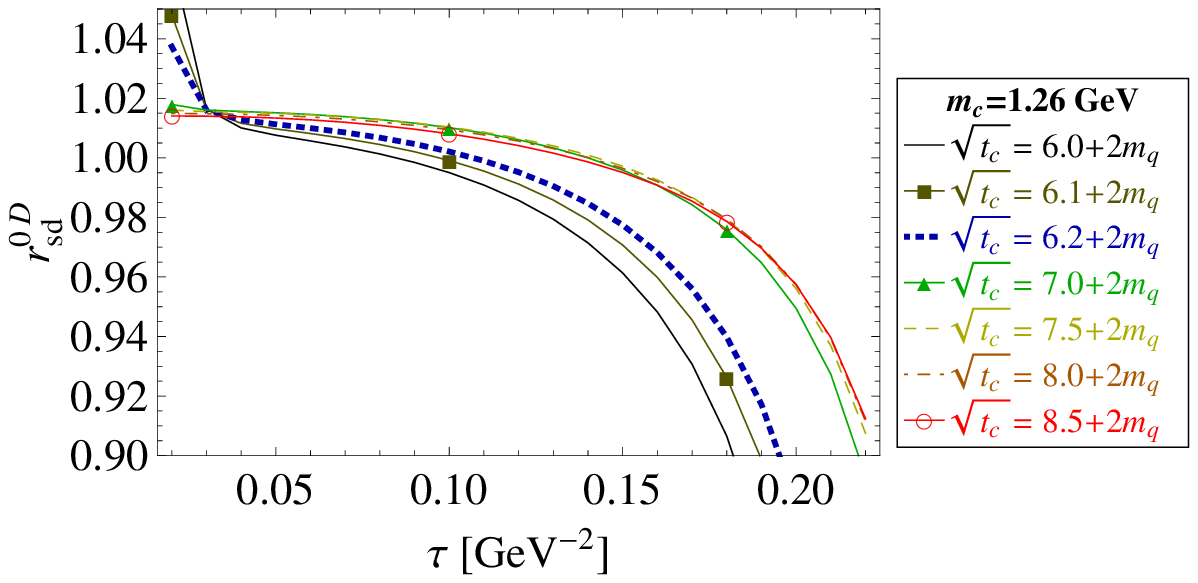}}\\
\centerline {\hspace*{-7cm} b) }\vspace{-0.5cm}
{\includegraphics[height=30mm]{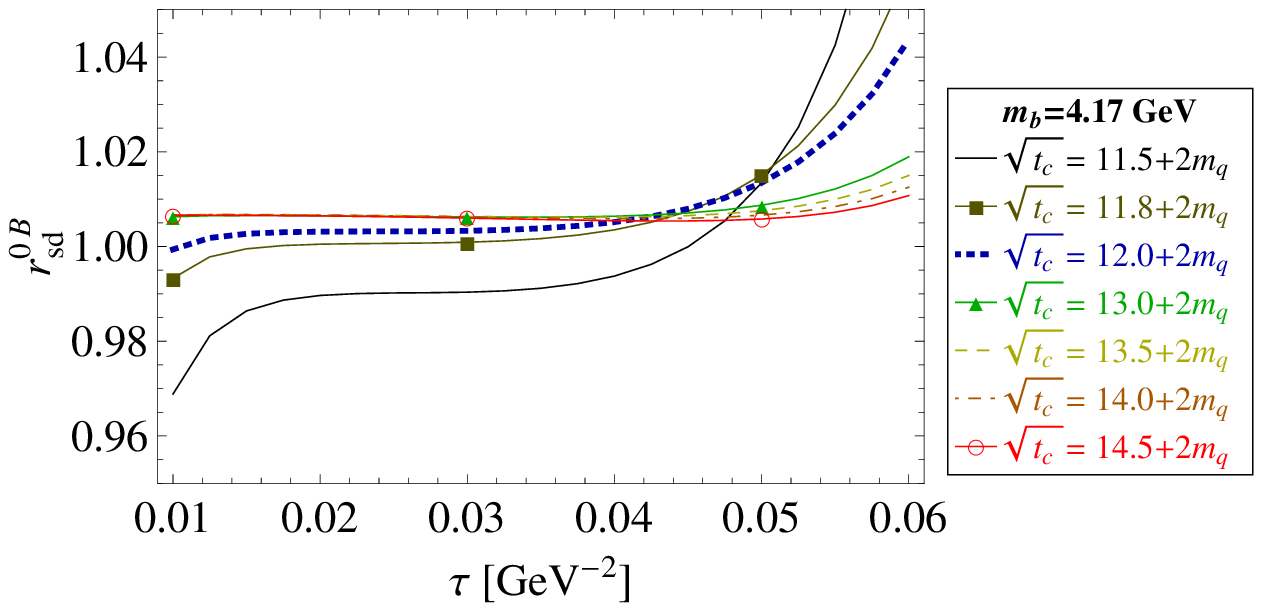}}\\
\centerline {\hspace*{-7cm} c) }\vspace{-.7cm}
{\includegraphics[height=30mm]{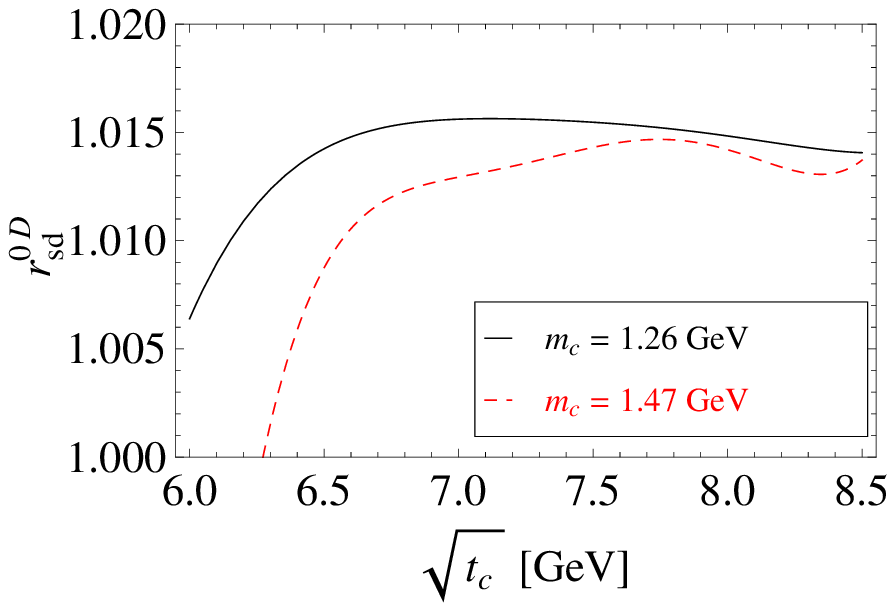}}\\
\centerline {\hspace*{-7cm} d) }\vspace{-0.7cm}
{\includegraphics[height=30mm]{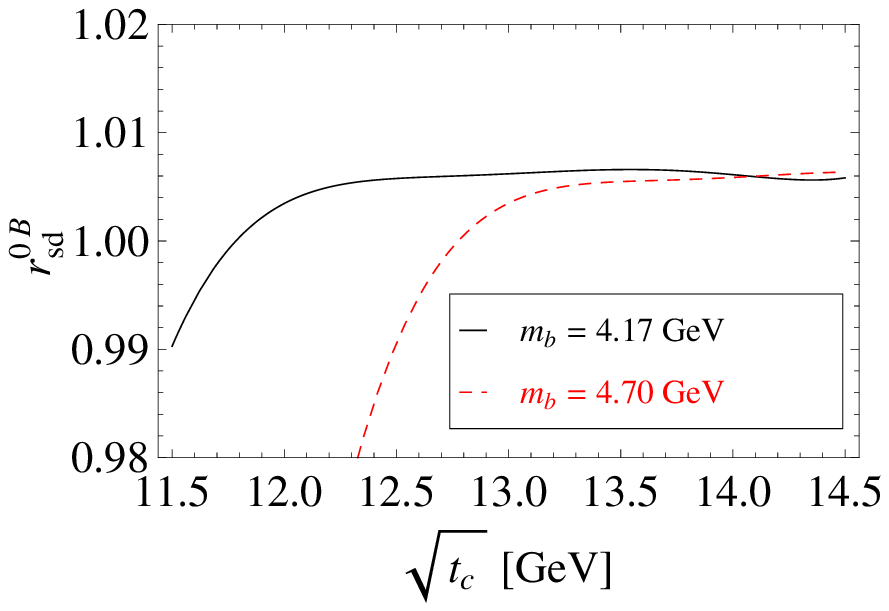}}
\caption{\scriptsize 
{\bf a)} $\tau$-behaviour of $r^D_{sd}$  for different values of $t_c$
and for $m_c=1.26$ GeV; {\bf b)} $\tau$-behaviour of $r^B_{sd}$  for different values of $t_c$
and for $m_b=4.17$ GeV; {\bf c)} $t_c$-behaviour of the inflexion points (or minimas) of $r^D_{sd}$  from Fig \ref{fig:rsd2}a; {\bf d)}the same for the $b$ quark using $r^B_{sd}$  from Fig \ref{fig:rsd2}b.}
\label{fig:rDtc} 
\label{fig:rD} 
\end{center}
\end{figure} 

{\scriptsize
\begin{table}[hbt]
\setlength{\tabcolsep}{0.7pc}
 \caption{
Masses of the four-quark and molecule states from the present analysis combining
Laplace (LSR) and Finite Energy (FESR). We have used double ratios (DRSR) of sum rules
for extracting the $SU(3)$ mass-splittings. The results correspond to the value of the running heavy quark masses but the $SU(3)$ mass-splittings are less affected by such definitions. As already mentioned in the text for simplifying notations, $D$ and $B$ denote the scalar $D^*_0$ and $B^*_0$ mesons. The errors do not take into account the unknown ones from PT corrections. }
    {\small
\begin{tabular}{llll}
&\\
\hline
States&& States &   \\
\hline
{\it Four-quarks}&\it $1^{--}$&&$0^{++}$\\
 $Y_{cd}$&4818(27)&$Y^0_{cd}$&6125(51) \\
$Y_{cs}$&4900(67)&$Y^0_{cs}$&6192(59) \\
$Y_{bd}$&11256(49)& $Y^0_{bd}$&12542(43)\\
$Y_{bs}$&11334(55)&$Y^0_{bs}$&12592(50) \\
{\it Molecules}&$1^{--}$&&$0^{++}$\\
$\bar D^*_dD_d$&5268(24)&$\bar D_dD_d$&5955(48)\\
$\bar D^*_sD_s$&5363(33)&$\bar D_sD_s$&6044(56)\\
$\bar B^*_dB_d$&11302(30)&$\bar B_dB_d$&11750(40)\\
$\bar B^*_sB_s$&11370(40)&$\bar B_sB_s$&11844(50)\\
{\it Singlet current}&$1^{--}$&{\it Octet current}&$1^{--}$\\
$J/\psi S_2$&5002(31)&&5118(29)\\
$J/\psi S_3$&5112(41)&&5231(40)\\
$\Upsilon S_2$&10015(33)&&10268(28)\\
$\Upsilon S_3$&10125(40)&&10371(45)\\
\hline
\end{tabular}
}
\label{tab:res}
\end{table}
}
\nin
\subsection*{\b $Y_{Qd}^0$ mass and decay constant from LSR and FESR}
\nin
We do the analysis  of the $Y_{cd}^0$ and $Y_{bd}^0$ masses using  LSR and FESR. We show the results  in Figs \ref{fig:Y0} for the current mixing parameter $b=0$ from  which we deduce in MeV, for the running quark masses, and respectively for $\sqrt{t_c}=6.5$ and 13.0 GeV where LSR and FESR match:
\bea
M_{Y^0_{cd}}&=& 6125(16)_{m_c}(7)_\Lambda (44)_{\bar uu}(12)_{G^2}(14)_{\rho}\nnb\\
&=&6125(51)~{\rm MeV}~,\nnb\\
M_{Y^0_{bd}}&=& 12542(22)_{t_c}(13)_{m_b}(1)_\Lambda (7)_{\bar uu}(34)_{G^2}(2)_{\rho}\nnb\\
&=&12542(43)~{\rm MeV}~.
\eea
One can notice that the splittings between the lowest ground state and the 1st radial excitation approximately given by $\sqrt{t_c}$ is in MeV:
\beq
M'_{Y^0_{cd}}-M_{Y^0_{cd}}\approx 375~,~~~~~M'_{Y^0_{bd}}-M_{Y^0_{bd}}\simeq 464~,
\eeq
which is larger  than the ones of the $1^{--}$ states, comparable with the ones of the $J/\psi$ and $\Upsilon$, and are (almost) heavy-flavour independent. We show in Fig \ref{fig:b0} the effect of the choice of $b$ operator mixing parameter on the mass predictions, indicating an optimal value at $b=0$. For completeness, we show in Fig. \ref{fig:fb0} the $\tau$ and $b$ behaviours of the decay constants from which we deduce:
\beq
f_{Y^0_{cd}} \simeq 0.12~{\rm MeV}~~~~{\rm and} ~~~~~f_{Y^0_{bd}} \simeq 0.03~{\rm MeV}~,
\label{eq:f0}
\eeq
which are comparable with the ones of the spin 1 case in Eq. (\ref{eq:f1}).
\subsection*{\b $SU(3)$ breaking for $M_{Y_{Qs} }^0$ from DRSR}
\nin
We show in Figs.\,\ref{fig:su3Y0} the $\tau$ and $t_c$ behaviours of the $SU(3)$
breaking ratios for the current mixing parameter $b=0$:
\beq
r^{0Q}_{sd}\equiv {Y^0_{Qs}\over Y^0_{Qd}}~:~~~~Q\equiv c,b~,
\eeq
from which we deduce:
\bea
r^{0c}_{sd}&=&1.011(2)_{m_c}(3.8){m_s}(1.4)_{\kappa}(1)_{\bar uu}(0.7)_{\rho}~,\nnb\\
r^{0b}_{sd}&=&1.004(1)_{m_c}(1.7){m_s}(0.3)_{\kappa}~,
\eea
leading (in units of MeV) to:
\beq
M_{Y^0_{cs}}= 6192(59)~,~~~~~~~~~
M_{Y^0_{bs}}= 12592(50)~,
\eeq
and the $SU(3)$ mass-splittings:
\beq
\Delta M_{sd}^{Y^0_c}\simeq 67\approx \Delta M_{sd}^{Y^0_b}\simeq 50~{\rm MeV}~.
\eeq
\subsection*{\b $M_{D_{d}D_{d}}$ and $M_{B_{d}B_{d}}$ from LSR and FESR}
\nin
We show the $\tau$ and $t_c$ behaviours of the masses $M_{D_{d}D_{d}}$ and $M_{B_{d}B_{d}}$ 
in Figs \ref{fig:Dd}. 
Like in previous sections, we consider as a final result (in units of MeV) the one corresponding to the running masses
for $\sqrt{t_c}=6.25(3)$ and 12.02 GeV:
\bea
M_{D_{d}D_{d}}&=&5955(24)_{t_c}(14)_{m_c}(5)_\Lambda (36)_{\bar uu}(4)_{G^2}(4)_{G^3}(12)_{\rho}\nnb\\
&=&5955(48)~,\nnb\\
M_{B_{d}B_{d}}&=& 11750(12)_{m_b}(4)_\Lambda (35)_{\bar uu}(7)_{G^2}(3)_{G^3}(12)_{\rho}\nnb\\
&=&11750(40)
\eea
One can notice that the splittings between the lowest ground state and the 1st radial excitation approximately given by $\sqrt{t_c}$ is in MeV:
\beq
M'_{D_{d}D_{d}}-M_{D_{d}D_{d}}\approx 290~,~~~~~~M'_{B_{d}B_{d}}-M_{B_{d}B_{d}}\approx 270~,
\eeq
which, like in the case of the $1^{--}$ states are smaller than the ones of the $J/\psi$ and $\Upsilon$, and almost  heavy-flavour independent. 
\subsection*{\b $SU(3)$ breaking for $M_{\bar D_{s}D_{s}}$ and $M_{\bar B_{s}B_{s}}$ from DRSR}
\nin
We show in Fig. \ref{fig:rD} the $\tau$- behaviour of the $SU(3)$ mass ratios for different values of $t_c$
and  the $t_c$ behaviour of their $\tau$-extremas. Therefore, we deduce:
\bea
r^{0D}_{sd}&\equiv &{M_{D_sD_s}\over M_{D_dD_d}}=1.015(1)_{m_c}(4){m_s}(2)_{\kappa}(1)_{\bar uu}(0.5)_{\rho}~,\nnb\\
r^{0B}_{sd}&\equiv& {M_{B_sB_s}\over M_{B_dB_d}}=1.008(1)_{m_c}(4){m_s}(2)_{\kappa}(1)_{\bar uu}(0.5)_{\rho}~.
\eea
Using the previous values of $M_{D_dD_d}$ and $M_{B_dB_d}$, we deduce in MeV:
\beq
M_{D_sD_s}=6044(56)~,~~~~~~~~M_{B_sB_s}=11844(50)~,
\eeq
which corresponds to a $SU(3)$ splitting:
\beq
\Delta M^{DD}_{sd}\approx 89~{\rm MeV} \approx \Delta M^{BB}_{sd}\approx 94~{\rm MeV}~.
\eeq
\section{Summary and conclusions}
We have studied the spectra of the $1^{--}$ and $0^{++}$ four-quarks and molecules states by combining Laplace (LSR) and finite energy (FESR) sum rules. The $SU(3)$ mass-splittings have been obtained using double ratios of sum rules (DRSR). We consider the present results as improvement of the existing ones in the literature extracted only from LSR where the criterion for fixing the value of the continuum thresholds are often ad hoc or based on the ones of the standard charmonium/bottomium systems mass-splittings which are not confirmed by the present analysis. Our  results are summarized in Table \ref{tab:res}. 
We find that :\\
\b The  three $Y_c(4260,~4360,~4660) $   $1^{--}$ experimental candidates are too low for being pure four-quark or/and molecule $\bar DD^*$ and $J/\psi S_2$ states but can result from their mixings.  The $Y_b(10890)$ is lower than the predicted values of the four-quark and $\bar BB^*$ molecule masses but heavier than the predicted $\Upsilon S_2$ and $\Upsilon S_3$ molecule states. Our results may indicate that some other natures (hybrids, threshold effects,...) of these states are not excluded. On can notice that our predictions for the masses are above the corresponding meson-meson thresholds indicating that these exotic states can be weakly bounded. \\
\b For the $1^{--}$, there is a regularity of about (250-300) MeV  for the value of  the mass-splittings between the lowest ground state and the 1st radial excitation roughly approximated by the value of the continuum threshold $\sqrt{t_c}$ at which the LSR and FESR match. These mass-splittings  are  (almost) flavour-independent
and are much smaller than the ones of 500 MeV of ordinary charmonium and bottomium states and do not support some ad hoc choice used in the literature for fixing the $t_c$-values when extracting the optimal results from the LSR. \\
\b  There is also a regularity of about 50--90 MeV for the $SU(3)$ mass-splittings of the different states which are also (almost) flavour-independent. \\
\b The spin 0  states are  much more heavier ($\geq$ 400 MeV) than the spin 1 states, like in the case of hybrid states \cite{SNB}. \\
\b The decay constants of the $1^{--}$ and $0^{++}$ four-quark states obtained in Eqs (\ref{eq:f1})and (\ref{eq:f0}) are much smaller than $f_\pi$, $f_\rho$ and $f_{D,B}$. Unlike $f_B$ expected to behave as $1/\sqrt{M_Q}$, the  four-quark states decay constants  exhibit  a $1/M_Q$ behaviour which can be tested using HQET or/and lattice QCD.\\
It is likely that some other non-perturbative approaches such as potential models, HQET, AdS/QCD and lattice calculations check the previous new features and values on mass-splittings, mass and decay constants derived in this paper. We also expect that present and future experiments (LHCb, Belle, Babar,...) can test our predictions. 
\vspace*{ -.2cm}
\subsection*{Acknowledgment}
\noindent
This work has been partly supported by the CNRS-IN2P3 
within the project Non-Perturbative QCD and Hadron Physics, 
by the CNRS-FAPESP program and by  CNPq-Brazil. S.N. has been partly supported
by the Abdus Salam ICTP-Trieste (Italy)  as an ICTP consultant for Madagascar. We thank the referee for his comments which lead to the improvements of the original manuscript.
\vspace*{ -.2cm}


\begin{thebibliography}{99}


\bibitem{X1} R.D. Matheus, S. Narison, M. Nielsen, J.M. Richard,
{\it Phys. Rev.}  {\bf D75} (2007) 014005; J. M. Dias, S. Narison, F.S. Navarra, M. Nielsen, J. M. Richard,
{\it Phys. Lett.} {\bf  B703} (2011) 274.
\bibitem{X2} 
S. Narison, F.S. Navarra, M. Nielsen, {\it Phys. Rev. } {\bf D83}  
 (2011) 016004.




\bibitem{SWANSON} For reviews, see e.g: E.~S.~Swanson,
 {\it  Phys.\ Rept.\ } {\bf 429}, 243 (2006); J.~M.~Richard, {\it Nucl. Phys (Proc. Suppl.)} {\bf  B164} (2007) 131; 
F. S. Navarra, M. Nielsen, S. H. Lee, 
{\it Phys. Rep. } {\bf 497} (2010) 41; N. Brambilla, et al., {\it Eur. Phys. J.} {\bf C71} (2011) 1534;
S. L. Zhu, {\it Int.  J.  Mod.  Phys.} {\bf  E 17}  (2008) 283.



\bibitem{SVZ} M.A. Shifman, A.I. and Vainshtein and V.I. Zakharov,
{\it Nucl. Phys.} {\bf B147} (1979) 385; M.A. Shifman, A.I. and Vainshtein and V.I. Zakharov,
{\it Nucl. Phys.} {\bf B147} (1979) 448.
\bibitem{SNB} For reviews, see
e.g., S. 
Narison, {\it QCD as a theory of hadrons,
Cambridge Monogr. Part. Phys. Nucl. Phys. Cosmol.} {\bf 17} (2002) 1
[hep-h/0205006]; ibid, {\it QCD
spectral sum rules ,  World Sci. Lect. Notes Phys.} {\bf 26}
(1989) 1;
ibid, { Acta Phys. Pol.} {\bf B26} (1995) 687; 
ibid, { Phys. Rept.} {\bf 84} (1982) 263; ibid, hep-ph/9510270 (1995).

\bibitem{RRY} L.J. Reinders, H. Rubinstein and S. Yazaki, {\it Phys. Rept. }
{\bf 127}  (1985) 1. 


\bibitem{DRSR}S. Narison, {\it Phys. Lett.}  {\bf B210} (1988)  238. 

\bibitem{SNGh} S. Narison, {\it Phys. Lett.} {\bf B387}  (1996) 162; 
ibid, {\it Phys. Lett.} {\bf B605}  (2005) 319;
ibid, {\it Phys. Lett.} {\bf B322}  (1994) 327;
ibid, {\it Phys. Rev.} {\bf D74} (2006) 034013; 
ibid, {\it Phys. Lett.} {\bf B358} (1995) 113; ibid {\it Phys. Lett.} {\bf B466} (1999) 34.
\bibitem{HBARYON} R.M. Albuquerque, S. Narison, {\it Phys. Lett.} {\bf B694} (2010) 217; R.M. Albuquerque, S. Narison, M. Nielsen, {\it Phys. Lett.} 
{\bf B684} (2010) 236.
\bibitem{SNFORM} S. Narison, {\it Phys. Lett.} {\bf B337} (1994) 166; {\it Phys. Lett.} 
{\bf B668} (2008) 308.

\bibitem{BELLEX}  S.-K. Choi  {\it et al.} [Belle Collaboration],   
{\it Phys. Rev. Lett.}  {\bf 91} (2003)  262001.

\bibitem{BABARX}  B.~Aubert {\it et al.}  [{\sc {\sc BaBar}} Collaboration],
 {\it Phys.\ Rev.}  {\bf D71} (2005)  071103.
  
\bibitem{CDF}D. Acosta {\it et al.} [CDF II Collaboration], {\it Phys. Rev. Lett. }
{\bf 93} (2004) 072001.

\bibitem{D0} 
 V.~M.~Abazov {\it et al.}  [D0 Collaboration],
 {\it  Phys.\ Rev.\ Lett.}  {\bf 93} (2004) 162002.
  
   \bibitem{ZHU} W. Chen and S. L. Zhu, {\it Phys. Rev.} {\bf  D83} (2011)  034010.
\bibitem{BABARY}  B.~Aubert {\it et al.}  [{\sc {\sc BaBar}} Collaboration],
	arXiv:0808.1543v2 [hep-ex]; 
 {\it  Phys.\ Rev.}  {\bf D98} (2007) 212001.

\bibitem{BELLEY}  X. L. Wang {\it et al.} [Belle Collaboration],   
{\it Phys. Rev. Lett.}  {\bf 99} (2007) 142002.

\bibitem{RAPHAEL1} R. M. Albuquerque and  M. Nielsen, {\it Nucl. Phys. } 
{\bf A815} (2009) 53; Erratum-ibid. {\bf A857} (2011) 48.
\bibitem{RAPHAEL2} R. M. Albuquerque and  M. Nielsen, {\it Phys. Rev.} {\bf  D84} (2011) 116004.
    
\bibitem{BABAR0}  B.~Aubert {\it et al.}  [{\sc {\sc BaBar}} Collaboration],
 {\it  Phys.\ Rev.\ Lett.}  {\bf 101} (2008) 082001.

\bibitem{BELLE0}  S.-K. Choi {\it et al.} [Belle Collaboration],   
{\it Phys. Rev. Lett.}  {\bf 94} (2005) 182002.

\bibitem{CDF0}T. Aaltonen {\it et al.} [CDF II Collaboration], {\it Phys. Rev. Lett. }
{\bf 102} (2009) 242002.

\bibitem{BELLEYb}  
K. -F. Chen et al. [ Belle Collaboration ], {\it Phys.
Rev.} {\bf  D82} (2010) 091106 .

\bibitem{MANNEL}E. Bagan, M. Chabab, H.G. Dosch, S. Narison, {\it Phys. Lett.} {\bf  B} 301 (1993) 243;
A. Khodjamirian, Ch. Klein, Th. Mannel and Y.-M. Wang, {\it JHEP} {\bf  1109} (2011) 106
 \bibitem{FNR}E.G. Floratos, S. Narison and E. de Rafael, {\it Nucl. Phys.} 
{\bf B155} (1979) 155.
\bibitem{TARRACH} S. Narison and R. Tarrach, {\it Phys. Lett.} {\bf B125} (1983) 217.
\bibitem{KREMER} M. Jamin and M. Kremer, {\it Nucl. Phys.} {\bf B 277} (1986) 349; V. Spiridonov and K.G. Chetyrkin, {\it Sov. J. Nucl. Phys.} {\bf 47} (1988) 522.
 

\bibitem{SNH10}S. Narison,  {\it Phys. Lett.} {\bf B693} (2010)  559; Erratum ibid 705 (2011) 544;
ibid, {\it Phys. Lett.} {\bf B706} (2011)  412; ibid, {\it Phys. Lett.} {\bf B707} (2012)  259. 
\bibitem{SNTAU}S. Narison, {\it Phys. Lett.} {\bf B673} (2009) 30.
\bibitem{BNP}E. Braaten, S. Narison and A. Pich, {\it Nucl. Phys.} {\bf B373} (1992) 581 ; S. Narison and A. Pich, {\it Phys. Lett.} {\bf  B211} (1988) 183.

 \bibitem{SNmass} For reviews, see e.g.: S. Narison, {\it  Phys.Rev.} {\bf D74}
 (2006) 034013; ibid,
arXiv:hep-ph/0202200; ibid, {\it Nucl.Phys.Proc.Suppl.} {\bf 86} (2000) 242; 
ibid, {\it Phys. Lett.} {\bf B216} (1989)  191; ibid, {\it Phys. Lett.} 
{\bf B358} (1995) 113; ibid, {\it Phys. Lett.} {\bf B466} (1999) 345; ibid, {\it Riv. Nuov. Cim.} {\bf 10N2} 
(1987) 1; S. Narison, H.G. 
Dosch, {\it Phys. Lett.} {\bf B417} (1998) 173; 
S. Narison, N. Paver, E. de Rafael and D. Treleani, {\it Nucl. Phys.} {\bf B212} 
 (1983) 365; S. Narison, E. de Rafael, {\it Phys. Lett.} {\bf B103} (1981) 57; C. 
Becchi, S. Narison, E. de Rafael, F.J. Yndurain, {\it Z. Phys.} {\bf C8} (1981)  335.

\bibitem{PDG} K. Nakamura et al. (PDG), {\it Journal Physics} {\bf G37}, 075021 (2010).
\bibitem{SNHmass}S. Narison, {\it Phys. Lett.} {\bf B197}(1987) 405 ; 
ibid, {\it Phys. Lett.} {\bf B341} (1994) 73 ; ibid, {\it Phys. Lett.} {\bf B520}  (2001) 115.
\bibitem{IOFFE} B.L. Ioffe and K.N. Zyablyuk, {\it Eur. Phys. J.} {\bf  C27}
(2003)  229 ; B.L. Ioffe, {\it Prog. Part. Nucl. Phys.} {\bf 56} (2006) 232.
\bibitem{JAMI2}Y. Chung et al.{\it Z. Phys.} {\bf C25} (1984)  151;  H.G. Dosch, 
Non-Perturbative Methods (Montpellier 1985);  
H.G. Dosch, M. Jamin and S. Narison, {\it Phys. Lett.} {\bf B220} (1989)  251.

\bibitem{HEID}B.L. Ioffe, {\it Nucl. Phys.} {\bf B188} (1981)  317; B.L. Ioffe, {\bf
B191} (1981) 591; A.A.Ovchinnikov and A.A.Pivovarov,
{\it Yad.\ Fiz.}  {\bf 48} (1988) 1135.

\bibitem{LNT}G. Launer, S. Narison and R. Tarrach, {\it  Z. Phys.} {\bf C26}
(1984) 433.
\bibitem{SNI}S. Narison, {\it Phys. Lett.} {\bf B300} (1993) 293; ibid {\bf B361} (1995) 121.
\bibitem{fesr} R.A. Bertlmann, G. Launer and E. de Rafael, 
{ Nucl. Phys.} {\bf B250} (1985) 61; R.A. Bertlmann et al., 
{ Z.\ Phys.}  {\bf C39} (1988) 231.
\bibitem{YNDU}F.J. Yndurain, hep-ph/9903457.
\bibitem{SNHeavy}S. Narison, {\it Phys. Lett.} {\bf B387} (1996) 162.
\bibitem{BELL}J.S. Bell and R.A. Bertlmann, {\it Nucl. Phys.} {\bf B227} (1983) 435;
R.A. Bertlmann, {\it Acta Phys. Austriaca} {\bf 53} (1981) 305;  R.A. Bertlmann 
and H. Neufeld, {\it Z. Phys.} {\bf C27} (1985)  437.
\bibitem{SNG} S. Narison, {\it Phys. Lett.} {\bf B361} (1995) 121;
S. Narison,  {\it Phys. Lett.} {\bf B624} (2005) 223; {\it Phys. Lett.} {\bf B520} (2001) 115.


\bibitem{CNZ} K. Chetyrkin, S. Narison and V.I. Zakharov, {\it Nucl. Phys.} 
{\bf B550} (1999)  353;
S. Narison and V.I. Zakharov, {\it  Phys. Lett.} {\bf B522} (2001) 266; 
S. Narison and V.I. Zakharov, {\it Phys. Lett.} {\bf B679} (2009) 355.
\bibitem{ZAK} For reviews, see e.g.: V.I. Zakharov, {\it Nucl. Phys. Proc. Suppl.} 
{\bf 164} (2007) 240; S. Narison,  {\it Nucl. Phys. Proc. Suppl.} {\bf 164} 
 (2007) 225. 
\bibitem{SHIF} M.A. Shifman, hep-ph/0009131.
\bibitem{PERIS}O. Cat\`a, M. Golterman, S. Peris, {\it Phys. Rev.} {\bf D 77} (2008) 0930064.
\bibitem{SNFB}S. Narison, {\it Phys. Lett.} {\bf B198} (1987) 104; 
 \bibitem{VENEZIA} S. Narison and G. Veneziano, {\it Int. J. Mod. Phys.}
 {\bf A4} (1989) 2751; S. Narison, {\it Nucl. Phys.} {\bf B509} (1998) 312.
 \bibitem{MINK} P. Minkowski, W. Ochs, {\it  Eur. Phys. J.} {\bf  C9} (1999) 283; G. Mennessier, S. Narison, W. Ochs, {\it Phys. Lett.} {\bf B665} (2008) 205; 
{\it Nucl. Phys. Proc. Suppl.} {\bf 238} (2008) 181; 
G. Mennessier, P. Minkowski, S. Narison, W. Ochs, HEPMAD 07 Conference,  SLAC Econf  C0709107, {\it arXiv: 0707.4511 [hep-ph] } (2007; R. Kaminski, G. Mennessier and S. Narison, {\it Phys. Lett.} {\bf B680} (2009) 148; G. Mennessier, S. Narison and X.-G. Wang, {\it Phys. Lett.} {\bf B688} (2010) 59;  ibid, {\it Phys. Lett.} {\bf B696} (2011) 40.

\end{thebibliography}
\end{document}